\documentclass[useAMS,usenatbib]{mn2e}

\usepackage{epsfig}

\usepackage{amsmath}

\begin{document}

\title[The structure of the Galactic halo]{The structure of the Galactic halo: SDSS versus SuperCOSMOS}

\author[Y. Xu, L. C. Deng and J. Y. Hu]{Y. Xu$^{1,2}$\thanks{E-mail:xuyan@bao.ac.cn(YX); licai@bao.ac.cn(LCD); hjy@bao.ac.cn(JYH)}, L.C.Deng$^{1}$\footnotemark[1] and J. Y. Hu$^{1}$\footnotemark[1]
\\$^{1}$National Astronomical Observatories, Chinese Academy of Sciences, Beijing 100012, P. R. China\\
$^{2}$Graduate University of Chinese Academy of Sciences, Beijing,
100049, P. R. China}

\pagerange{\pageref{firstpage}--\pageref{lastpage}} \pubyear{2002}

\maketitle

\label{firstpage}

\begin{abstract}

The halo structure at high Galactic latitudes near both the north
and south poles is studied using SDSS and SuperCOSMOS data. For the
south cap halo, the archive of the SuperCOSMOS photographic
photometry sky survey is used. The coincident source rate between
SuperCOSMOS data in $B_J$ band from $16^m.5$ to $20^m.5$ and SDSS
data is about 92\%, in a common sky area in the south. While that in
the $R_F$ band is about 85\% from $16^m.5$ to $19^m.5$. Transformed
to the SuperCOSMOS system and downgraded to the limiting magnitudes
of SuperCOSMOS, the star counts in the northern Galactic cap from
SDSS show up to an $16.9\pm6.3\%$ asymmetric ratio (defined as
relative fluctuations over the rotational symmetry structure) in the
$B_J$ band, and up to $13.5\pm6.7\%$ asymmetric ratio in the $R_F$
band. From SuperCOSMOS $B_J$ and $R_F$ bands, the structure of the
southern Galactic hemisphere does not show the same obvious
asymmetric structures as the northern sky does in both the original
and downgraded SDSS star counts. An axisymmetric halo model with
n=2.8 and q=0.7 can fit the projected number density from
SuperCOSMOS fairly well, with an average error of about 9.17\%. By
careful analysis of the difference of star counts between the
downgraded SDSS northern halo data and SuperCOSMOS southern halo
data, it is shown that no asymmetry can be detected in the south
Galactic cap at the accuracy of SuperCOSMOS, and the Virgo
overdensity is likely a foreign component in the Galactic halo.
\end{abstract}


\section{Introduction}

The modern use of star counts in the study Galatic structure began
with Bahcall \& Soneira (1980). In Bahcall's standard model, the
structure of the Galaxy is assumed to be an exponential disk and a
de Vaucouleurs spheriodal halo. A lot of work has been done to
constrain and examine this theoretical model, as summarised in Xu,
Deng \& Hu (2006) (XDH06 here after), most of them using only a
small sky area. The global structure of the Galactic halo can only
be inferred by different observations of small sky areas with
different magnitude limits, photometric passbands and different
original observational goals. SDSS provides us with the opportunity
to examine the large scale structure of the Galaxy from optical
photometry thanks to its deep photometry and large sky coverage.
From SDSS observational data it is clear that the stellar halo of
the Galaxy is aymmetric, contrary to what has been generally
assumed. From colour star counts it is obvious that the asymmetric
projected stellar number density is produced by halo stars. There
are two possible explanations for such a halo structure.  Firstly,
that there are some large scale star streams embedded in the
axi-symmetric smooth structure of the Galactic halo (Juri\'c et al.
2005). Secondly, that the galactic stellar halo is intrinsically not
axi-symmetric (Newberg \& Yanny 2005; XDH 2006). Based on the data
we have so far, some combination of the two might also be possible .
In Paper I, we tested the second option and fitted the observational
data with triaxial halo. The triaxial halo model fits fairly well
the projected number density near the northern cap of the Galactic
stellar halo. However, in some sky areas, the triaxial halo model
cannot reproduce the actual star counts. The multi-solutions that
are intrinsic in fitting the observational data with the triaxial
halo model make the interpretation of the data somewhat difficult.
On the other hand, the alternative option where the asymmetry of the
halo is caused by large scale star streams also has some problems,
even if the overwhelmingly large Virgo overdensity that covers
nearly a quarter of the northern hemisphere can be explained by a
large scale star stream.  The observed underdensity near Ursa Major
with respect to the assumed axi-symmetric halo still challenges such
a picture. Nevertheless, the conservation of such a huge structure
in the gravitational well of the Galaxy certainly needs to be
verified. Mart\'inez-Delgado et al. (2006) show that the Virgo
overdensity can be reproduced by the dynamical evolution of the Sgr
stream. Assuming a certain structure of the stellar halo (an oblate
ellipse), their numerical simulation can predict an overdensity on a
few hundred square degree scale.

Although it is the most advanced photometric sky survey in terms of
depth and data quality, SDSS does not have good data coverage near
the southern cap of the Galaxy which is, of course, crucial in
understanding the overall structure of the stellar halo. Limited to
the sky coverage of SDSS photometry database, it is probably
premature to draw a firm conclusion on the stellar halo structure.
Assuming that the Galactic stellar halo is non axi-symmetric, and
can be described by a triaxial model, there must be some
corresponding evidence in the southern hemisphere similar to what is
found in XDH06 for the northern cap.  In the axisymmetric halo
model, the maximum star counts should be at longitude $l=0^\circ$
(due to the location of the observer). In the case of a triaxial
halo, however, the maximum projected number density also depends on
a certain parameters of the halo including azimuth angle, axial
ratios and the limiting magnitudes of the observations. In the
simplest case, the plane defined by the primary and the middle axis
of the triaxial halo stays in the Galactic disk, the azimuth angle
is only related with the angle between the primary axis and the
direction of the Galactic center from the Sun, therefore the
expected star counts and asymmetric ratio of northern and southern
sky ought to be mirror symmetric with respect to the Galactic plane,
i.e. what was found in the north cap should also be found in the
south under such a halo model. If the two planes do not overlap, the
situation will be more complicated, but similar results should still
hold.

It is also interesting to examine archived sky survey data that has
the good coverage and reasonable quality in the southern Galactic
halo: the photographic photometry of SuperCOSMOS is ideal for this
purpose. As reviewed by Hambly et al. (2001a), photographic
observations for the Galaxy started in the late nineteenth century.
In the 1930s, the development of Schmidt telescopes with wide fields
of view further advanced photographic surveys. The 1.2-m Palomar
Oschin, 1.0-m ESO and 1.2-m UK Schmidt telescopes finished the
photographic whole sky survey in the last century, such surveys form
a legacy library for examining the structure of the Galaxy. In the
late twentieth century, the photographic plates were eventually
digitized using microdensitometry and digital electronics machines.
There are several major programs to digitize the photographic
plates, of which SuperCOSMOS is one. In Hambly et al. (2001a), a
general overview of these programmes (APM, APS, COSMOS, DSS, PMM,
SuperCOSMOS) is presented. The digitized photographic sky survey of
SuperCOSMOS provides a catalog of three bands, namely blue($B_J$),
red($R_F$) and near-infrared($I_{VN}$), which have deeper detection
limit for the same detection completeness compared to other similar
survey programs (see fig2 of Hambly et al. 2001a). We therefore
adopt the SuperCOSMOS data archive for our present study.

In section 2, the observational data are described and the stellar
source cross identification between the SuperCOSMOS data and the
SDSS data is carried out, and the viability of using SuperCOSMOS
data to study the structure of southern Galactic stellar halo is
discussed. In section 3,downgraded SDSS and SuperCOSMOS
observational of star counts results are presented. In section 4,
the model fits to the SuperCOSMOS star counts are introduced. In
section 5, the SuperCOSMOS observational data and theoretical models
are compared, and SuperCOSMOS southern sky star counts and SDSS
downgraded northern sky star counts also compared and analyzed. In
section 6, the result of star counts is summarized.

\section{The observational data}

\subsection{SuperCOSMOS photometric data}

The SuperCOSMOS Sky Survey is a digitized photography sky survey. It
is described in detail in a series of papers by Hambly and
collaborators (Hambly et al. 2001a; Hambly, Irwin \& MacGillivray
2001b; Hambly, Davenhall \& Irwin 2001c).

The SuperCOSMOS photography atlas of the SuperCOSMOS sky survey
includes blue($B_J$), red($R_F$), and near infrared($I_{VN}$)
passband photometric survies carried out by UK Schmidt survey for
$-90^{\circ} < Dec < +2.5^{\circ}$, ESO Red Survey of $-90^{\circ} <
Dec < -17.5^{\circ}$, and Palomar surveies including, POSS-I Red
Survey for $-20.5^{\circ} < Dec < +2.5^{\circ}$, POSS-II Blue Survey
for $-2.5^{\circ} < Dec < +90.0^{\circ}$, POSS-II Red Survey for
$-2.5^{\circ} < Dec < +90.0^{\circ}$. Data of $B_J$ band has 90\% or
about detection completeness from $16^m.5$ to $20^m.5$, and that of
$R_F$ band has same completeness from $16^m.5$ to $19^m.5$ mag. The
photometric data has a magnitude error of $0^m.3$, but color is
accurate to about $0^m.16$(Hambly et al. 2001a).

There are two interface applications of SuperCOSMOS: the SuperCOSMOS
Sky Survey (http:// www-wfau.roe.ac.uk/sss, SSS hereafter) and the
SuperCOSMOS Sky Archive (http:// surveys.roe.ac.uk/ssa, SSA
hereafter).

Images of small sky areas and catalogs from the SuperCOSMOS sky
survey can be downloaded. We thank the SuperCOSMOS working group who
made all the data available to the community. The SSA only includes
photometric data from UKST and ESO. As made clear by Hambly et
al.(2001a), although the entire sky is digitized, the data in this
archive is released progressively.
The total amount of data is enormous, only F type stars (selected by
0.504 $\le B_J-R_F \le$ 0.8236 ) from $20^m.4$ to $20^m.415$ are
adopted to show the sky coverage which is in the upper panel of
figure 1. The survey covers most of the high latitude southern sky,
and a little of the northern hemisphere. The clump at
(l,b)=($302.616^{\circ},-44.580^{\circ}$) is the SMC, and the clump
at (l,b)=($280.085^{\circ},-30.430^{\circ}$) is the LMC.

Except for the difference in sky coverage, the two interface
applications use different selection standards.  The SSA SQL
selection is much more configurable (private communication by email
with Hambly) than that of the SSS. For example, there are 4 kinds of
B magnitude in the SSA, namely classMagB (B band magnitude selected
by B image class), gCorMagB (B band magnitude assuming the object is
a galaxy), sCorMagB(B band magnitude assuming the object is a star),
classB (image classification from B band detection). The most
appropriate attribute for point sources is sCorMag, while the most
possible class of an object from all three bands is provided by
parameter ``meanclass''. The SSS only includes selection parameters
applied to the primary passband, coresponding to classB of the SSA
in the example. Our aim is to count the stars in each selected sky
area, and using classB will lose some stars due to not synthesizing
information of all the three bands. This will influence the result
of star counts seriously. So ``sCorMagB" of the SSA data is selected
to carry out the study and the ``meanclass" is limited to 2(star
label). Because the SSA only covers limited sky areas of high
latitude northern galactic hemisphere (upper panel of figure 1) we
cannot directly compare SuperCOSMOS star counts of northern sky with
those of SDSS.

The SSA includes $R_F$ band data from both UKST and ESO. However
data from the $R_F$ band of UKST is deeper than that of ESO.
Therefore, only UKST is adopted. The detailed instrumental
specifications of UKST can be found
 in Cannon (1984), the main parameters of the survey telescope and instruments
are listed in table 1.

\begin{table*}

 \centering

 \begin{minipage}{140mm}

\caption{Parameters of telescope of UKST}\label{tab1}

\begin{tabular}{rr}

  \hline

site & Siding Spring Mountain, -31$^{\circ}$S\\

aperture & 1.24m \\

focal, focal ratio & 3.07m, f/2.5\\

photographic plates & Kodak IIIa-J emulsion, 356$mm^2$, 67.1 arcsec $mm^{-1}$, 6.5$^{\circ}\times$6.5$^{\circ}$ \\

primary pointing accuracy & $\pm$6 arcsec r.m.s.\\

\hline

\end{tabular}

\end{minipage}

\end{table*}

As demonstrated in the upper panel of figure 1, the UKST atlas of
SuperCOSMOS covers most of the high Galactic latitude southern
hemisphere.  The structure of the Galactic halo near the southern
cap can be studied using a stellar photometry catalog selected in a
similar way as we did for the northern sky in XDH06, shown here in
the lower panel of Fig 1. The selected sky area for this work is
shown in lower panel of figure 1, the Lambert projection of southern
hemisphere. Each of the squares represents a rectangular sky area of
about $2^{\circ}\times2^{\circ}$. Some of the selected sky areas may
be trimmed if sitting near the survey's edge, or the region is
masked by contaminats such as saturated bright stars, or clumps such
as the dwarf galaxy IC1613 in ($130^{\circ},-60^{\circ}$). The first
group of sky areas are along a circle of $b=-60^{\circ}$, equally
spaced by $10^{\circ}$. The other 12 groups are a selection of sky
areas along longitudinal directions equally spaced by $30^{\circ}$.
At a given longitude, the sky areas are selected by a step of
$5^{\circ}$. This selection of sky areas can evenly cover the
southern Galactic cap, so that the global structure of the halo near
the southern pole can be examined.

\subsection{Cross checking of SuperCOSMOS and SDSS data sets}

In our previous work (XDH06), SDSS data is used to study the
structure of the Galactic stellar halo near the Northern Galactic
pole, The SDSS catalog providing a uniform and accurate photometric
data set. The five broadband filters, u,g,r,i,z are 95\% complete to
$22^m.0, 22^m.2, 22^m.2, 21^m.3, 20^m.5$ respectively, and the
uncertainty in the photometry is about 3\% at g=$19^m$(Chen et al.
2001).

Compared to the high-quality photometry data of SDSS, the
SuperCOSMOS data has a narrower dynamic range, lower magnitude limit
and larger photometric error, due to photographic photometry. To
evaluate any uncertainties due to misclassifications and the
relatively less accurate photometry of SuperCOMOS, a comparison in
areas common to both surveys is needed.

The photometric calibration between SDSS and SuperCOSMOS has been
made available by the 2dF Galaxy Redshift Survey(2DFGRS) Final Data
Release Photometric Calibration which defines a set of color
equations in its final data
\footnote{http://magnum.anu.edu.au/$\sim$TDFgg/Public/Release/PhotCat/
photcalib.html}.  The $B_J$ band is correlated with SDSS g and r
band, $B_J=0.15+0.13\times(g-r)$, while $R_F$ is very similar to
SDSS r band, $R_F$=r-0.13. The results of such color calibration are
shown in figure 2 .

The two small sky areas with superpositions of SDSS and SuperCOSMOS
surveys in both the northern and southern sky are chosen to examine
the color equations and the classification of SuperCOSMOS objects.
The northern area is located at ($l,b$)=($280^{\circ},60^{\circ}$)
with $2^{\circ}\times2^{\circ}$ field of view (FOV), the southern
area is at ($l,b$)=($62^{\circ},-59^{\circ}$) with
$1^{\circ}\times4^{\circ}$ FOV. The equinox of SuperCOSMOS data
associated with the photometric image library is J2000.0. The
position accuracy of SuperCOSMOS is $\pm$0.2arcsec at $B_J=
19^m$,$R_F=18^m$, $\pm$0.3arcsec at $B_J=22^m$,$R_F=21^m$ (Hambly et
al 2001c). Taking into account proper motion, cross-identification
is carried out between SDSS and SuperCOSMOS in the two superimposed
sky areas in a identification criterion box of $0^m.3$ and 10
arcsec. In such a box, multiple sources can be present, the pair of
stars with the nearest coordinates and magnitudes are identified as
the same source. We take the SDSS data as the ``true'' values of
both position and magnitude. Based on the matched star list in the
two areas, uncertainties in the magnitude of SuperCOSMOS photometry
for each object can be measured. The systematic error calculated
this way infers  the error of the color equations from 2DFGRS
calibration; while the scatter can be used to measure the error in
SuperCOSMOS photometric data. Fitting the systematic error with a
2nd order polynomial, the color equations are refined. Using the
modified color equations, $B_J$ and $R_F$ magnitude of SDSS data is
defined as $B_{JSDSS}$=g+0.15+0.13$\times$(g-r)+ $\bigtriangleup
{mod}$, $R_{FSDSS}$=r-0.13+$\bigtriangleup {mod}$. Iterating the
cross-identification procedure reduces the systematic error. The
error in the SuperCOSMOS data in the $B_J$ band is found to be
$\epsilon_{B_J}$ = $B_J$ - $B_{JSDSS}$, and that in $R_F$
$\epsilon_{R_J}$ = $R_F$ - $R_{FSDSS}$. The variance of the errors
as functions of magnitude is obtained from fitting the scatter with
a gaussian.

After such modification, and repeating the cross-identification, the
source matching ratios between the two surveys are improved. In the
end, the SuperCOSMOS data matches that of  SDSS in the $B_J$ band
magnitude limits by 92-93\% in a 10 arcsec and $0^m.3$ box. For the
$R_F$ band, the matching ratio can be raised to 85\% or larger in
the $16^m.5-19^m.5$ magnitude range. The matching ratio of the $R_F$
band is not as good as that of the $B_J$ band, this is likely due to
the lower sensitivity in the $R_F$ band. A 85\% is still lower than
the intrinsic completeness estimated for different surveys in the
SuperCOSMOS atlas (see figure 12b of Hambly et al. 2001b). This is
possibly caused by the brighter magnitude limit of the SSA compared
to that of SDSS as the bright stars are saturated, therefore
influencin more neighbours.

In the upper and lower panel of figures 3, the contours in the
color-color diagram represent the SuperCOSMOS data in the three
bands that are cross-identified in the SDSS data. Black points
over-plotted on the contours are the matched stellar sources, while
the crosses represent SuperCOSMOS sources which are unmatched.

\section{Observational star counts}

\subsection{Star counts from downgraded SDSS data}

The examination of halo structure through star counts depends
critically on the depth of the photometry. The SSA has a narrower
dynamic range and shallower detection limit than SDSS. A test is
carried out to check if the asymmetric structure found in XDH06 is
still present with the shallower limit of SSA data. The data used in
XDH06 is downgraded by applying the SuperCOSMOS magnitude limits,
photometric errors of SuperCOSMOS are also added to the SDSS data. A
Monte-Carlo method is used to reproduce the photometric errors as of
SuperCOSMOS $\epsilon_{B_J}$, $\epsilon_ {R_F}$ (Rockosi private
communication). Gaussian errors similar in size to those of the
SuperCOSMOS data are added to the magnitude of each star before
measuring the star counts.  We find that the results of XDN06 are
recovered, with the average fluctuation raised only by about 3.7\%.
After transforming into the SuperCOSMOS system, the errors are
$16^m.5<B_{JSDSS}<20^m.5$ and $16^m.5<R_{FSDSS}<19^m.5$
respectively.

Figure 4 and Figure 5 show the star counts from the SDSS data with
same sky areas as in XDH06 but downgraded to the SuperCOSMOS
magnitude limits, for $B_{JSDSS}$ and $R_{FSDSS}$ respectively. From
the present SDSS public data release, the sky area $l=210^{\circ}$
is now added. Panel a) is for star counts in sky areas along the
$b=60^{\circ}$ circle. Panels b)-f) are for star counts of sky areas
along the selected longitudinal directions paired by mirror symmetry
on the both sides of the $l=0^\circ$ meridian. The asymmetric
structure still appears clearly with the  magnitude limit of the
downgraded SDSS data (especially in figure 4a).  The asymmetric
structure is not so prominent as with the original SDSS magnitude
limits, but we can still see that the star counts in $l
\in[180^{\circ},360^{\circ}]$ are systematically higher than in $l
\in [0^{\circ},180^{\circ}]$. The largest asymmetry of star counts
appear in panels b), c) and d). In panels e) and f), the errors are
so large at the downgraded limits that the asymmetric differences
between sky areas found in XDH06 are only marginally visible. As in
figure 4, figure 5 shows the results of star counts from the
$R_{FSDSS}$ data. Again, the most prominent excess over mirror
symmetry is found in panels b), c) and d). However, the $R_{FSDSS}$
band magnitude limit is fainter than that of the $B_{JSDSS}$ band,
which leads to weaker features of asymmetry than figure 4. Tables 2
and 3 describe the asymmetric ratio and its uncertainty in the
downgraded SDSS data. Columns 1--4 are the Galactic coordinates ($l$
and $b$), counted numbers and the corresponding errors for sky areas
with $l\leq 180^{\circ}$, and columns 5--8 are the same quantities
for sky areas on the other side of the $l=0^{\circ}$ meridian.
Comparison is between sky areas paired with mirror symmetry with
respect to the $l=0^{\circ}$ meridian.  The asymmetric ratios are
defined by: $asymmetric \: ratio=(number \: density_2 - number \:
density_1)/(number \: density_1)\times 100\%$ which are given in
column 9; column 10 gives the uncertainties in the ratios which are
inferred from the error of the number densities
(tables~\ref{tab3}--\ref{tab5} all have the same entries, but for
different data). The asymmetric ratios measured from downgraded SDSS
data are all positive with one exception which is very near to zero,
this means that all the sky areas in $l\in[180^{\circ},360^{\circ}]$
have higher projected number densities than those in
$l\in[0^{\circ},180^{\circ}]$. The largest asymmetric ratio is
$16.9\pm6.3\%$ in the $B_{JSDSS}$ band and $13.5\pm6.7\%$ in the
$R_{FSDSS}$ band.







Therefore if there are similar levels of asymmetric structure in the
southern Sky, they should be visible even with the SuperCOSMOS
magnitude limit.

\begin{table*}

 \centering

 \begin{minipage}{140mm}

\caption{The relative deviations of the sky field pairs for
$B_{JSDSS}$$\in[16^m.5,20^m.5]$ mag. }\label{tab2}
\begin{tabular}{@{}rrrrrrrrrr@{}}
  \hline
 $\ell_1$  & $b$      & number & error of & $\ell_2$  & $b$       & number & error of & asymmetry ratio & uncertainty of
 \\$(^{\circ})$& $(^{\circ})$ & $density_1$ &  $density_2$ & $(^{\circ})$  &  $(^{\circ})$ & $density_1$ & $density_2$  &  ($\%$)         & asymmetry ratio ($\%$)\\
\hline
  10 & 60 & 1597.500  &  48.330 &350  &60 & 1685.949  & 21.825 &  5.536 & 4.391\\
  40 & 60 & 1389.890  &  13.915 &320  &60 & 1595.569  & 19.651 & 14.798 & 2.415\\
  50 & 60 & 1420.219  &  75.928 &310  &60 & 1581.060  & 38.915 & 11.325 & 8.086\\
  60 & 60 & 1377.050  &  38.606 &300  &60 & 1387.020  & 74.542 &  0.724 & 8.216\\
  70 & 60 & 1258.550  &  31.836 &290  &60 & 1428.900  & 61.433 & 13.535 & 7.410\\
  80 & 60 & 1124.489  &  49.622 &280  &60 & 1283.959  & 42.167 & 14.181 & 8.162\\
  90 & 60 & 1081.709  &  44.143 &270  &60 & 1264.489  & 24.530 & 16.897 & 6.348\\
 100 & 60 & 1074.750  &  56.961 &260  &60 & 1133.510  & 11.626 &  5.467 & 6.381\\
 110 & 60 & 1010.599  &  31.438 &250  &60 & 1049.510  & 16.462 &  3.850 & 4.739\\
 120 & 60 &  961.590  &   4.174 &240  &60 & 1056.290  & 37.484 &  9.848 & 4.332\\
 130 & 60 &  908.416  &  44.612 &230  &60 &  989.323  & 39.663 &  8.906 & 9.277\\
 150 & 60 &  870.270  &  41.345 &210  &60 &  911.882  & 17.599 &  4.781 & 6.773\\
 170 & 60 &  866.552  &  30.676 &190  &60 &  871.720  & 34.395 &  0.596 & 7.509\\
  30 & 65 & 1303.510  &  60.955 &330  &65 & 1476.130  & 50.739 & 13.242 & 8.568\\
  30 & 70 & 1219.079  &  24.728 &330  &70 & 1266.099  & 43.532 &  3.857 & 5.599\\
  30 & 75 & 1101.650  &  38.571 &330  &75 & 1146.250  & 13.678 &  4.048 & 4.742\\
  60 & 65 & 1195.140  &  37.210 &300  &65 & 1326.010  & 35.114 & 10.950 & 6.051\\
  60 & 70 & 1110.910  &  14.796 &300  &70 & 1220.300  & 42.086 &  9.846 & 5.120\\
  60 & 75 & 1005.109  &  33.589 &300  &75 & 1100.689  & 58.191 &  9.509 & 9.131\\
  90 & 55 & 1229.219  &  54.959 &270  &55 & 1347.400  & 30.259 &  9.614 & 6.932\\
  90 & 65 & 1023.700  &  26.978 &270  &65 & 1166.369  & 56.959 & 13.936 & 8.199\\
  90 & 70 &  955.830  &  39.693 &270  &70 & 1098.619  & 18.859 & 14.938 & 6.125\\
  90 & 75 &  913.866  &  15.131 &270  &75 & 1017.349  & 31.349 & 11.323 & 5.086\\
 120 & 55 & 1028.329  &  28.122 &240  &55 & 1090.479  & 54.862 &  6.043 & 8.069\\
 120 & 65 &  919.445  &  50.675 &240  &65 &  991.031  & 56.564 &  7.785 &11.663\\
 150 & 65 &  828.664  &  24.510 &210  &65 &  901.859  & 33.080 &  8.832 & 6.949\\
 150 & 70 &  806.091  &  18.853 &210  &70 &  841.559  & 19.937 &  4.399 & 4.812\\
 150 & 75 &  837.734  &  51.541 &210  &75 &  851.314  & 22.645 &  1.621 & 8.855\\
\hline

\end{tabular}

\end{minipage}

\end{table*}

\begin{table*}

 \centering

 \begin{minipage}{140mm}

\caption{The relative deviations of the sky field pairs for
$R_{FSDSS}$$\in[16^m.5,19^m.5]$ mag.}\label{tab4}
\begin{tabular}{@{}rrrrrrrrrr@{}}
\hline
$\ell_1$  & $b$      & number & error of & $\ell_2$  & $b$       & number & error of & asymmetry ratio & uncertainty of
\\$(^{\circ})$& $(^{\circ})$ & $density_1$ & $density_1$ & $(^{\circ})$  &  $(^{\circ})$ & $density_2$ & $density_2$  &  ($\%$)         & asymmetry ratio ($\%$)\\

  \hline

 10  &60  &1726.890  &  50.382 &350  &60  &1811.989  & 31.156  &  4.927   &4.721  \\

  40 & 60 & 1544.520 &   17.300& 320 & 60 & 1730.890 &  27.820 &  12.066  & 2.921 \\

  50 & 60 & 1549.510 &   97.815& 310 & 60 & 1695.439 &  58.618 &   9.417  &10.095 \\

  60 & 60 & 1458.390 &   25.113& 300 & 60 & 1493.660 &  75.721 &   2.418  & 6.914 \\

  70 & 60 & 1338.709 &   40.464& 290 & 60 & 1508.959 &  52.391 &  12.717  & 6.936 \\

  80 & 60 & 1246.390 &   55.123& 280 & 60 & 1374.219 &  37.407 &  10.256  & 7.423 \\

  90 & 60 & 1174.689 &   54.628& 270 & 60 & 1333.349 &  23.945 &  13.506  & 6.688 \\

 100 & 60 & 1155.189 &   66.160& 260 & 60 & 1209.199 &   9.641 &   4.675  & 6.561 \\

 110 & 60 & 1103.040 &   32.209& 250 & 60 & 1141.750 &  19.037 &   3.509  & 4.645 \\

 120 & 60 & 1043.449 &   30.933& 240 & 60 & 1138.829 &  34.655 &   9.140  & 6.285 \\

 130 & 60 &  985.525 &   19.785& 230 & 60 & 1043.609 &  65.759 &   5.893  & 8.680 \\

 150 & 60 &  947.528 &   54.534& 210 & 60 &  973.794 &  15.702 &   2.772  & 7.412 \\

 170 & 60 &  950.013 &   29.894& 190 & 60 &  954.971 &  48.084 &   0.521  & 8.208 \\

  30 & 65 & 1425.650 &   58.515& 330 & 65 & 1575.650 &  49.862 &  10.521  & 7.601 \\

  30 & 70 & 1322.020 &   12.712& 330 & 70 & 1377.750 &  30.668 &   4.215  & 3.281 \\

  30 & 75 & 1186.050 &   34.852& 330 & 75 & 1217.760 &  13.473 &   2.673  & 4.074 \\

  60 & 65 & 1291.579 &   29.649& 300 & 65 & 1402.810 &  40.511 &   8.611  & 5.432 \\

  60 & 70 & 1214.660 &   22.258& 300 & 70 & 1290.349 &  36.804 &   6.231  & 4.862 \\

  60 & 75 & 1094.579 &   24.285& 300 & 75 & 1161.560 &  32.224 &   6.119  & 5.162 \\

  90 & 55 & 1359.380 &   48.864& 270 & 55 & 1461.239 &  52.372 &   7.493  & 7.447 \\

  90 & 65 & 1093.819 &   37.443& 270 & 65 & 1207.569 &  59.053 &  10.399  & 8.822 \\

  90 & 70 & 1036.369 &   20.023& 270 & 70 & 1167.410 &  20.154 &  12.644  & 3.876 \\

  90 & 75 & 1006.340 &    4.486& 270 & 75 & 1076.699 &  15.964 &   6.991  & 2.032 \\

 120 & 55 & 1124.869 &   42.921& 240 & 55 & 1236.449 &  23.534 &   9.919  & 5.907 \\

 120 & 65 &  979.656 &   34.514& 240 & 65 & 1067.819 &  57.995 &   8.999  & 9.443 \\

 150 & 65 &  893.812 &   35.045& 210 & 65 &  959.431 &  42.456 &   7.341  & 8.670 \\

 150 & 70 &  867.317 &   21.781& 210 & 70 &  895.692 &  25.912 &   3.271  & 5.498 \\

 150 & 75 &  919.247 &   32.209& 210 & 75 &  917.749 &  21.114 &  -0.163  & 5.800 \\

\hline

\end{tabular}

\end{minipage}

\end{table*}

\begin{table*}

 \centering

 \begin{minipage}{140mm}

\caption{The relative deviations of the sky field pairs for
$B_J$$\in[16^m.5,20^m.5]$ mag. }\label{tab3}

\begin{tabular}{@{}rrrrrrrrrr@{}}

  \hline

 $\ell_1$  & $b$      & number & error of & $\ell_2$  & $b$       & number & error of & asymmetry ratio & uncertainty of  \\

   $(^{\circ})$& $(^{\circ})$ & $density_1$ & $density_1$ & $(^{\circ})$  &  $(^{\circ})$ & $density_2$ & $density_2$  &  ($\%$)         & asymmetry ratio ($\%$)\\

  \hline

  10 &-60 & 1552.579  &  76.917 &350 &-60  &1584.079 &   61.902 &   2.028 &   8.941\\

  20 &-60 & 1544.079  &  44.300 &340 &-60  &1439.569 &   47.212 &  -6.768 &   5.926\\

  30 &-60 & 1515.079  &  66.156 &330 &-60  &1409.569 &   58.819 &  -6.963 &   8.248 \\

  40 &-60 & 1523.579  &  43.333 &320 &-60  &1453.579 &   80.638 &  -4.594 &   8.136 \\

  50 &-60 & 1422.069  &  63.642 &310 &-60  &1380.069 &   53.733 &  -2.953 &   8.253 \\

  60 &-60 & 1370.569  &  33.600 &300 &-60  &1307.069 &   46.027 &  -4.633 &   5.809 \\

  70 &-60 & 1179.930  &  27.163 &290 &-60  &1351.569 &   71.606 &  14.546 &   8.370 \\

  80 &-60 & 1132.560  &  49.012 &280 &-60  &1247.060 &   87.287 &  10.109 &  12.034 \\

  90 &-60 & 1053.050  &  34.337 &270 &-60  &1191.060 &   31.689 &  13.105 &   6.270 \\

 100 &-60 & 1006.549  &  66.823 &260 &-60  &1089.060 &   41.794 &   8.197 &  10.791 \\

 110 &-60 &  876.658  &  50.321 &250 &-60  &1051.550 &   43.258 &  19.949 &  10.674 \\

 120 &-60 &  938.547  &  57.981 &240 &-60  & 967.549 &   65.758 &   3.090 &  13.184 \\

 130 &-60 &  905.307  &  43.079 &230 &-60  & 916.546 &   34.188 &   1.241 &   8.534 \\

 140 &-60 &  895.546  &  34.853 &220 &-60  & 948.549 &   47.782 &   5.918 &   9.227 \\

 150 &-60 &  859.044  &  34.550 &210 &-60  & 940.049 &   18.947 &   9.429 &   6.227 \\

 160 &-60 &  919.547  &  18.449 &200 &-60  & 908.546 &   64.258 &  -1.196 &   8.994 \\

 170 &-60 &  933.547  &  57.718 &190 &-60  & 838.543 &   32.895 & -10.176 &   9.706 \\

  30 &-55 & 1756.609  & 62.784  & 330&-55  &1732.640 &   34.151 &  -1.364 & 5.518\\

  30 &-65 & 1336.380  & 25.611  & 330&-65  &1302.069 &   36.909 &  -2.567 & 4.678\\

  30 &-70 & 1190.050  & 115.05  & 330&-70  &1225.869 &   55.002 &   3.009 &14.289\\

  30 &-75 &  978.531  & 76.464  & 330&-75  &1014.270 &   20.205 &   3.652 & 9.879\\

  30 &-80 &  904.171  & 96.671  & 330&-80  &1026.550 &   49.882 &  13.534 &16.208\\

  60 &-55 & 1443.650  & 54.077  & 300&-55  &1470.670 &   46.232 &   1.871 & 6.948\\

  60 &-65 & 1205.050  & 68.259  & 300&-65  &1147.079 &   28.443 &  -4.810 & 8.024\\

  60 &-70 & 1079.670  & 37.970  & 300&-70  & 976.601 &   28.756 &  -9.546 & 6.180\\

  60 &-75 & 1093.479  & 42.780  & 300&-75  &1015.239 &   50.576 &  -7.155 & 8.537\\

  60 &-80 &  918.567  & 53.861  & 300&-80  &1048.150 &  164.492 &  14.106 &23.771\\

  90 &-65 & 1057.750  & 37.072  & 270&-65  & 992.672 &   48.298 &  -6.152 & 8.070\\

  90 &-70 &  983.179  & 49.343  & 270&-70  & 950.286 &   28.261 &  -3.345 & 7.893\\

  90 &-75 &  985.293  & 81.426  & 270&-75  & 903.184 &   56.136 &  -8.333 &13.961\\

  90 &-80 & 1002.070  & 76.915  & 270&-80  & 928.646 &   60.223 &  -7.327 &13.685\\

 120 &-65 &  858.383  & 62.833  & 240&-65  & 929.966 &   73.010 &   8.339 &15.825\\

 120 &-70 &  859.643  & 48.598  & 240&-70  & 921.046 &   57.155 &   7.142 &12.302\\

 120 &-75 &  881.934  & 33.591  & 240&-75  & 925.401 &   15.142 &   4.928 & 5.525\\

 120 &-80 &  894.091  & 32.333  & 240&-80  & 833.622 &   57.494 &  -6.763 &10.046\\

 150 &-65 &  885.005  & 24.511  & 210&-65  & 909.851 &   44.259 &   2.807 & 7.770\\

 150 &-70 &  836.981   &64.097  & 210&-70   &828.940  &  44.485  & -0.960 &12.973\\

 150 &-75 &  906.083   &72.184  & 210&-75   &835.567  &  49.243  & -7.782 &13.401\\

 150 &-80 &  810.585   &25.940  & 210&-80   &814.905  &  49.143  &  0.532 & 9.262\\

\hline

\end{tabular}

\end{minipage}

\end{table*}

\begin{table*}

 \centering

 \begin{minipage}{140mm}

\caption{The relative deviations of the sky field pairs for
$R_F$$\in[16^m.5,19^m.5]$ mag. }\label{tab5}

\begin{tabular}{@{}rrrrrrrrrr@{}}

  \hline

 $\ell_1$  & $b$      & number & error of & $\ell_2$  & $b$       & number & error of & asymmetry ratio & uncertainty of  \\

   $(^{\circ})$& $(^{\circ})$ & $density_1$ & $density_1$ & $(^{\circ})$  &  $(^{\circ})$ & $density_2$ & $density_2$  &  ($\%$)         & asymmetry ratio ($\%$)\\

  \hline

 10 &-60 & 1746.089  &  71.268 &350&-60 & 1848.599  & 110.406 &   5.870  & 10.404 \\

 20 &-60 & 1751.089  &  29.558 &340&-60 & 1617.079  &  68.217 &  -7.652  &  5.583 \\

 30 &-60 & 1761.089  &  91.829 &330&-60 & 1644.089  & 143.147 &  -6.643  & 13.342 \\

 40 &-60 & 1802.089  &  55.405 &320&-60 & 1663.089  &  61.489 &  -7.713  &  6.486 \\

 50 &-60 & 1693.589  &  96.918 &310&-60 & 1591.079  &  23.517 &  -6.052  &  7.111 \\

 60 &-60 & 1562.079  &  49.526 &300&-60 & 1523.079  &  91.257 &  -2.496  &  9.012 \\

 70 &-60 & 1444.229  &  42.711 &290&-60 & 1539.579  &  80.699 &   6.602  &  8.545 \\

 80 &-60 & 1335.069  &  58.455 &280&-60 & 1432.069  &  93.267 &   7.265  & 11.364 \\

 90 &-60 & 1217.560  &  39.018 &270&-60 & 1472.579  &  46.217 &  20.945  &  7.000 \\

100 &-60 & 1176.060  &  49.969 &260&-60 & 1348.569  &  69.336 &  14.668  & 10.144 \\

110 &-60 &  911.591  &  39.793 &250&-60 & 1270.069  &  55.966 &  39.324  & 10.504 \\

120 &-60 & 1188.680  &  81.417 &240&-60 & 1152.560  &  83.339 &  -3.038  & 13.860 \\

130 &-60 & 1133.849  &  59.553 &230&-60 & 1070.560  &  23.034 &  -5.581  &  7.283 \\

140 &-60 & 1096.560  &  36.696 &220&-60 & 1176.060  &  72.195 &   7.249  &  9.930 \\

150 &-60 &  997.552  &  66.039 &210&-60 & 1140.560  &  34.719 &  14.335  & 10.100 \\

160 &-60 & 1146.060  &  46.602 &200&-60 & 1115.060  & 166.630 &  -2.704  & 18.605 \\

170 &-60 & 1155.560  &  35.725 &190&-60 & 1058.050  &  25.871 &  -8.438  &  5.330 \\

  30 &-55 & 1959.729  &  64.605 & 330&-55  &1961.910 &  30.947  &  0.111  &  4.875 \\

  30 &-65 & 1537.520  &  61.215 & 330&-65  &1549.349 &  41.863  &  0.769  &  6.704 \\

  30 &-70 & 1509.489  & 116.808 & 330&-70  &1416.660 & 114.977  & -6.149  & 15.355 \\

  30 &-75 & 1197.810  &  52.558 & 330&-75  &1249.969 &  10.672  &  4.354  &  5.278 \\

  30 &-80 & 1121.569  &  67.932 & 330&-80  &1216.599 &  87.526  &  8.472  & 13.860 \\

  60 &-55 & 1693.839  &  63.233 & 300&-55  &1695.589 &  38.849  &  0.103  &  6.026 \\

  60 &-65 & 1524.510  &  43.793 & 300&-65  &1363.000 &  41.841  &-10.594  &  5.617 \\

  60 &-70 & 1308.469  &  62.136 & 300&-70  &1183.469 &  24.054  & -9.553  &  6.587 \\

  60 &-75 & 1321.449  &  30.234 & 300&-75  &1273.150 &  41.728  & -3.655  &  5.445 \\

  60 &-80 & 1151.810  &  68.150 & 300&-80  &1254.030 & 125.622  &  8.874  & 16.823 \\

  90 &-65 & 1262.430  &  45.872 & 270&-65  &1193.810 &  57.514  & -5.435  &  8.189 \\

  90 &-70 & 1222.209  &  53.918 & 270&-70  &1130.109 &  31.735  & -7.535  &  7.008 \\

  90 &-75 & 1195.869  &  97.713 & 270&-75  &1134.050 &  66.108  & -5.169  & 13.698 \\

  90 &-80 & 1238.199  &  56.400 & 270&-80  &1076.939 &  72.794  &-13.023  & 10.434 \\

 120 &-65 & 1106.849  &  66.695 & 240&-65  &1087.329 & 132.024  & -1.763  & 17.953 \\

 120 &-70 & 1116.219  &  28.901 & 240&-70  &1128.650 &  22.175  &  1.113  &  4.575 \\

 120 &-75 & 1111.839  &  20.551 & 240&-75  &1251.900 &  52.029  & 12.597  &  6.528 \\

 120 &-80 & 1140.290  &  20.054 & 240&-80  &1102.859 &  82.898  & -3.282  &  9.028 \\

 150 &-65 & 1044.729  &  23.094 & 210&-65  &1021.659 &  31.659  & -2.208  &  5.240 \\

 150 &-70 & 1057.739  &  70.615 & 210&-70  &1006.570 &  41.154  & -4.837  & 10.566 \\

 150 &-75 & 1220.020  &  62.724 & 210&-75  &1073.199 &  20.037  &-12.034  &  6.783 \\

 150 &-80 & 1062.540  &  31.4566& 210&-80  &1025.109 & 112.344  &-3.522   &13.533\\

\hline

\end{tabular}

\end{minipage}

\end{table*}

\subsection{Southern Galactic cap: star counts from SuperCOSMOS data}

In XDH06, star counts from SDSS data show a prominent asymmetric
structure in the northern Galactic hemisphere through comparing the
projected number densities of sky area pairs with mirror symmetry on
both sides of the $l=0^{\circ}$ meridian. We will use the same
method to examine the structure of stellar halo in the southern sky,
in particular to check whether the halo structure has the same
features or is different from its northern counterpart.

Star counts for southern sky from the SuperCOSMOS data are shown in
figures 6 and 7. Star counts in each sky area are plotted using
triangles and squares. Panel a) shows the results of star counts for
the selected sky areas along a circle of $b=-60^{\circ}$. Panels b),
c), d), e) and f) are for sky areas along the longitudinal
directions, also paired with mirror symmetry on the both side of the
$l=0^{\circ}$ meridian. Each of the sky areas is divided into four
subfields to account for the fluctuation of star counts over the
average value of the area. The fluctuations calculated for all sky
areas this way are used as error bars in the plots. The error bars
actually measure the intrinsic fluctuations of the projected number
density, and the uncertainties in classification and photometry. The
average error of star counts will be discussed in section 6.

Dividing the southern cap into two halves by the $l=0^{\circ},
180^{\circ}$ meridian, the data for both $B_J$ and $R_F$ bands show
that the structures of the two halves are basically symmetric within
statistical errors. This is clearly shown in panel b)-f) of figures
6 and 7 .

The $B_J$ band data shows smaller error bars and obvious smoother
structure in the projected number density distribution than the
$R_F$ band data does. Sizable fluctuations over a axisymmetric
structure do exist in the $B_J$ band data. In two pairs of data,
i.e. ($150^{\circ},-60^{\circ}$)\& ($210^{\circ},-60^{\circ}$) and
($90^{\circ},-60^{\circ}$)\&($270^{\circ},-60^{\circ}$), the
projected number density at $l>180^{\circ}$ side is higher than the
other side ($l<180^{\circ}$). While the pair of
($60^{\circ},-70^{\circ}$)\& ($300^{\circ},-70^{\circ}$) shows a
reversed excess.

The star counts from the $R_F$ band have a larger scatter than that
of the $B_J$ band. The $R_F$ band data also has less coincidence in
classification with SDSS data than the $B_J$ band data. Moreover,
its limiting magnitude is shallower than the $B_J$ band by about one
magnitude. For example, for F0-type stars, the distance limits given
by the $B_J$ band are from 5.23Kpc to 32.98Kpc, while those defined
by the $R_F$ band are from 6.46Kpc to 25.72Kpc; and for F8-type
stars, they are 2.68Kpc to 16.89Kpc for the $B_J$ band, and 3.79Kpc
to 15.09Kpc for the $R_F$ band. Selecting redder stars from a
shallower box in the $R_F$ band, star counts show larger deviations.

In both the $B_J$ and $R_F$ bands, there is a odd data point at
($130^{\circ},-60^{\circ}$), which has a projected number density
obviously lower than its neighbor sky areas. Because this sky area
is near the edge of the survey, it is very likely that this is a
boundary effect.

Table 4 (for $B_J$) and table 5 (for $R_F$) list the projected
number densities and their corresponding errors, and the asymmetric
ratio measured in the SuperCOSMOS data. Comparing the uniform
positive asymmetric ratios in the downgraded SDSS data (tables 2 and
3), the values given by SuperCOSMOS are quite irregular, with
apparantly random positive and negative values.

\section{The theoretical model}

From SuperCOSMOS star counts there is no obvious asymmetric
structure in the southern hao. A theoretical axisymmetric halo model
is therefore adopted here. Because the $R_F$ band data is shallower
and less consistent with SDSS data, only $B_J$ band data is used to
constrain the model parameters.

From the Bahcall standard model (Bahcall \& Soneira 1980), the
projected number density in a certain apparent magnitude interval
along a fixed direction can be described by the integral of a
density profile and a luminosity function.

\begin{equation}
A(m_1,m_2, \ell, b) = \int_{m_1}^{m_2} {\rm d}m^{'} \int_0^\infty
R^2\,{\rm d}R\,\rho(\mathbf{r})\, \phi(M)\, {\rm d}\Omega,
\label{eq1}
\end{equation}

where $m_1,m_2$ are the limits of the given magnitude interval, $R$
is the heliocentric distance of a star, $\rho$ is the density
profile of each stellar population, and $\phi(M)$ is the luminosity
function of the population.

For the thin and thick disk components, the density profile is
assumed to be exponential,

\begin{equation}
\rho(\mathbf{r})=\exp[-|z|/H-x/h], \label{eq2}
\end{equation}

where $|z|$ is absolute value of height of a star above the Galactic
plane, x is distance between the Galactic centre and the projected
point of that star on the Galactic plane, $H$ is the scale hight,
and $h$ is the scale length of the exponential disk.

The power-law halo density profile in Reid(1993) is adopted,

\begin{equation}
\rho(\mathbf{r})=\frac{a_0^n+r_0^n}{a_0^n+r^n}, \label{eq3}
\end{equation}

where $r_0$ is the distance from the Sun to the Galactic centre, and
$a_0=1000$ is a normalisation constant.  $r$ is the distance from
the star to the Galactic centre. $r=\sqrt{x^2+(y/p)^2+(z/q)^2}$ as
in XDH06. x, y, z define the position vector in three axes of
coordinate frame adopted in paper I. p, q are the axial ratios of
the middle axis and shortest axis to the major axis respectively.
The triaxial halo model naturally degenerates to asymmetric halo
model when p=1 The luminosity functions of the halo, thick and thin
disk components in $B_J$, $R_F$ bands are transformed from the
luminosity functions of Robin \& Cr\'ez\'e(1986), with
$B_J$=B-0.304$\times$(B-V) and $R_F=R+0.163\times(V-R)$ as provided
by the photometric calibration of the final data release of 2DFGRS.
Given the $R_F$ luminosity function, star counts in the $R_F$ band
can also be obtained, but these are not used for model fitting for
the reasons given above.

The three dimensional extinction model of the Milky Way derived from
COBE observations is adopted four our model. Directly correcting the
observational data for extinction is not possible due to the lack of
distance information for individual stars (Drimmel, Cabrera-Lavers
\& L\'opez-Corredoira 2003), however we solve this problem by
applying COBE extinction data to the theoretical model.

Using equation (1), the projected number density of each sky area
can be obtained. To reveal the distribution of star counts in
apparent magnitude, the $B_J$ band magnitude is divided into 8 bins
($16^m.5$ to $20^m.5$, in steps of $0.5^m$), the number density in
each magnitude bin is then calculated. Constraining star counts in
$B_J$ band magnitude limits of $16^m.5$ to $20^m.5$, a non-neligible
number of stars have no corresponding $R_F$ band data, therefore
color counts need to be treated with special care, and that will be
discussed later in section 5.2.

As a continuation of XDH06, the present work is focused on halo
structure near the southern cap of the Galaxy. For a better
comparison between the present work and XDH06, the parameters of the
thin and thick disks are fixed with the values used in XDH06, which
were taken from Chen et al.(2001). Only the halo parameters are
adjusted to fit the SuperCOSMOS observations, the scope of
parameters is listed in table 5. In an axisymmetric halo model there
are only two parameters: n is power law index of halo density
profile, and q is the axial ratio z/x.

A $\chi^2$ minimization is adopted to compare the theoretical
results and the observational data sets (Press et al. 1992). For a
non-Gaussian distribution of discrete data, Pearson's $\chi^2$ is
used,

\begin{equation}
\chi^2=\sum_{i=1}^N \frac{(R_i-S_i)^2}{(R_i+S_i)(N-m)}, \label{eq4}
\end{equation}

The meanings of all the symbols are described in XDH06. $\chi^2$ is
calculated in order to evaluate the similarity between the
theoretical projected number density and the observational data.
${\chi^2_{bin}}$ describes the difference between the distribution
of the theoretical star counts in apparent magnitude bins and that
of observations for each sky area. $\overline{\chi_{bin}^2}$ is the
average value of ${\chi^2_{bin}}$ in all sky areas.

\section{Results and analysis}

\subsection{Fitting SuperCOSMOS star counts with the axisymmetric model}

The theoretical projected surface number densities are calculated
using the axisymmetric model described in section 5, with extinction
included. The theoretical model that best fits the SuperCOSMOS
observational data in both $B_J$ band and $R_F$ band is shown in
figure 6 and 7 as the solid and dashed-lines respectively . The
model parameters are n=2.8, q=0.7. This is one of the best fitting
models in the provided parameter space. An axisymmetric model can
fit SuperCOSMOS data reasonably well within the statistical error
bars. The best fit theoretical model (solid line) and the
observational data (diamonds with error bars)for l=-60$^{\circ}$
fields are shown in figure 6a in which the data shows an irregular
pattern of deviations from the symmetric model. The projected number
densities at ($40^{\circ},-60^{\circ}$),
($60^{\circ},-60^{\circ}$),($270^{\circ},-60^{\circ}$),
($290^{\circ},-60^{\circ}$) are higher than the model, while the
observational data at ($110^{\circ},-60^{\circ}$),
($340^{\circ},-60^{\circ}$),($350^{\circ},-60^{\circ}$) are lower
than the model. It is clear from figures 6b--6f that the two
theoretical lines do not overlap perfectly due to different
extinctions. In figure 6e of b=-70$^{\circ}$, l=120$^{\circ}$, the
value of the theoretical curve has a dip because that sky area
(120$^{\circ}$,-70$^{\circ}$) has a large extinction from COBE, such
that when the distance is larger than 550PC, the extinction takes
$Av=0^m.174$. While in other areas around this, the extinction
ranges only $0^m.07 \sim 0^m.08$.

The above model parameter set is used to calculate the theoretical
$R_F$ band star count, and the results are plotted in figure 7. In
figure 7a, star counts from ($40^{\circ},-60^{\circ}$) to ($110
^{\circ}, -60^{\circ}$),($270^{\circ},-60^{\circ}$) to ($340
^{\circ}, -60^{\circ}$) also fluctuate around the theoretical value.
In figures 7b--7f, the pairs of ($120^{\circ},-75^{\circ}$) \&
($240^{\circ},-75^{\circ}$) and
($150^{\circ},-75^{\circ}$)\&($210^{\circ},-75^{\circ}$) show counts
higher than the theoretical line, while counts in all other areas
are random around the theoretical prediction.

\begin{table*}

 \centering

 \begin{minipage}{140mm}

\caption{Input range of parameters of theoretical model}\label{tab6}

\begin{tabular}{@{}rrrr@{}}

  \hline

 Parameter  & Lower limit & Upper limit  & step   \\

  \hline

  n         &  2 & 4   &  0.1   \\

  q         & 0.4 & 1.0 & 0.1   \\

\hline

\end{tabular}

\end{minipage}

\end{table*}

\begin{table*}

 \centering

 \begin{minipage}{140mm}

\caption{The best-fiting parameters}\label{tab7}

\begin{tabular}{@{}rrrr@{}}

  \hline

  n & q  & $\chi^2$  & $\overline{\chi_{bin}^2}$ \\

  \hline

  2.5 &  0.6 &  1.714 & 0.990  \\

  2.6 &  0.6 &  1.658 & 0.964 \\

  2.7 &  0.6 &  1.615 & 0.948  \\

  2.8 &  0.6 &  1.583 & 0.940  \\

  2.8 &  0.7 &  1.561 & 1.064  \\

  2.9 &  0.6 &  1.559 & 0.939  \\

  2.9 &  0.7 &  1.587 & 1.044  \\

  3.0 &  0.6 &  1.543 & 0.944  \\

  3.0 &  0.7 &  1.621 & 1.032  \\

  3.1 &  0.6 &  1.534 & 0.955\\

  3.1 &  0.7 &  1.662 & 1.026  \\

  3.2 &  0.6 &  1.530 & 0.971  \\

  3.2 &  0.7 &  1.708 & 1.025  \\

  3.3 &  0.7 &  1.760 & 1.029  \\

  3.4 &  0.7 &  1.815 & 1.037  \\

  3.5 &  0.7 &  1.874 & 1.050  \\

\hline

\end{tabular}

\end{minipage}

\end{table*}

As well as calculating the stellar projected number density, we also
compare the theoretical and the observational star counts in
apparent magnitude bin for each sky area.
Figure 8 shows, as an example, the observational (gray diamond) and
the theoretical (dark line) star counts in 12 sky areas of
$b=-60^{\circ}$, the Galactic coordinates of each sky area are
indicated in the corresponding panel. As shown in these plots, the
theoretical model can fit observation data fairly well, but with a
few exceptions. In ($90^{\circ},-60^{\circ}$) and
($330^{\circ},-60^{\circ}$) at the bin of $20^m$ to $20^m.5$, the
theoretical value is higher than the observational one. Similar to
what is shown in figures 6 and 7, the distribution of star counts in
apparent magnitude also fits fairly well a homogeneous axisymmetric
structure.

Using equation (4), $\chi^2$ and $\overline{\chi_{bin}^2}$ for each
parameter grid can be obtained. The contour plots of $\chi^2$ and
$\overline{\chi_{bin}^2}$ in the n-q plane are presented in figure
9. The minimum value of $\chi^2$ is 1.53, and the maximum is 11.915,
while the values for $\overline{\chi_{bin}^2}$ are 0.939 and 3.424
respectively, 20 levels of contours are used. The inner-most
(smallest values of $\chi^2$ and $\overline{\chi_{bin}^2}$) contour
indicates the best fitting combinations n and q. The open diamonds
in both panels of figure 9 indicate the most favorable parameters
given by both $\chi^2$ and $\overline{\chi_{bin}^2}$ minimizations,
which are listed in table 7. The fitting of the observed projected
number density using one of the best combinations (n=2.8, q=0.7) is
shown in figure 6.

\subsection{Comparison between star counts of the northern and southern Galactic caps}

In the previous subsection, the southern sky projected surface
number density of SuperCOSMOS $B_J$ band data is fitted by an
axisymmetric stellar halo model. As discussed above, star counts of
the northern sky show asymmetric structure due to an excess of halo
stars for $l>180^\circ$ (see XDH06 for details). The presence of the
same feature in the southern sky is the main concern of this paper.

To answer this question, we need to compare the distribution of
number density in the north from the downgraded SDSS data and that
of SuperCOSMOS data in the south. Singling-out the halo population
from star counts is now required. With the data we have, the halo
and disc populations can only be roughly distinguished through
colors based on photometric data. SuperCOSMOS $R_F$ band data has
only an 85\% coincidence with SDSS data, which makes our analysis
somewhat less accurate. However, this factor only affects the total
number of stars that can be used in statistics in color, and will
raise the level of random error in the final result. Further to this
aim, $R_F$ band data is still again used to obtain the star counts
in color.

Figure~\ref{fig10} shows the projected number density of SDSS
downgraded data of $b=60^\circ$ and SuperCOSMOS data of
$b=-60^\circ$. Both data sets are constrained by $B_J$ and $R_F$
band magnitude limits ($16^m.5<B_{JSDSS},B_J<20^m.5$,
$16^m.5<R_{FSDSS},R_F<19^m.5$). Black points and gray points
represent SuperCOSMOS data for $b=-60^\circ$ and SDSS downgraded
data for $b=60^\circ$ respectively. To show the difference clearly,
a 6th order polynomial function is used to fit for each data set.
The SDSS downgraded data are systematically higher than SuperCOSMOS
data. There are two possible reasons for this: firstly, a systematic
deviation between the two systems; secondly, an intrinsic difference
between the north and the south. From $l=0^\circ$ to $240^\circ$,
the two curves have similar shape, showing a possible systematic
deviation between the two systems. While from $l=240^\circ$ to
$360^\circ$, data set for $b=60^\circ$ shows an obvious excess over
that of $b=-60^\circ$ after considering the systematic deviation.
The largest excess appear around $l=330^\circ$, coincident with the
Virgo overdensity (Newberg \& Yanny. 2005; Juri\'c et al. 2005;
XDH06).

Figure~\ref{fig11} shows the projected surface number density in
$B_J-R_F$ color space for $(90^\circ, 60^\circ)$ (the gray line of
upper panel) \& $(270^\circ,60^\circ)$ (the black line of upper
panel), and $(90^\circ,60^\circ)$ (the gray line of lower panel) \&
$(90^\circ,-60^\circ)$ (the black line of lower panel). In the upper
panel, the distribution of SDSS downgraded data in color shows the
same property as that in XDH06, the halo populations (blue peak) in
the sky areas $l>180^\circ$ have an excess over those $l<180^\circ$,
while the disk populations (the red peak) are basically the same.
The Lower panel shows that both the SDSS downgraded data and the
SuperCOSMOS data sitting at two opposite sides of the Galactic plane
have a double peak structure in color space. The disk population in
the two sky areas has similar number density while the northern sky
star counts of halo population have larger numbers than those in the
southern sky. In figure~\ref{fig10}, the systematic deviation
between the two curves
 is caused by the difference in photometric sensitivity limits
between the two systems. The reason is quite straightforward: the
fainter stars between the photometric limits of SuperCOMOS and that
of SDSS are surely absent from SuperCOSMOS statistics, while
possibly being present in SDSS catalog.

The lowest number density of star counts in color appears for
$B_J-R_F=1.6$. The disk population and the halo population can be
roughly separated by this limit ( $B_J-R_F>1.6$ for the disk
population, and $B_J-R_F<1.6$ for the halo population).
Figure~\ref{fig12} demonstrates the difference between the selected
populations in the north (downgraded SDSS data) and in the south
(SuperCOSMOS data). The upper panel of figure~\ref{fig12} shows the
difference between the density of the halo population of sky areas
along the $b=60^\circ$ circle and that of the $b=-60^\circ$ circle.
The lower panel is the same as the upper one but for the disk
population. It is clear that the difference in disk population in
the lower panel (the southern Galactic cap) has a random
distribution around 0, the amplitude of such fluctuations is lower
than about 40 with no systematic feature; while the difference in
halo population in the upper panel has obvious features of over 200.
The systematic deviations between the SDSS downgraded data and the
SuperCOSMOS data are clearly caused by halo stars, i.e. the halo
population in the north has a certain amount of excess over that in
the south. Clearly, there is a prominent excess in the range of
$l=300^\circ$--360$^\circ$. This shows that there is an overdensity
only in the north, while no such features are found in the southern
SuperCOSMOS data.

\section{Discussion and Conclusions}

From SDSS data covering the northern cap, it has been found that the
northern halo is not axisymmetric (Newberg \& Yanny 2005; Juri\'c et
al. 2005; XDH06). This feature is also visible at shallower
magnitude limits (ie. closer halo stars) in SDSS data downgraded to
the limit of SuperCOSMOS. The main goal of this work was to examine
the halo structure near the southern cap of the Galaxy using
SuperCOSMOS data. We show that the southern halo structure does not
have a similar asymmetry to the northern galactic cap for the same
magnitude limits.

In XDH06, using very deep SDSS photometry from 15mag- 22mag, the
asymmetry ratio goes up to 23\%. The magnitude limit of SuperCOSMOS
data is from $16^m.5$-$20^m.5$ for $B_J$ band and $16^m.5$-$19^m.5$
for $R_F$ band. Converting the SDSS data to the same photometry
system and considering $B_{JSDSS}$ in the same magnitude range, the
asymmetric structure is weakened but still detectible, as
demonstrated by the asymmetry ratios and their errors in
tables~\ref{tab2} (also see \ref{tab4} for $R_{FSDSS}$), the
asymmetric ratio only pickes up to $16.9\% \pm 6.3\%$ . From
SuperCOSMOS data in the south, star counts shows no asymmetry
feature, as shown in tables~\ref{tab3} (also \ref{tab5} for $R_F$),
this is of course linked to the uncertainties in the data. The RMS
is over 7.8\% for asymmetry ratio measured in $B_J$.

Concerning the error of star counts of SuperCOSMOS $B_J$ band data,
there are three sources contributing. Firstly, the SuperCOSMOS data
of $B_J$ band has 92$\sim$93\% identification rate when
cross-correlating with SDSS data, this gives a error of 8\%, at the
worst case, in number counts. Secondly, the SuperCOSMOS data has an
overall photometry uncertainty of $0^m.3$, which creates an error of
3.7\%
 in the results of star counts, as derived from Monte Carlo
simulations. Thirdly, the SDSS photometry is far more accurate than
that of SuperCOSMOS, therefore it can be regarded as the precise
system to compare with the later one. Therefore we assume that the
statistical fluctuations measured in XDH06 are the intrinsic stellar
density fluctuations in the halo, which are 2.53\% on average for
number counts. Putting these factors together, we can estimate the
average error in SuperCOSMOS star counts as,

\begin{equation}
\sigma=\sqrt{8^2+3.7^2+2.53^2}=9.17 \label{erroranalysis}
\end{equation}

Having such an uncertainty in star counts for SuperCOSMOS data, and
considering the level of asymmetry of $16.9\% \pm 6.3\%$, it is not
possible to draw a firm conclusion for the symmetry issue for the
stellar halo near the south cap, when there is only SuperCOSMOS data
available.

However, when analyzing the population statistics using colors,
distinct properties of stellar halo structures in the north and
south can be found. As shown in figure~\ref{fig12}, the halo
population shows an apparent excess around $l=330^\circ$ in the
north (the upper panel) as from the downgraded SDSS data, while the
same plot for the south gives only random fluctuations of the same
level as statistical errors.

We attempt to fit triaxial halo models to both downgraded SDSS and
SuperCOSMOS data. By directly applying models in XDH06, no good fit
can be derived, because no obvious overdensity such as the Virgo one
in the north is found in the south. However, this does not exclude
the possibility to have a triaxial halo after removing the large
scale star streams. Due to large photometric uncertainties and low
sensitvity of SuperCOSMOS, an error in star counts around 9.17\%
prevents us from making a clear conclusion on this point.

Therefore, the present work can be concluded as the following:

\begin{enumerate}

\item SuperCOSMOS data (SSA) has been used to study the structure of
stellar halo covering the southern Galactic cap. Direct star counts
reveal that the structure can be fitted by a axi-symmetric halo
model. Limited by the photometric error and depth of the survey, no
asymmetry can be detected by star counts.

\item An Asymmetric structure, very similar to what have been found
using SDSS survey data (Newberg \& Yanny 2005; Juri\'c et al. 2005;
XDH06) can be detected by downgrading SDSS data to the limiting
magnitudes and photometric error of SuperCOSMOS.

\item A halo population excess, defined by ($B_J-R_F<1.6$), is
responsible for the asymmetry structure found in the north in
downgraded SDSS data, as revealed by both direct star counts
(figures~\ref{fig4}--\ref{fig8}, \ref{fig10}) and statistics in
color (figures~\ref{fig11} and \ref{fig12}). While for the southern
cap, no such features are present.

\item Considering the overall symmetry of the Galactic halo, the
asymmetry discovered in the north (the Virgo overdensity) is likely
to be a foreign component in the stellar halo of the Galaxy.
 However, due to a lack of good photometric data, an asymmetry in
the stellar halo near the south cap beyond SuperCOSMOS limits cannot
be ruled out. It is still an open question if we have a triaxial
halo with large scale star streams embedded.

\item For the structure of stellar halo near the southern cap,
SuperCOMOS data cannot go any further. Better quality survey data of
the SDSS quality is needed to adress these issues.

\end{enumerate}

\section*{Acknowledgments}

We would like to thank Richard Pokorny, Nigel Hambly, Constance
Rockosi, Liu Chao, for valuable suggestions and fruitful
discussions. This work is supported by the National Natural Science
Foundation of China through grants: 10573022, 10333060, 10403006. We
would like to express our gratitude to Dr. Simon Goodwin for proof
reading the manuscript and fruitful discussions.

{}

\begin{figure*}
\includegraphics[scale=0.8]{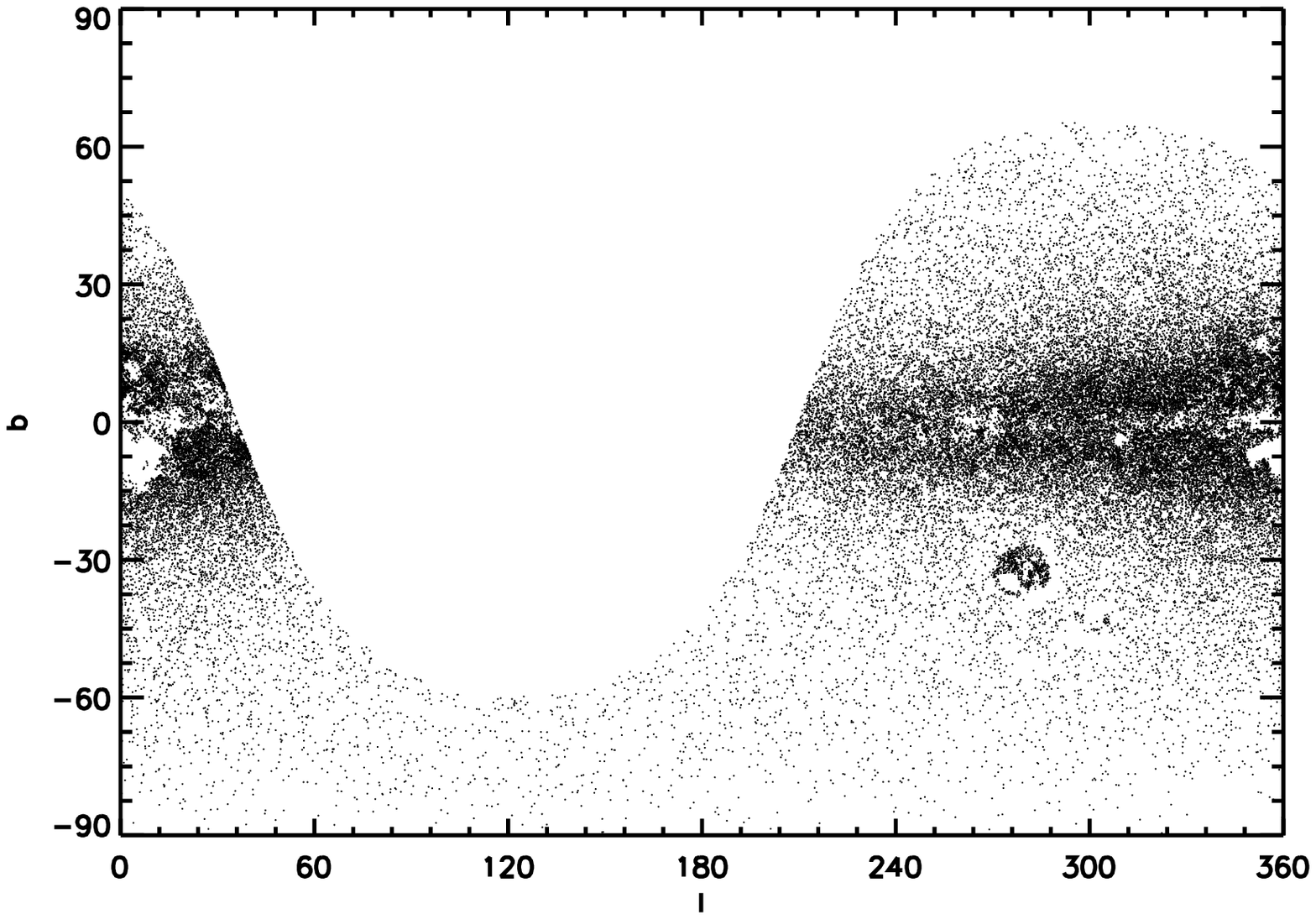}
\hfill
\includegraphics[scale=0.8]{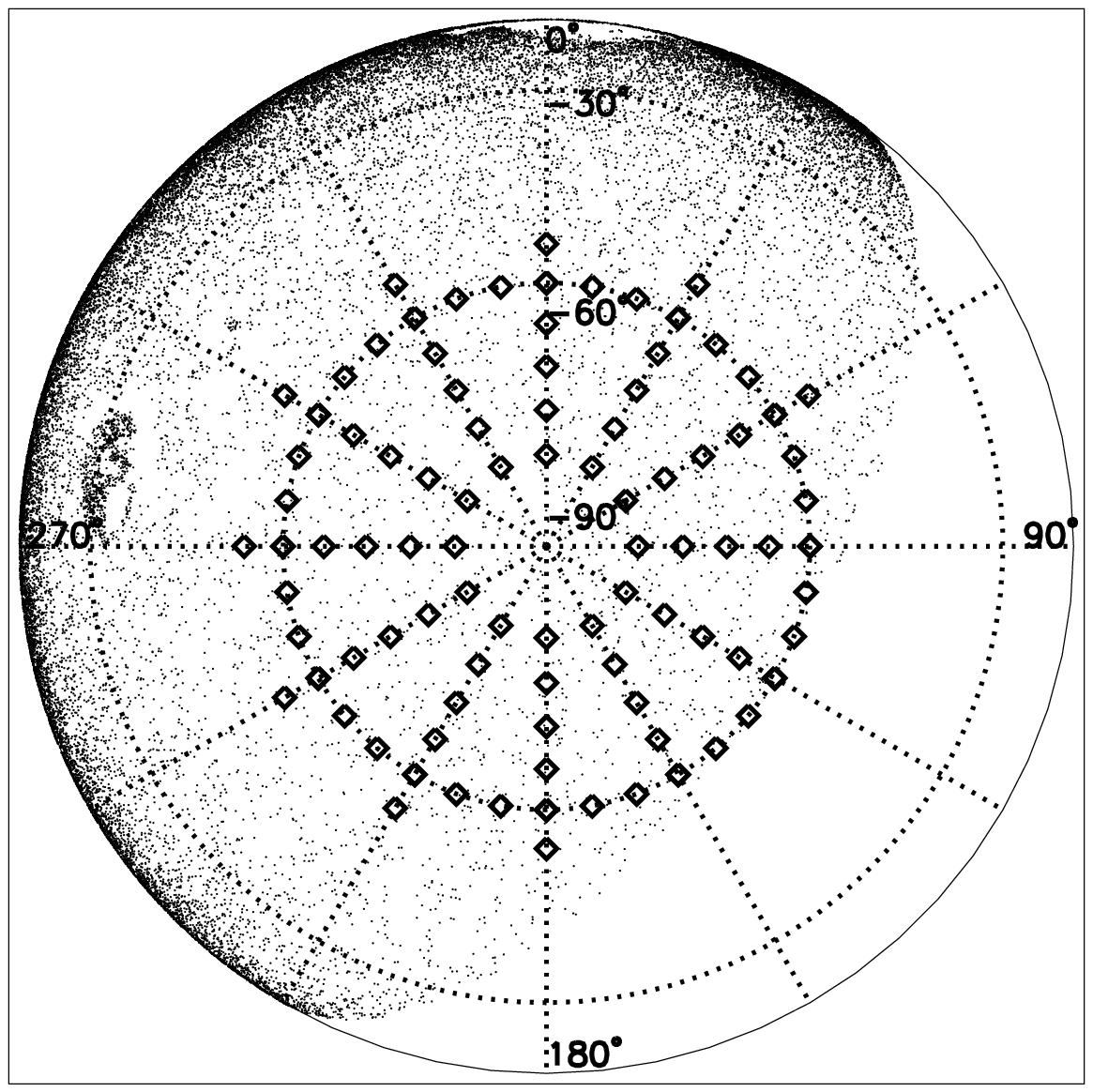}
\vspace{20pt} \caption{Upper panel: the sky coverage of the
SuperCOSMOS archive in Galactic coordinates as shown by F type stars
in the $B_J$ band from $20^m.4$ to $20^m.415$. Lower panel: Lambert
projection of the sky coverage of SuperCOSMOS (shown by tiny dots of
the same selection of stars in the upper panel), and the sky areas
selected for this study (squares).}\label{fig1}
\end{figure*}

\begin{figure*}
\includegraphics[scale=0.8]{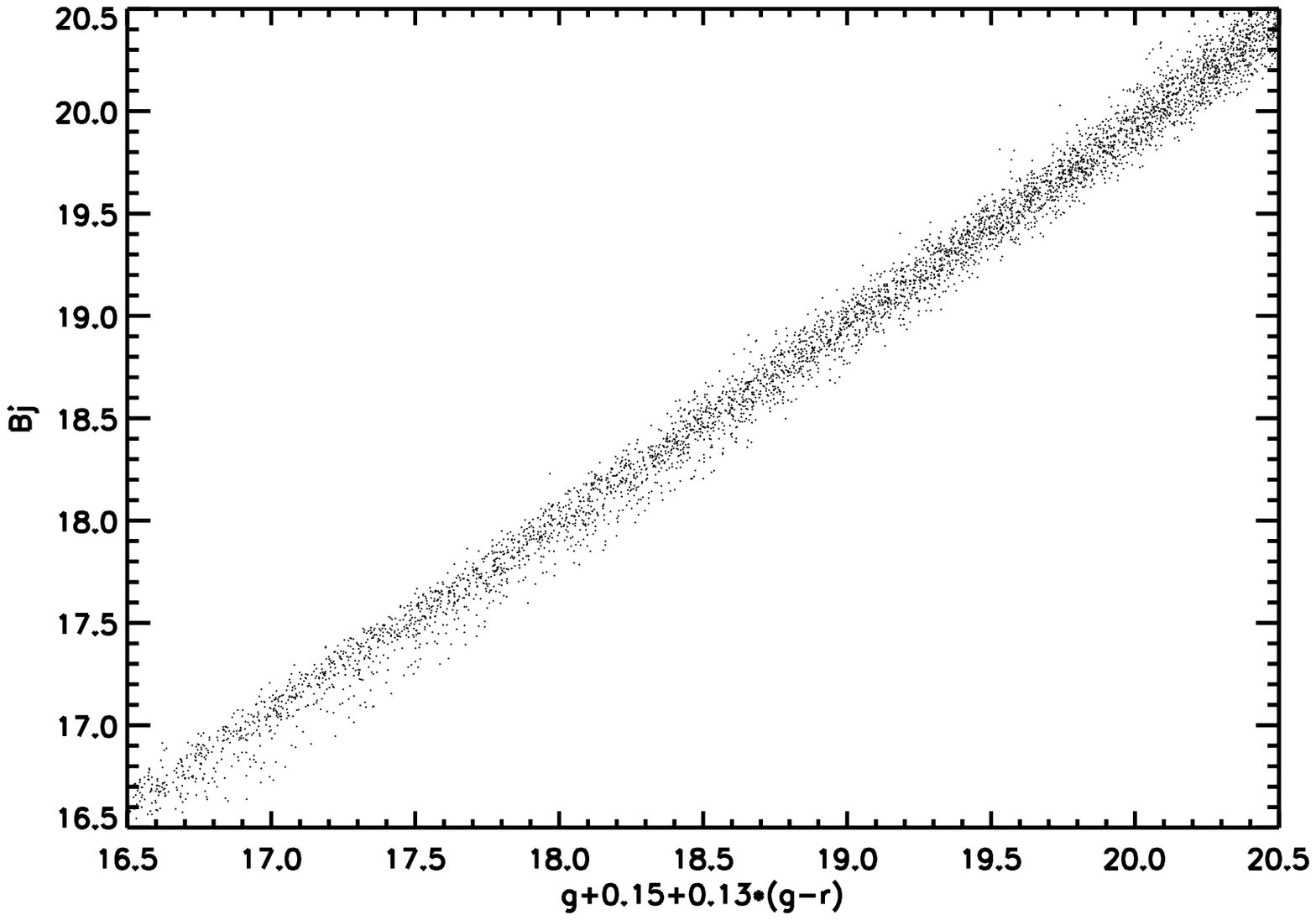}
\hfill
\includegraphics[scale=0.8]{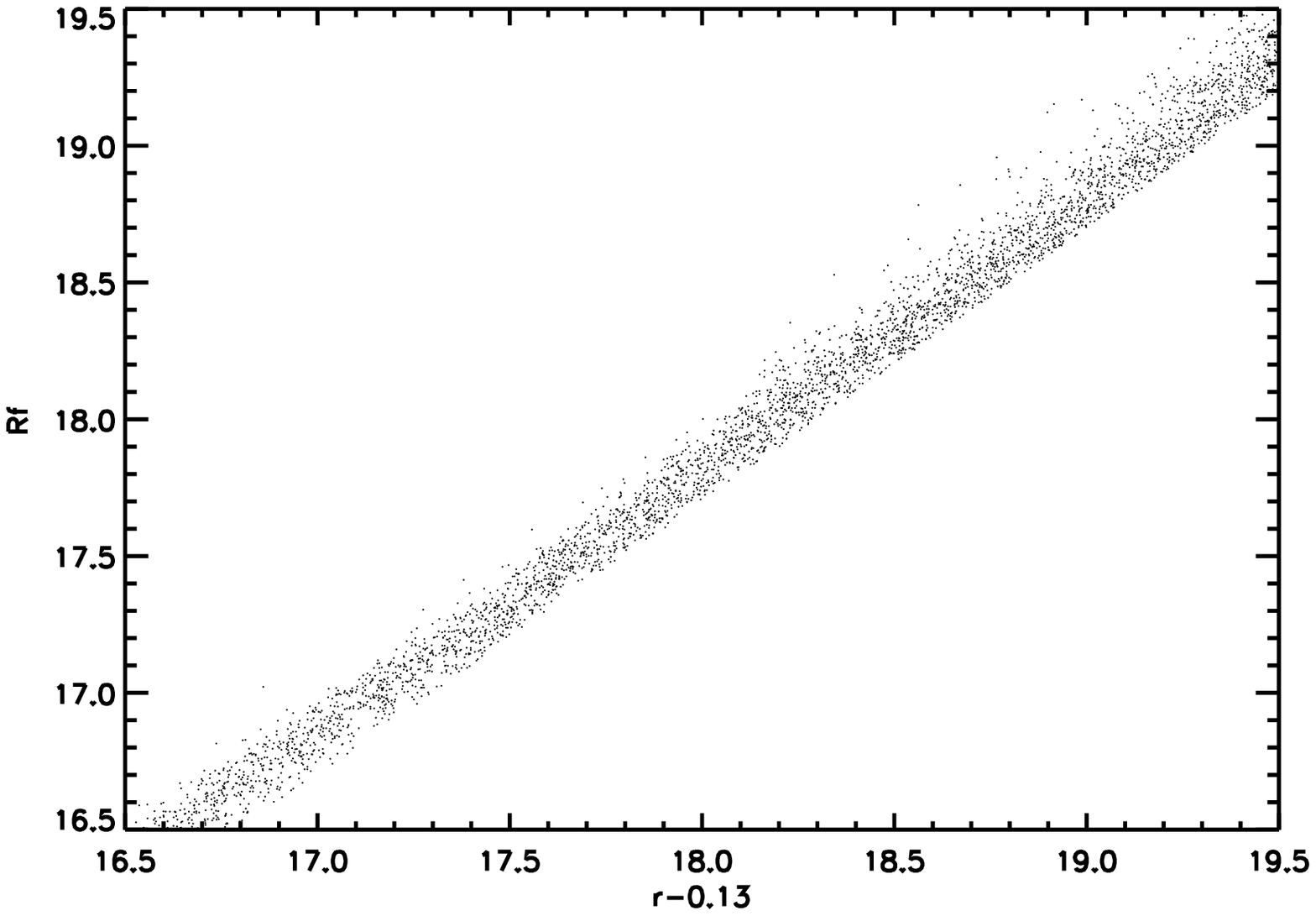}
\vspace{20pt} \caption{The color transformation between SuperCOSMOS
$B_J$, $R_F$ bands and SDSS g,r bands. The obvious offset between
the two systems infers that a systematic correction is needed. See
text for details.}\label{fig2}
\end{figure*}

\begin{figure*}
\includegraphics[scale=0.8]{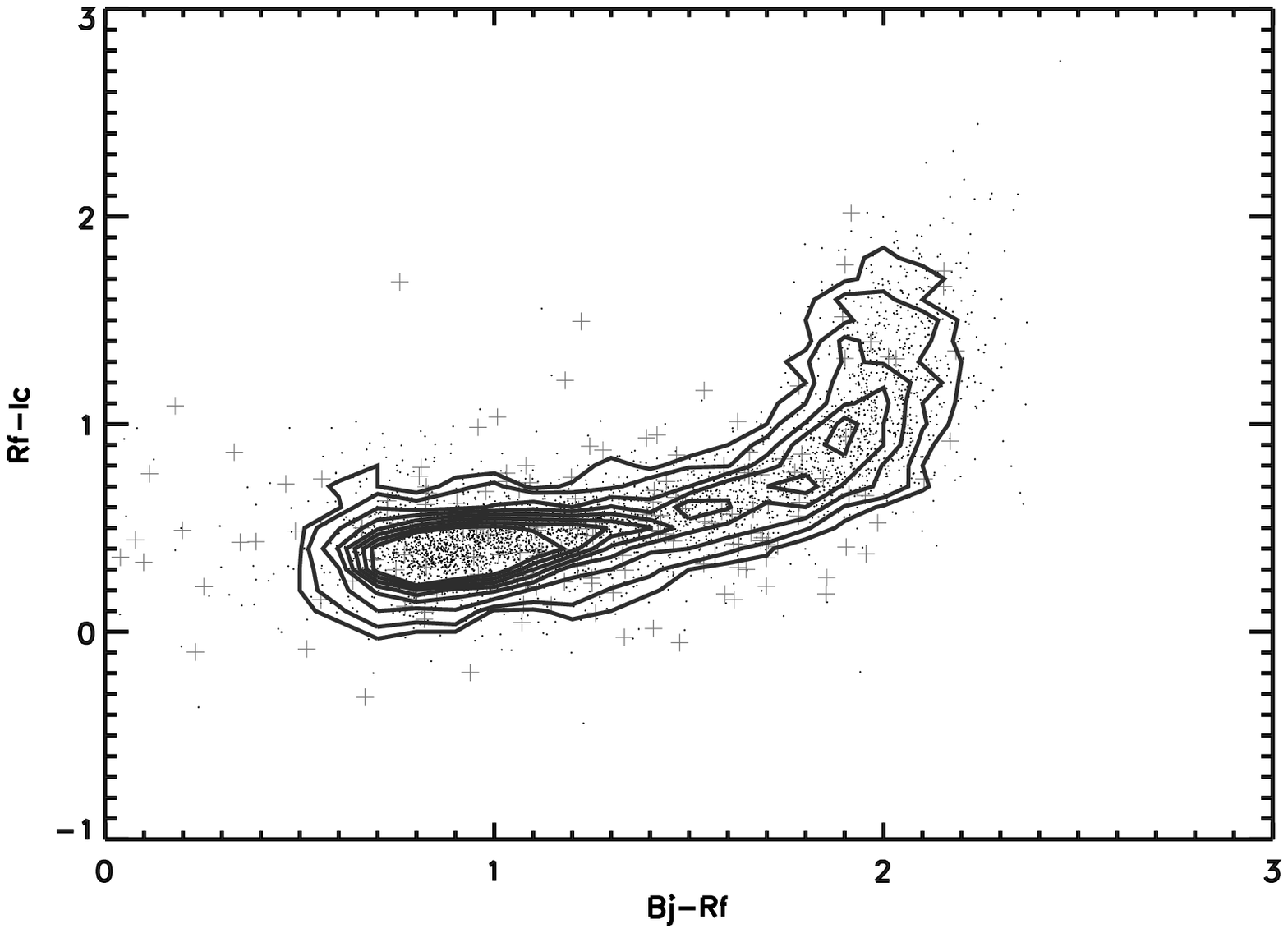}
\hfill
\includegraphics[scale=0.8]{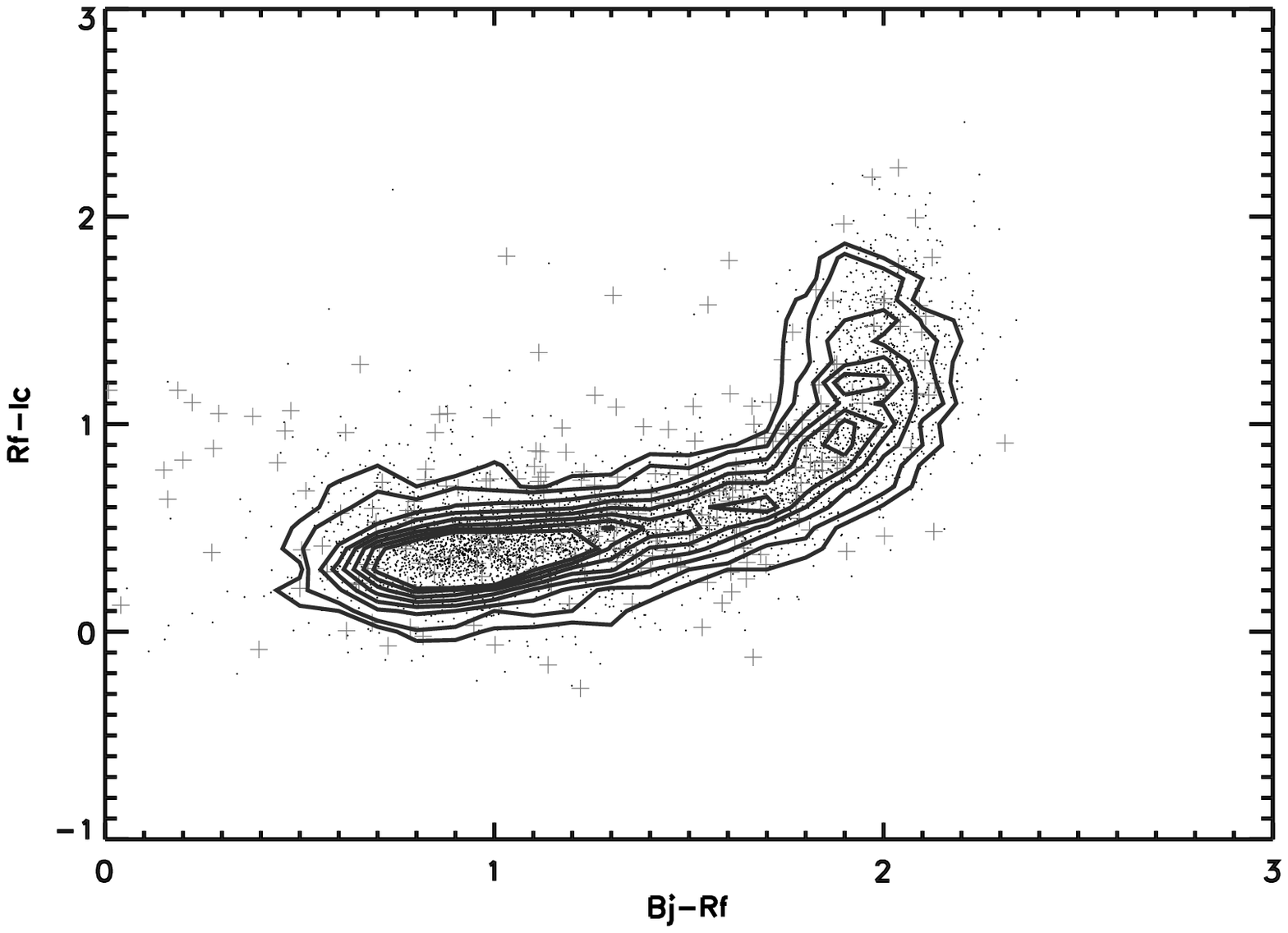}
\vspace{20pt} \caption{The cross identification between the
SuperCOSMOS and SDSS data sets in color-color space. The contour and
the black points on the color-color diagram represent the $B_J$ band
matched sources in the same overlapping sky areas, limited in
$0^m.3$ in magnitude and 10 arcsec in angular distance box and a
magnitude interval of $16^m.5<B_J,B_{JSDSS}<20^m.5$. The crosses
represent the unmatched sources. The upper panel: The northern
overlapped sky area around $(280^{\circ},60^{\circ})$; The lower
panel: the southern one around $(62^{\circ},-59^{\circ})$.}
\label{fig3}
\end{figure*}

\begin{figure*}
  \includegraphics[scale=0.4]{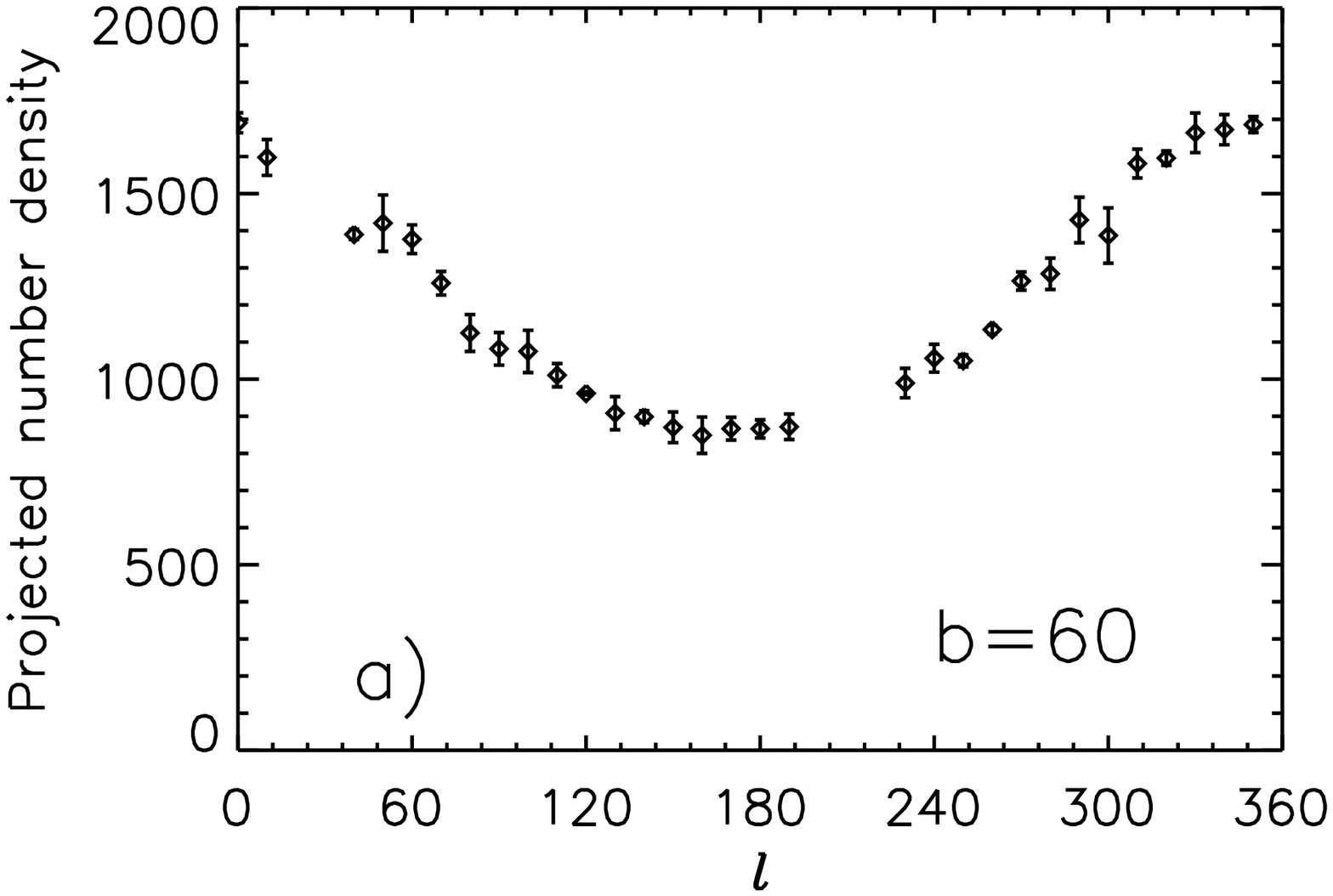}
  \hfill
  \includegraphics[scale=0.4]{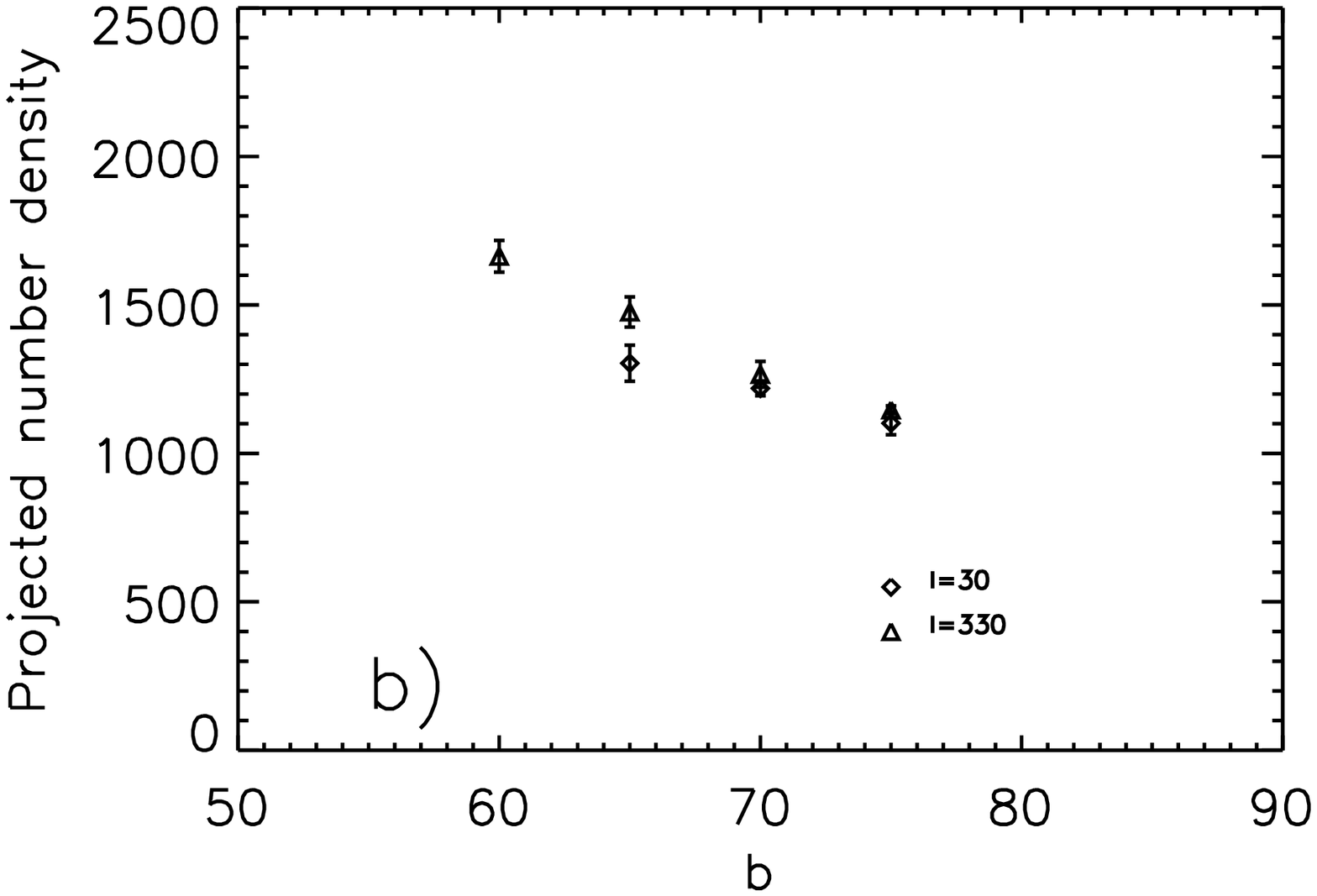}
  \vfill
  \includegraphics[scale=0.4]{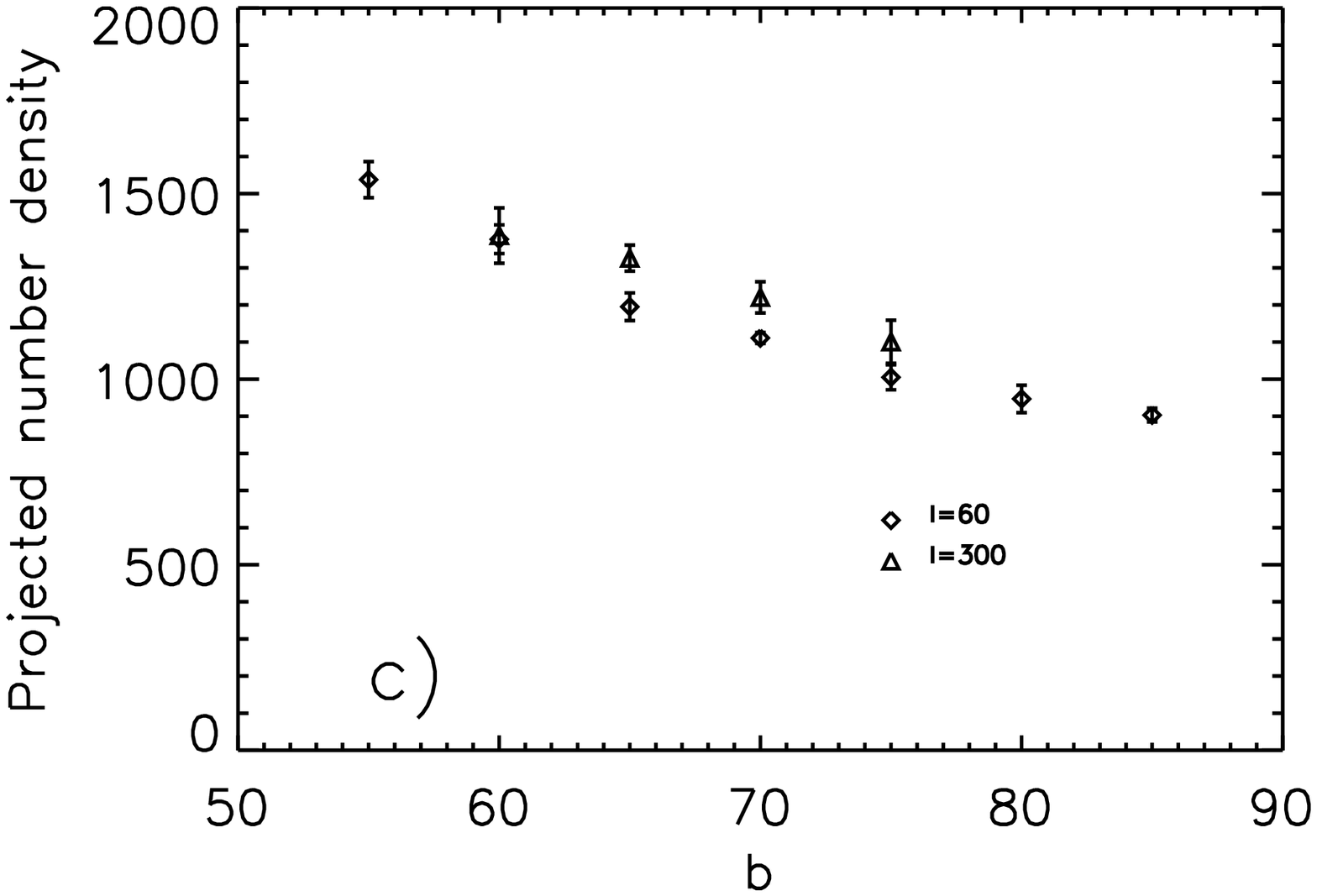}
  \hfill
  \includegraphics[scale=0.4]{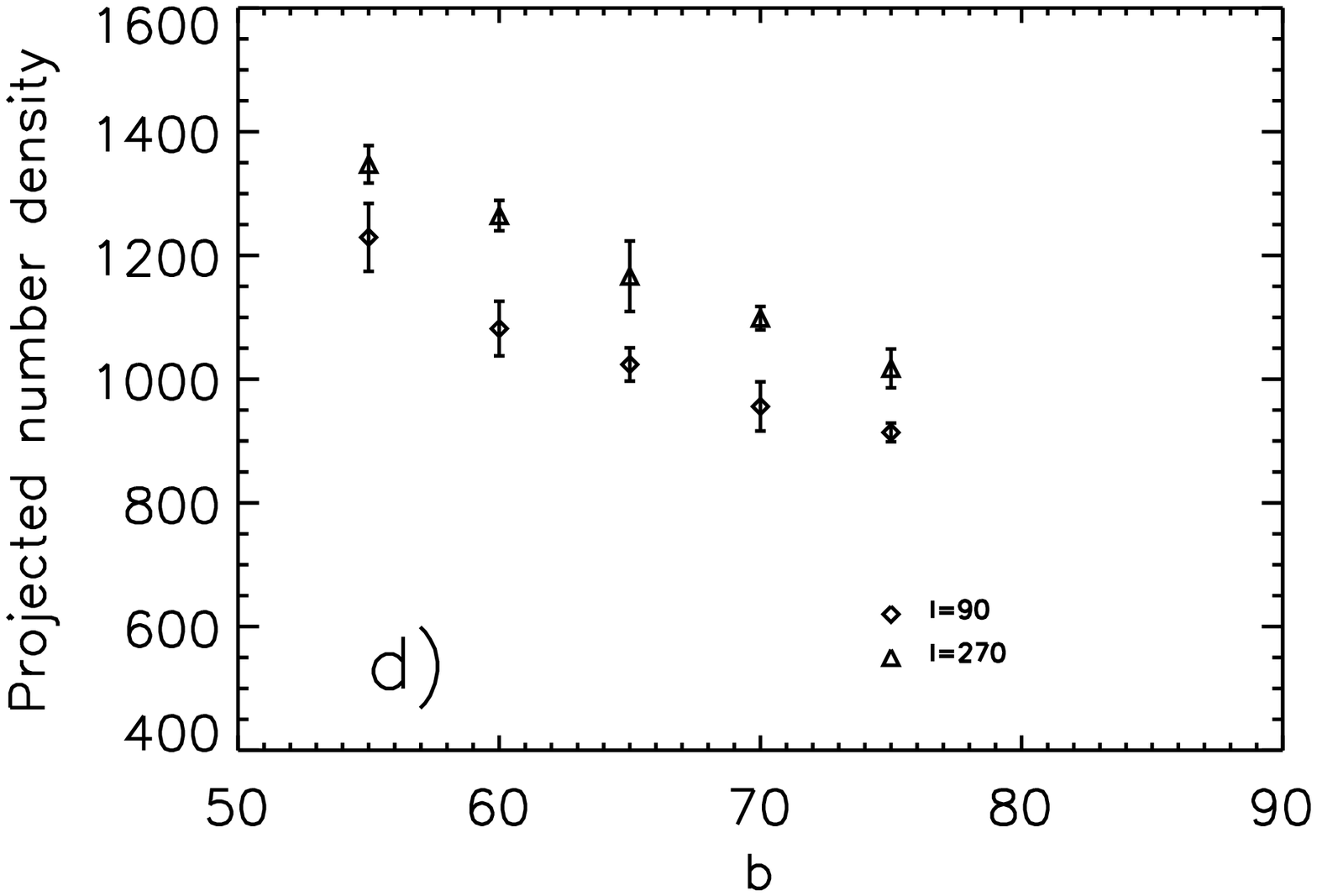}
  \vfill
  \includegraphics[scale=0.4]{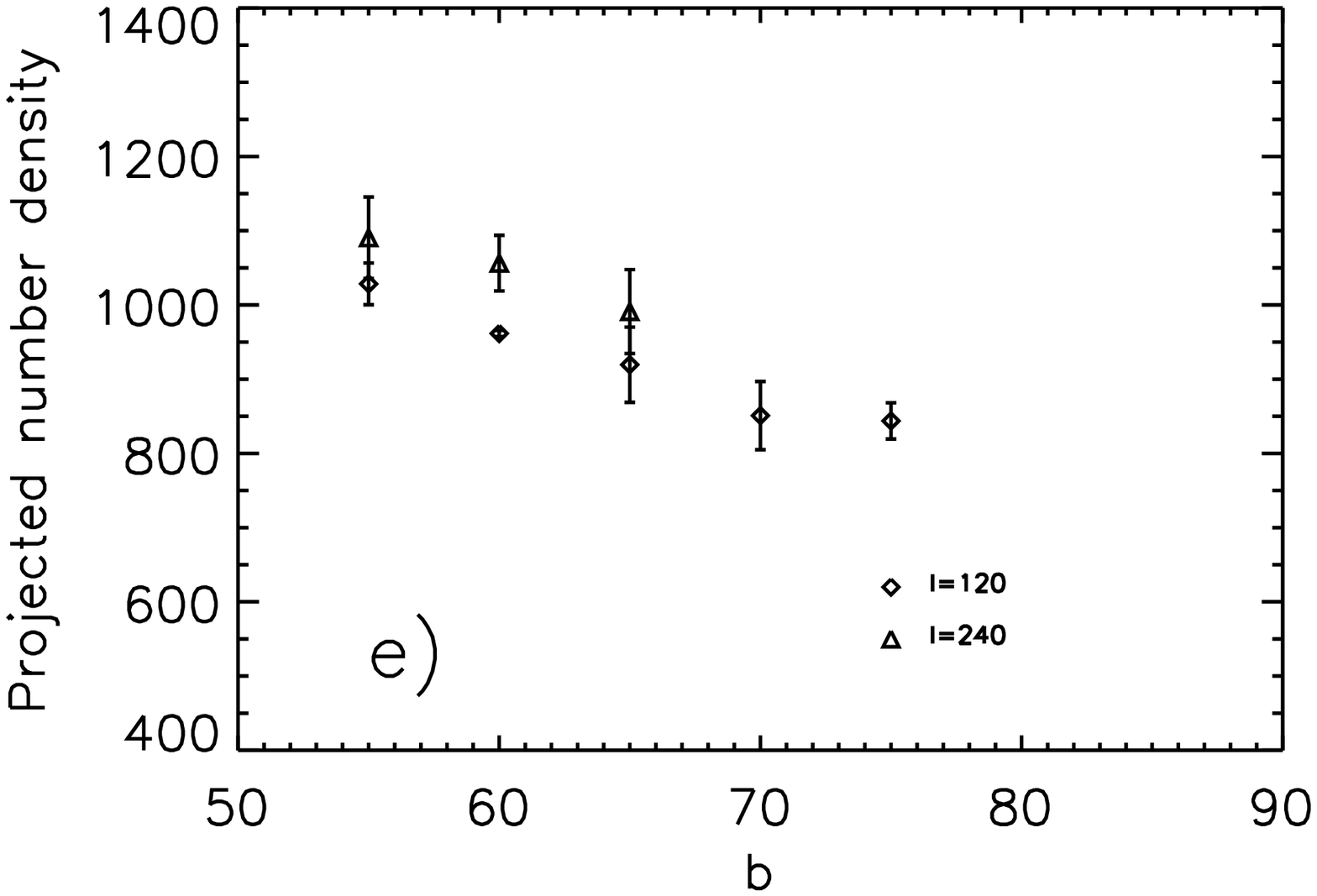}
  \hfill
  \includegraphics[scale=0.4]{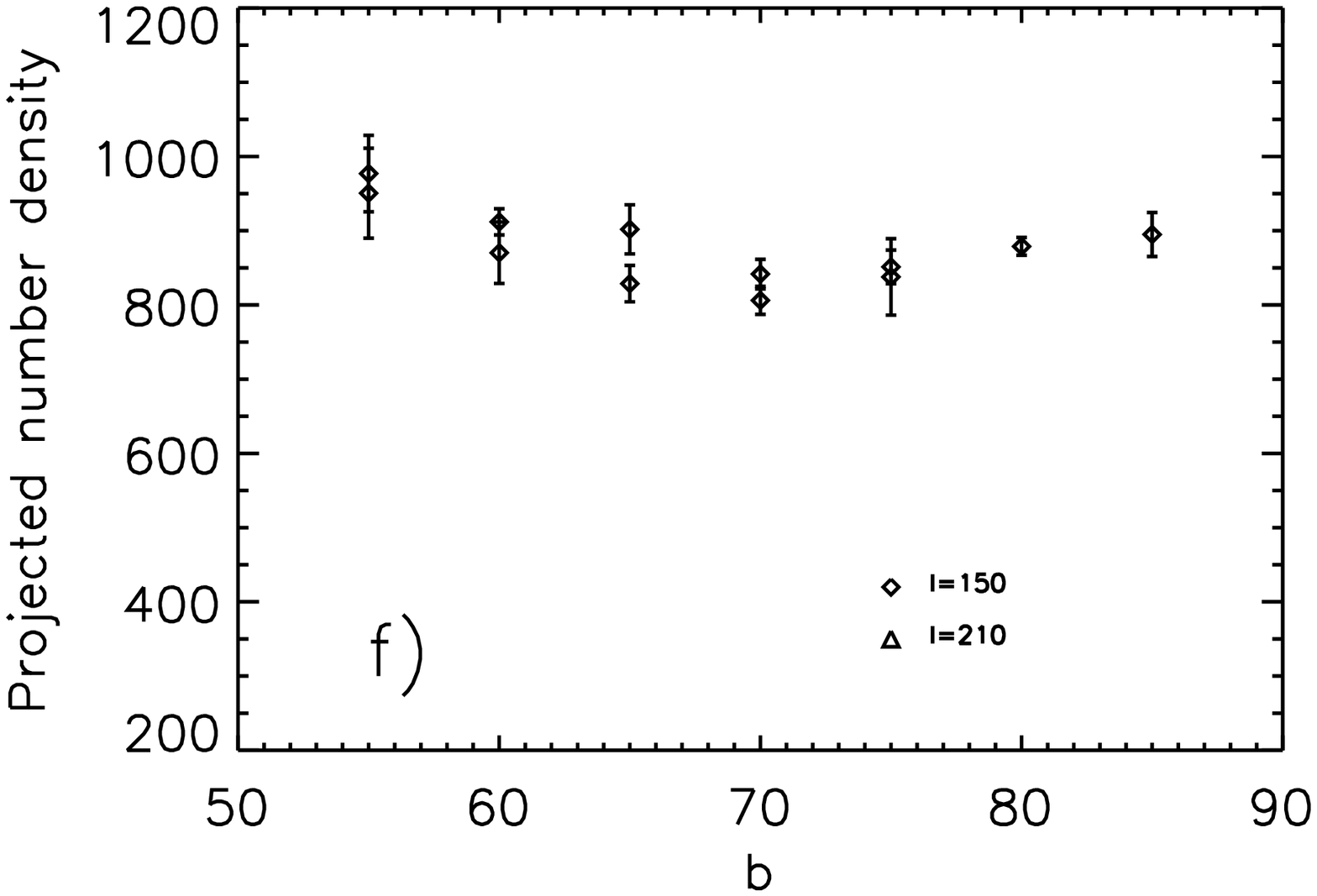}
\vspace{20pt} \caption{Projected surface number density of SDSS data
in the same selected sky areas as in XDH06, but downgraded to
$16^m.5<B_{JSDSS}<20^m.5$. Panel a). is for the selected areas along
a circle of $b=60^{\circ}$, the horizontal axis is the Galactic
longitude in degrees; while the others are for the ones along
different paired longitudinal directions, with the longitudes
indicated in the inlet of each panel, all the horizontal axis in
panels b)-f) are the Galactic latitude in degrees.}\label{fig4}
\end{figure*}

\begin{figure*}
\includegraphics[scale=0.4]{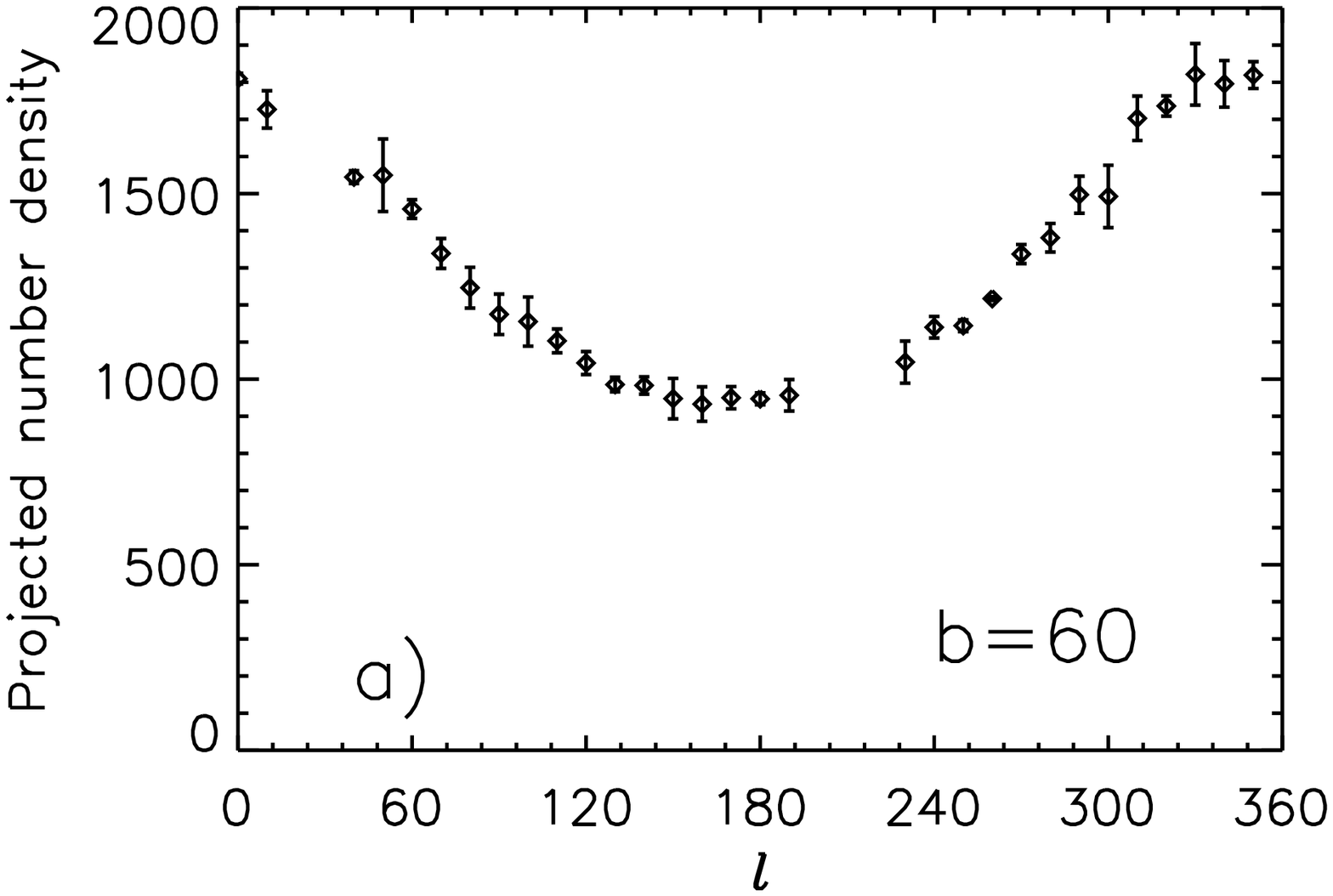}
\hfill
\includegraphics[scale=0.4]{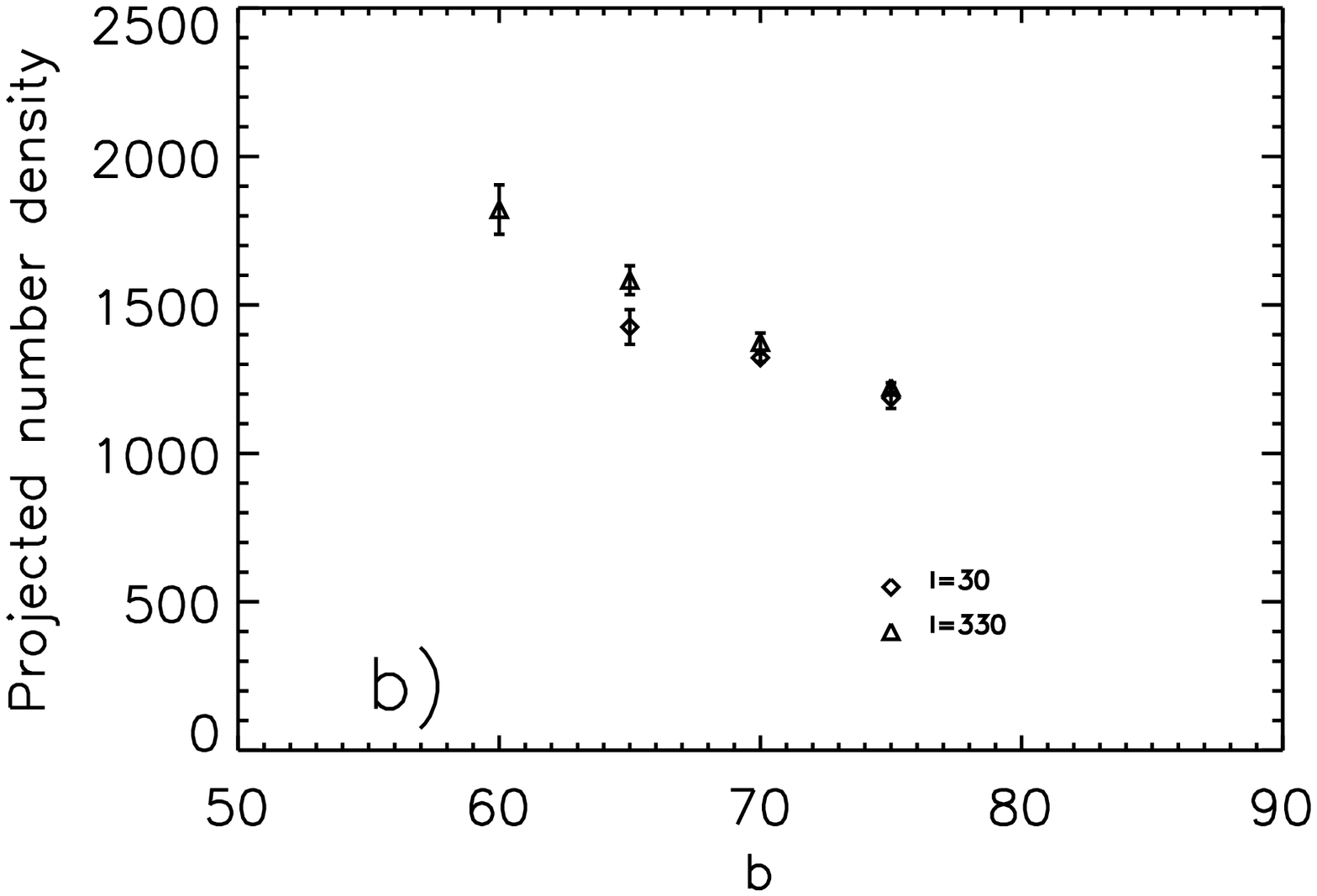}
\vfill
\includegraphics[scale=0.4]{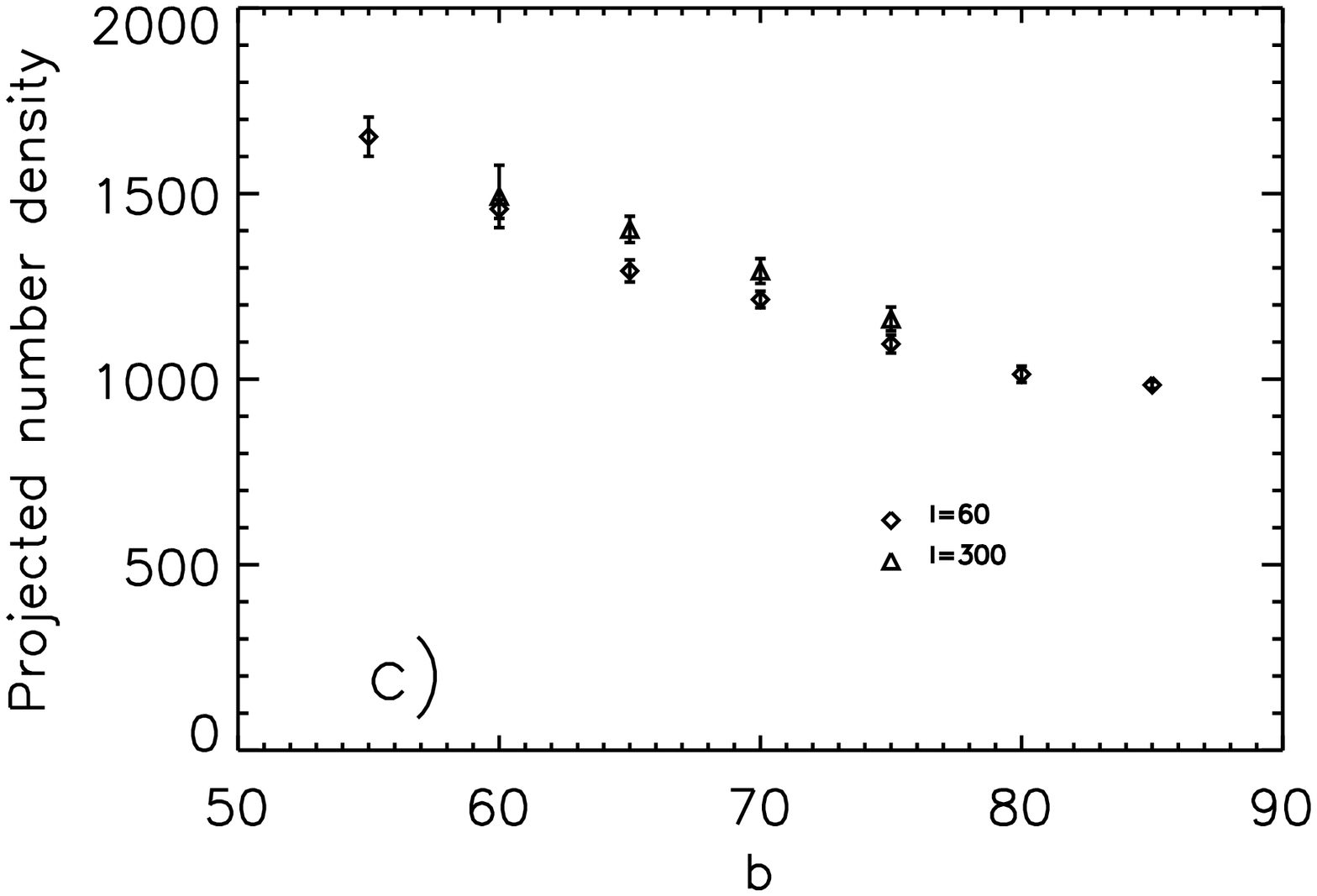}
\hfill
\includegraphics[scale=0.4]{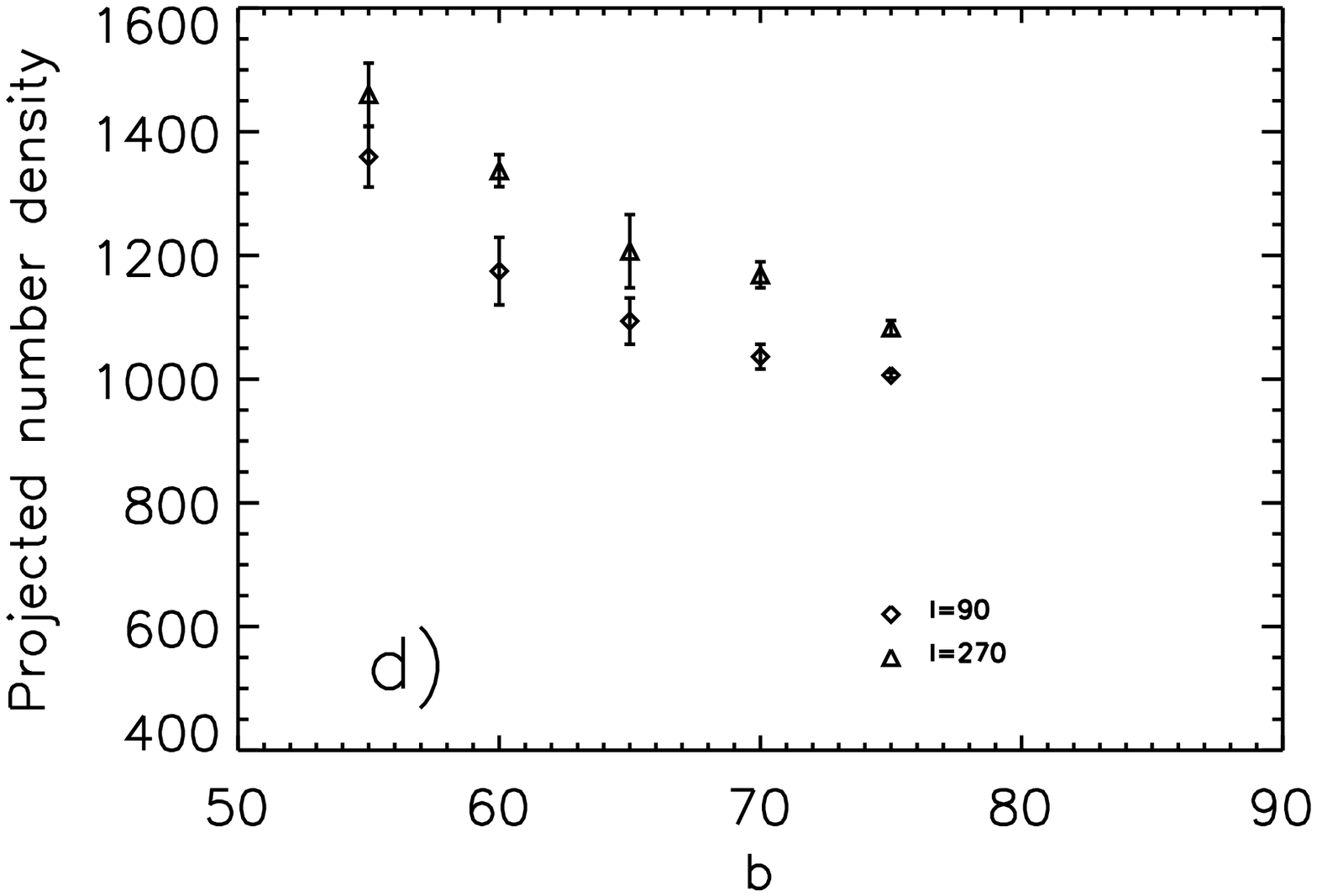}
\vfill
\includegraphics[scale=0.4]{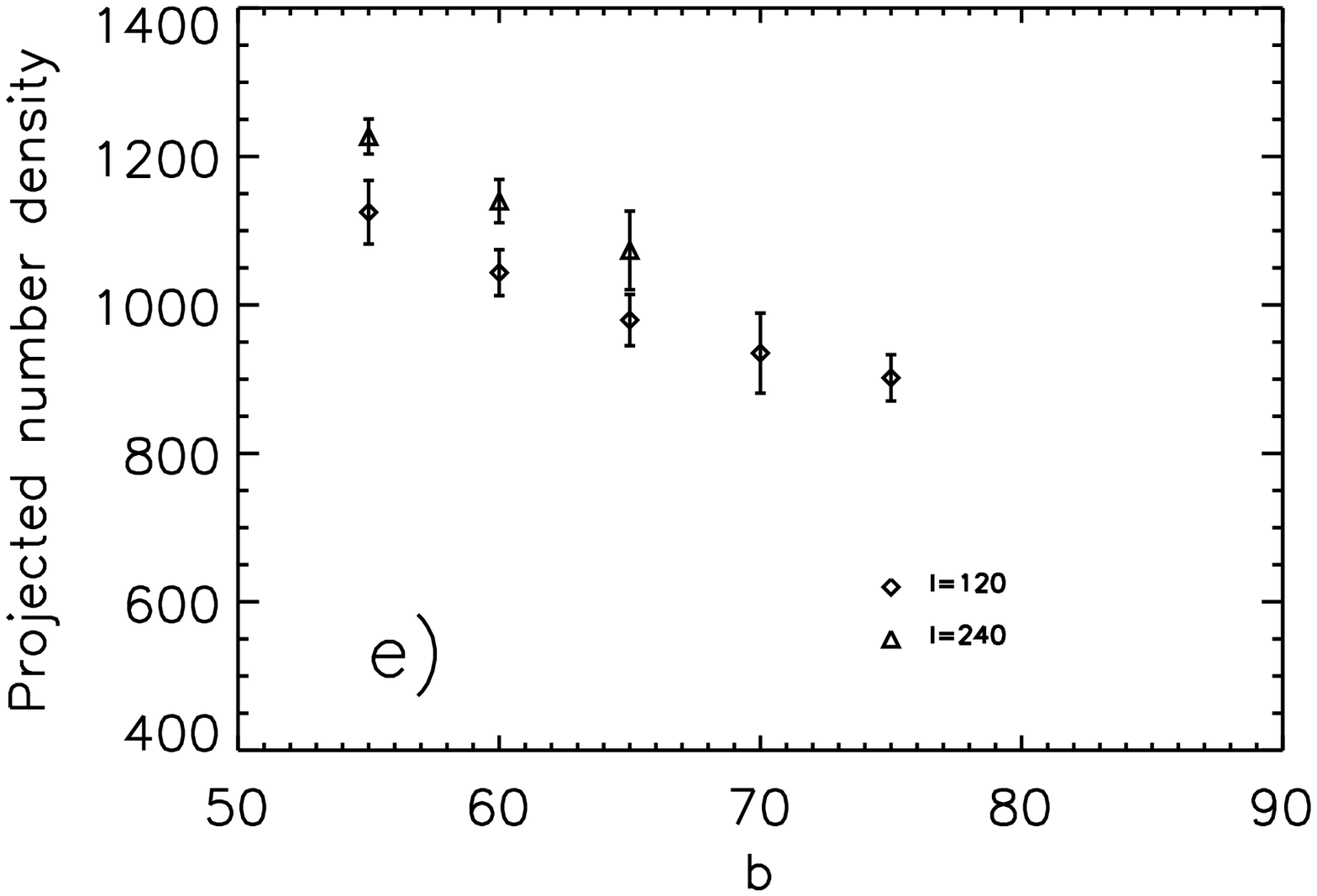}
\hfill
\includegraphics[scale=0.4]{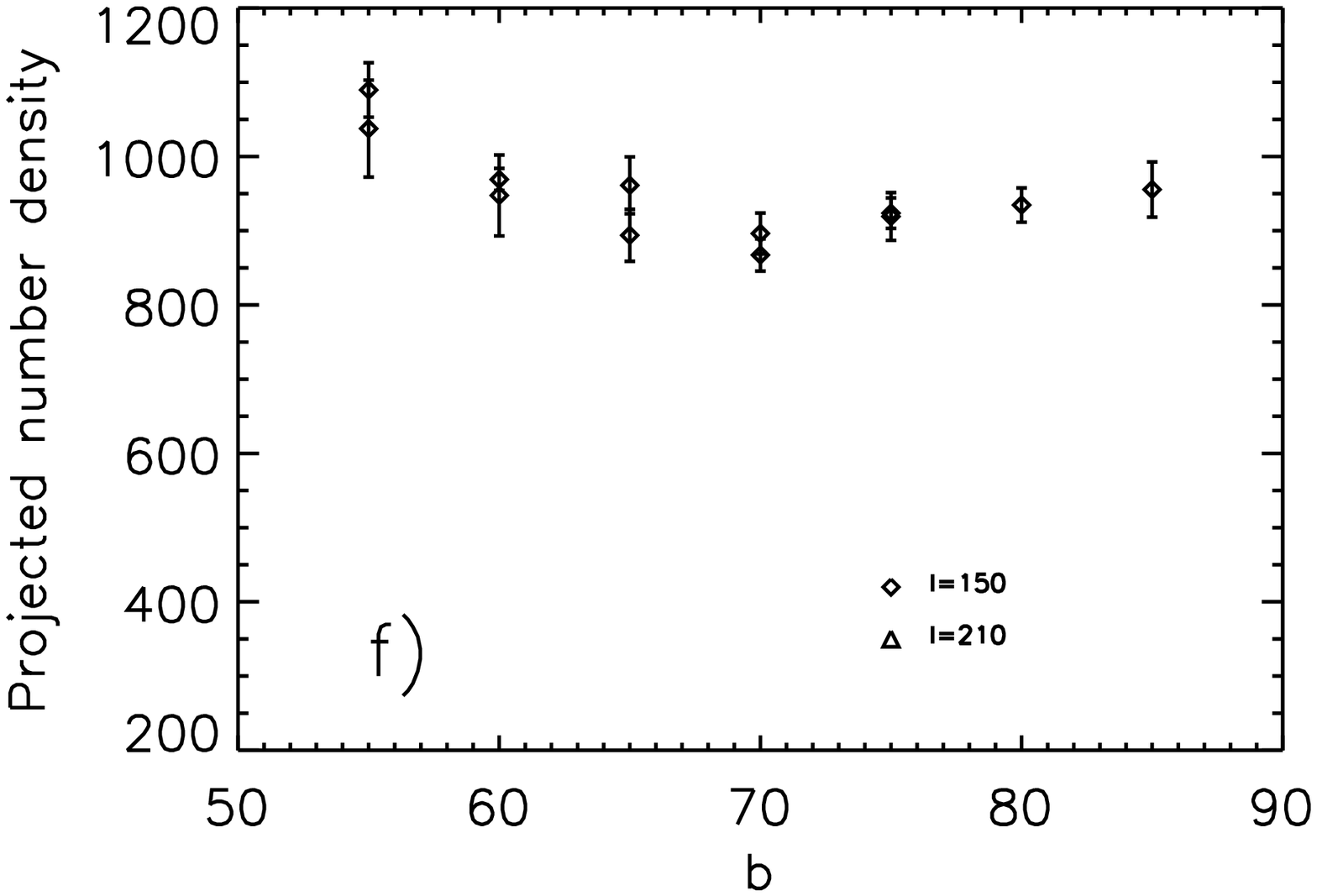}
\vspace{20pt} \caption{The same as figure~\ref{fig4} but for the
projected surface number density of SDSS data in $R_{JSDSS}$
downgraded to $16^m.5<R_{FSDSS}<19^m.5$}\label{fig5}
\end{figure*}

\begin{figure*}
\includegraphics[scale=0.4]{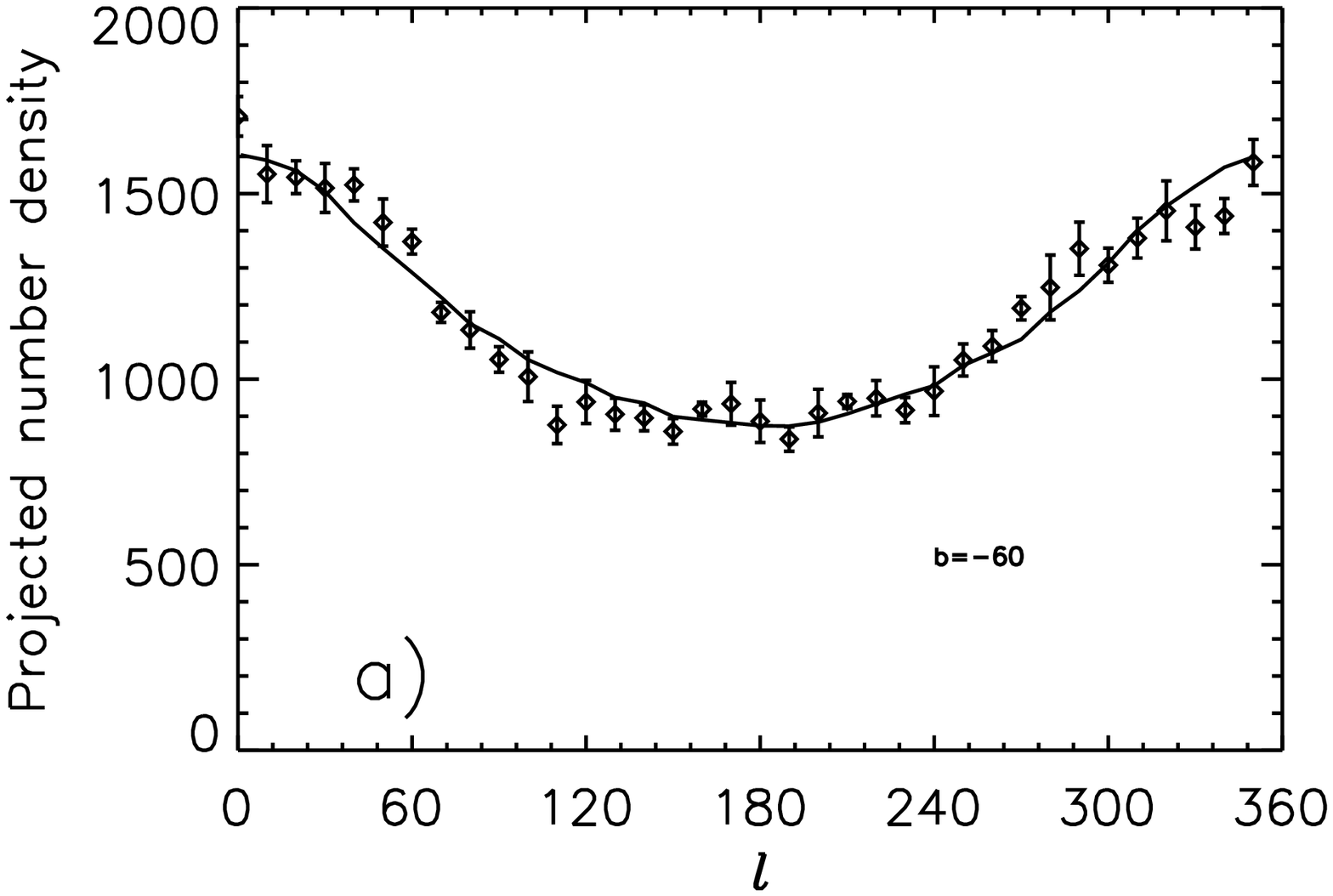}
\hfill
\includegraphics[scale=0.4]{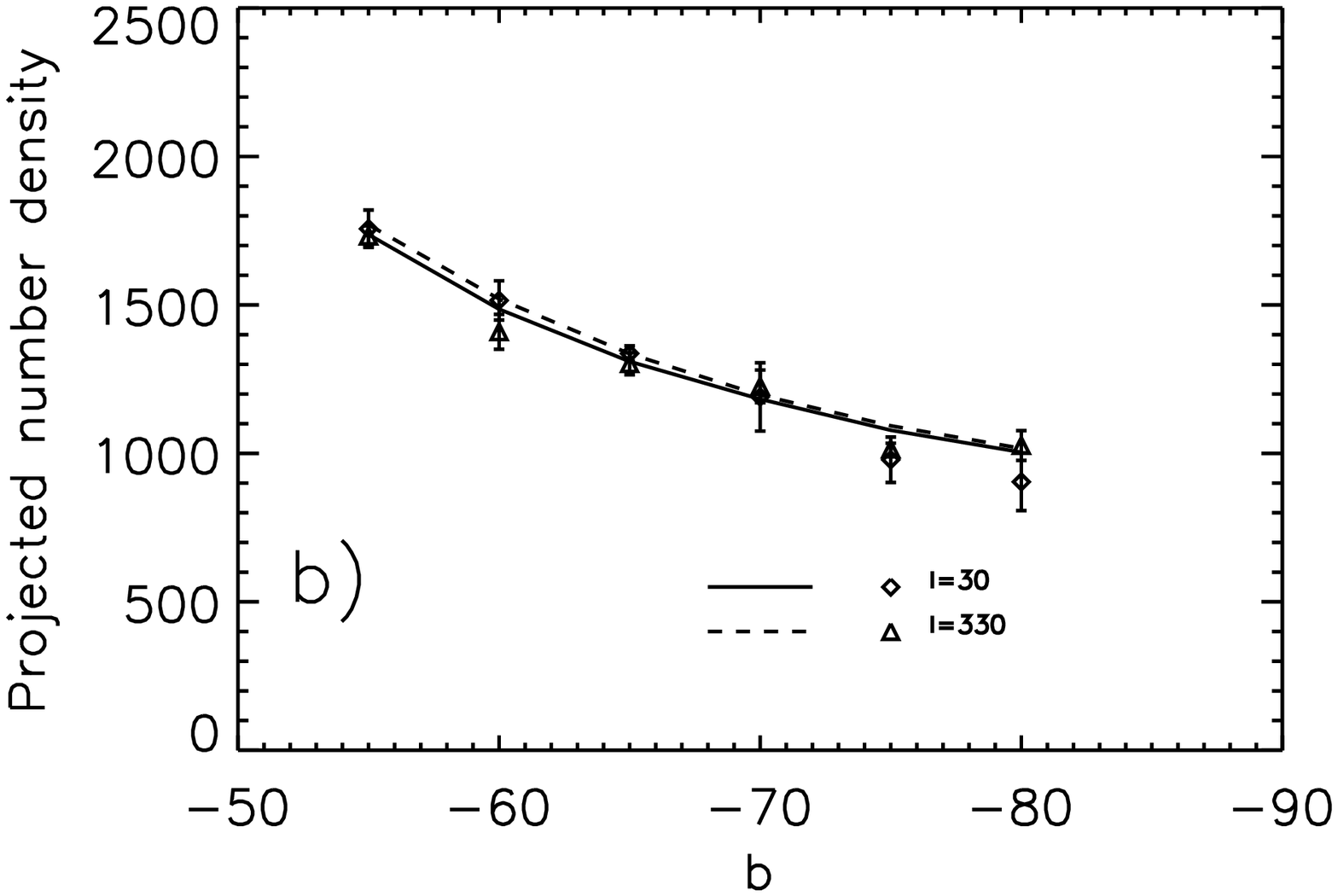}
\vfill
\includegraphics[scale=0.4]{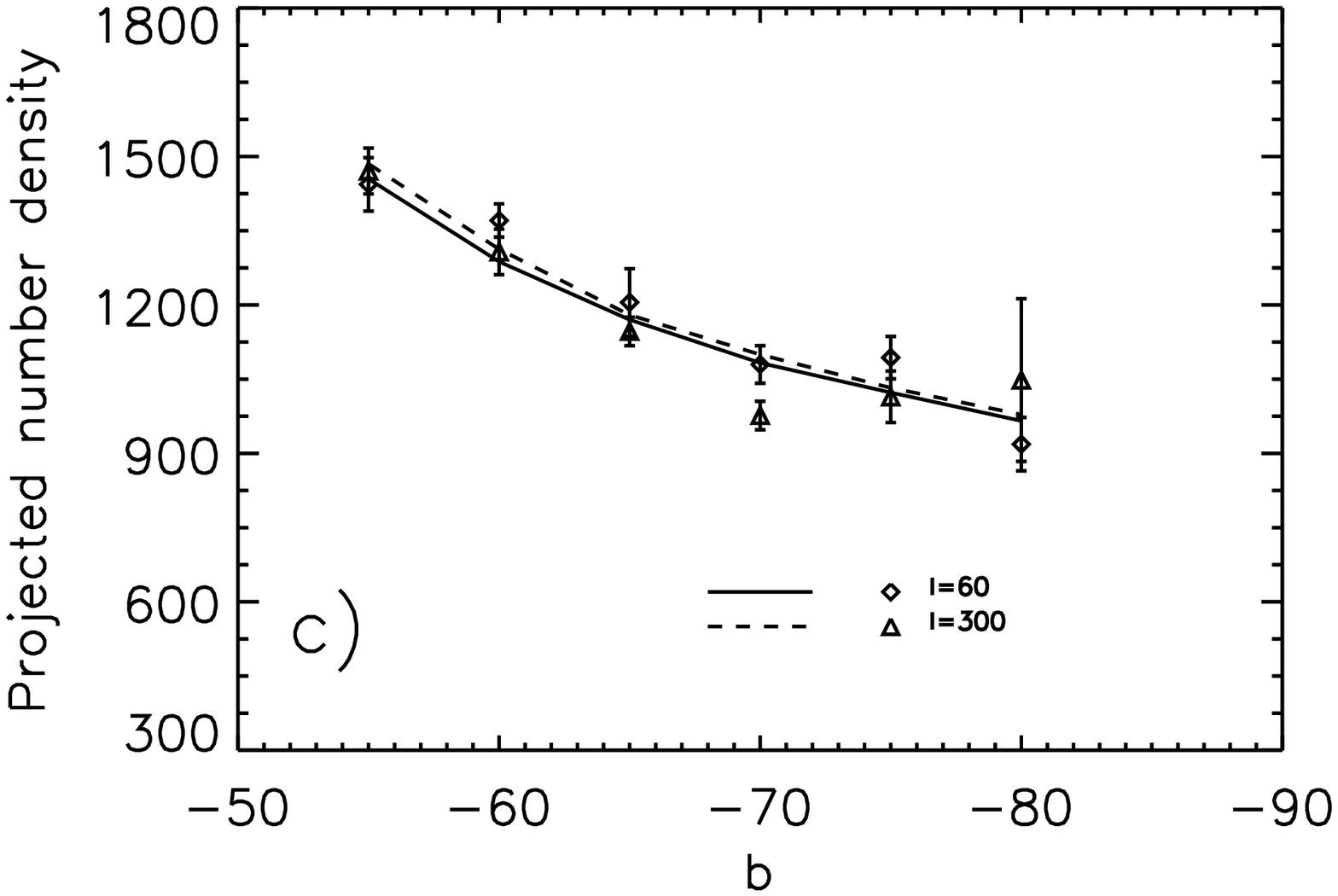}
\hfill
\includegraphics[scale=0.4]{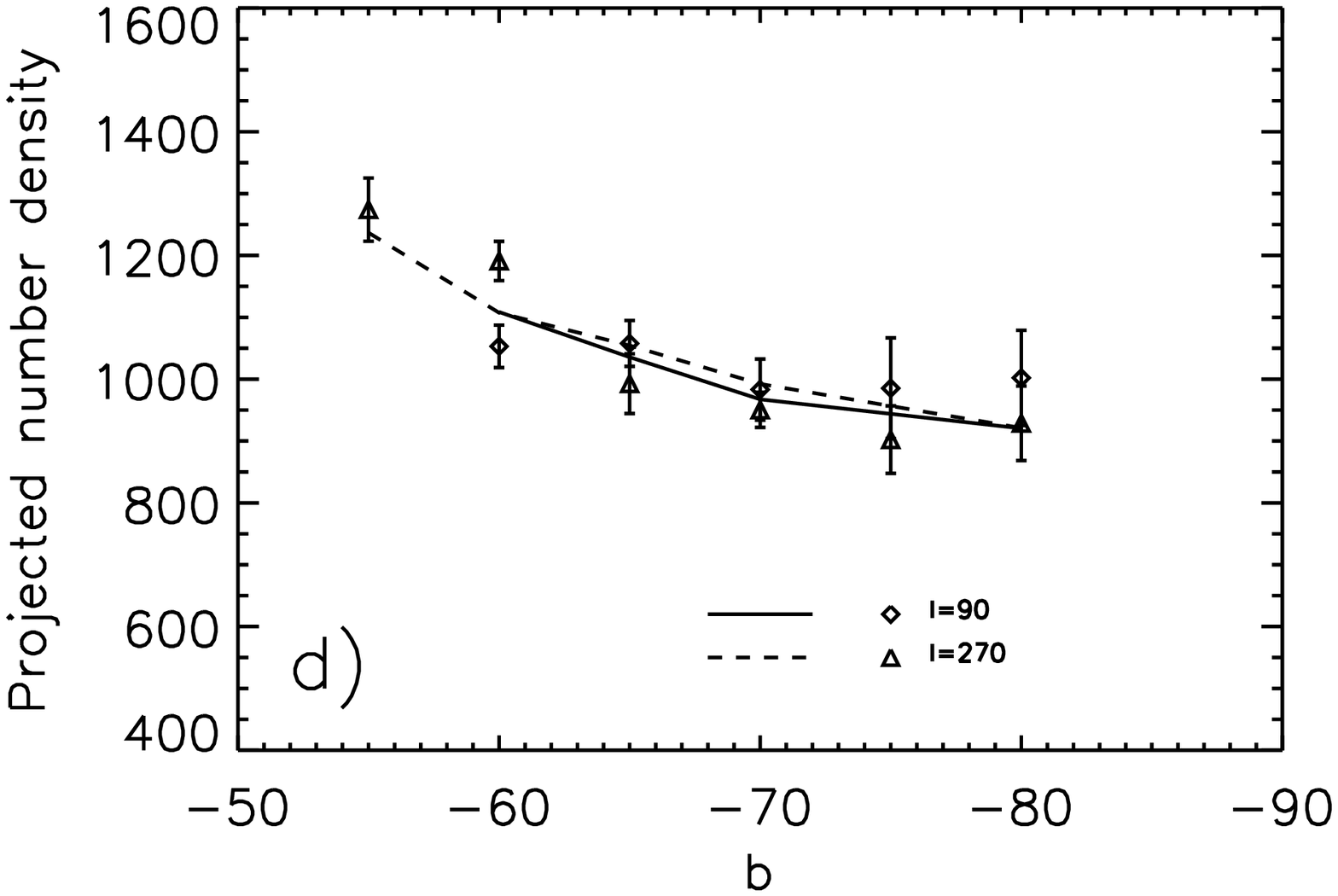}
\vfill
\includegraphics[scale=0.4]{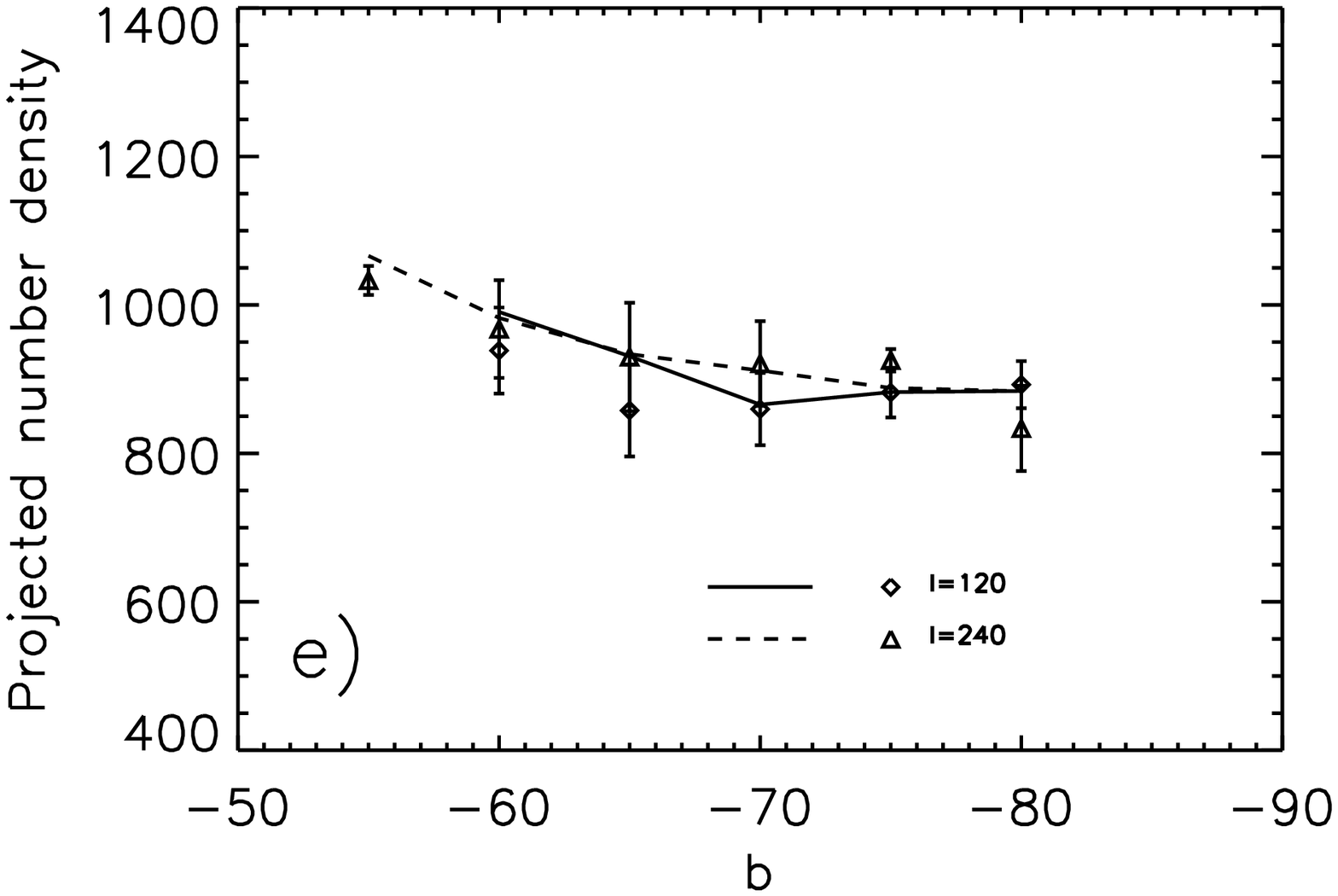}
\hfill
\includegraphics[scale=0.4]{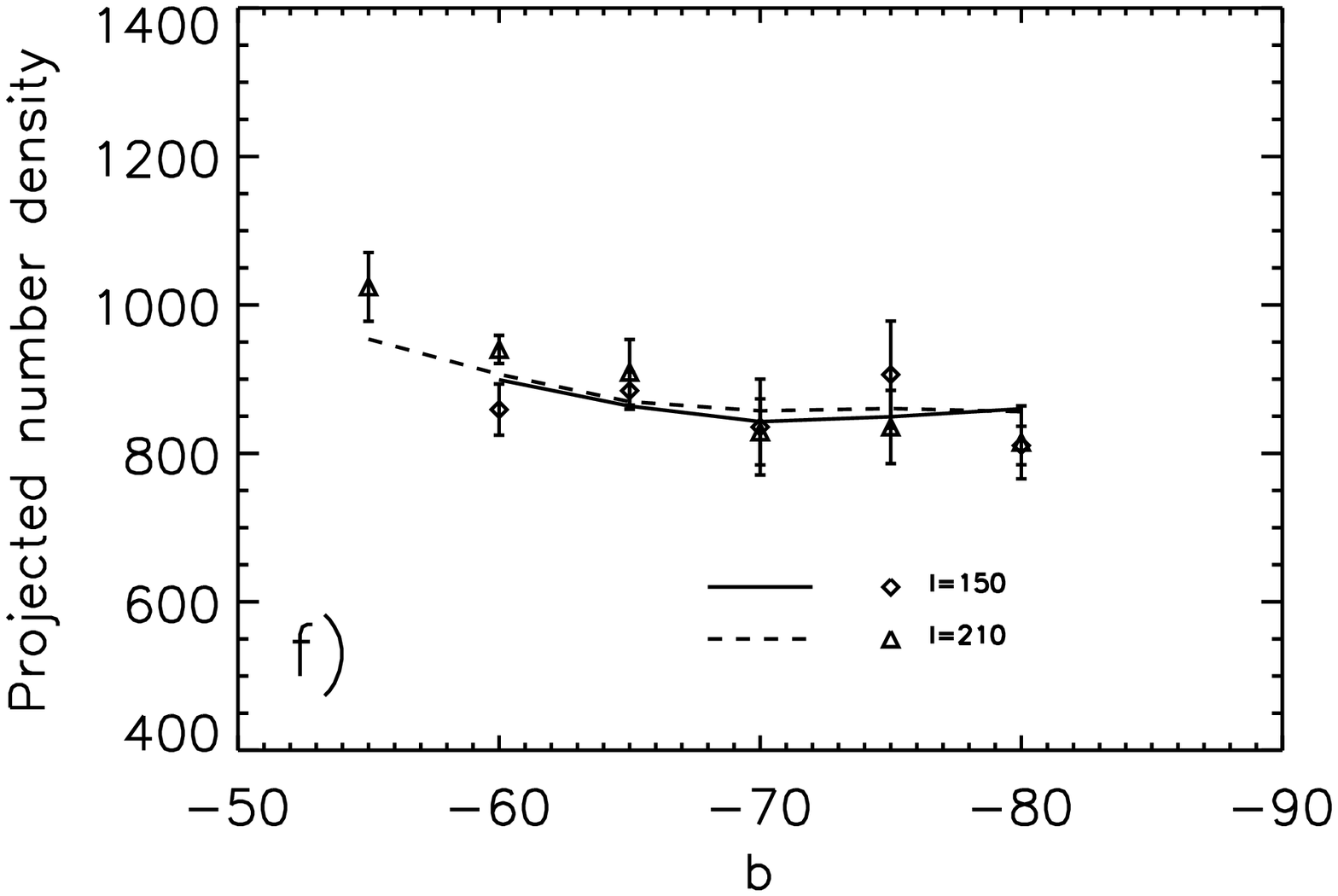}
\vspace{20pt} \caption{The same as figure~\ref{fig4}, but for the
fitting of the surface number density counted from SuperCOSMOS $B_J$
band data. The solid and dashed lines are the theoretical
predictions, while the diamonds and triangles show the observational
data.}\label{fig6}\end{figure*}

\begin{figure*}

\includegraphics[scale=0.4]{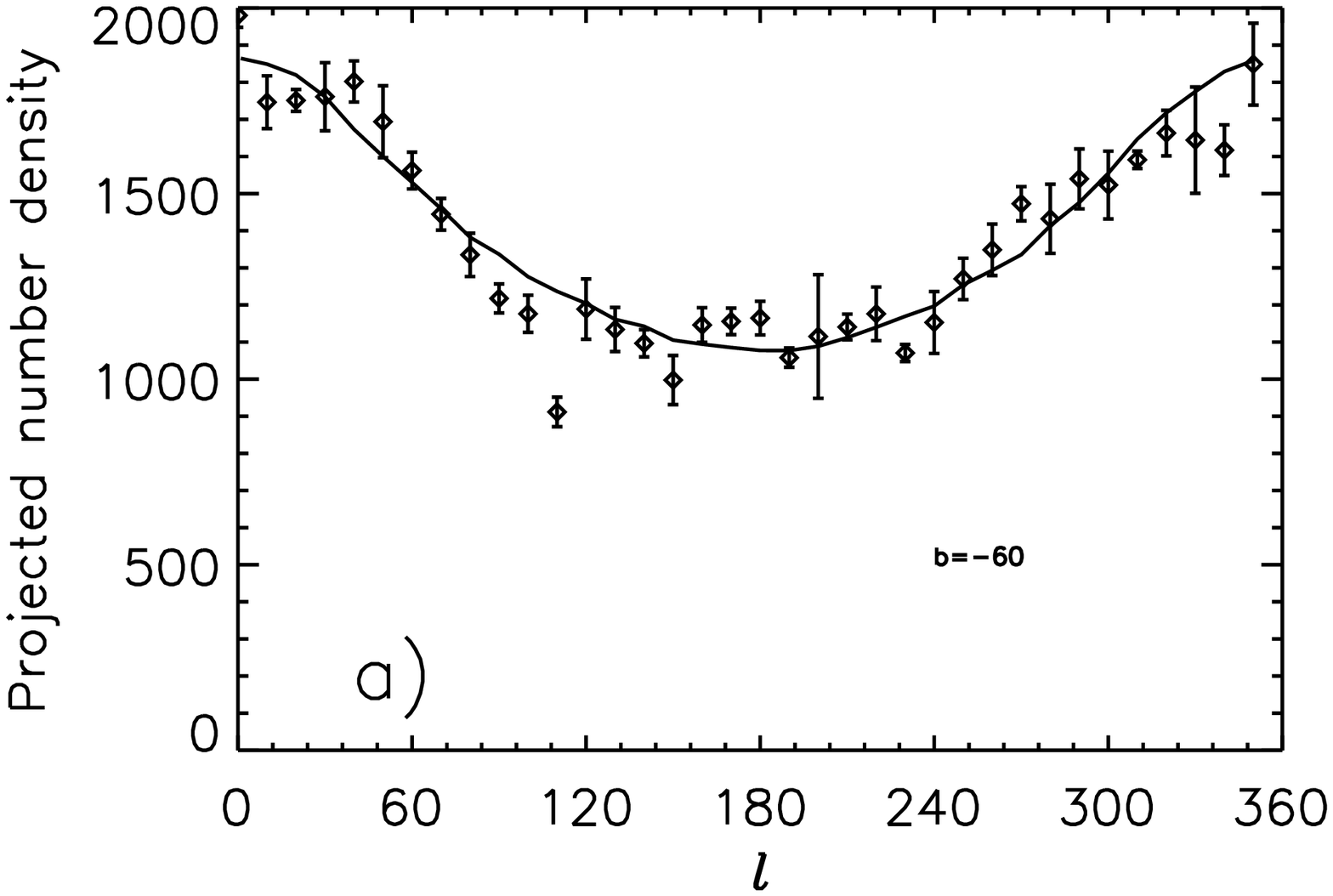}
\hfill
\includegraphics[scale=0.4]{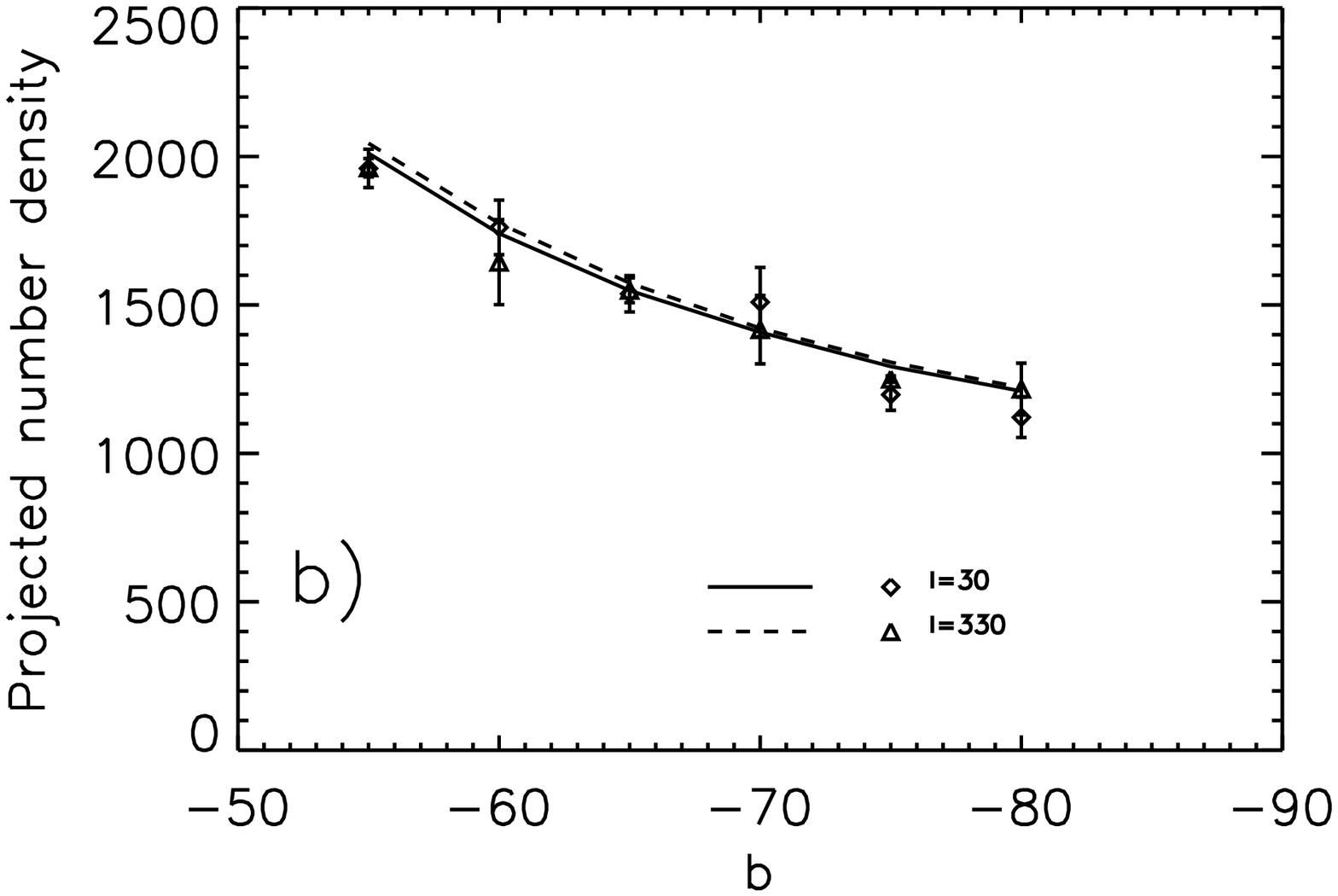}
\vfill
\includegraphics[scale=0.4]{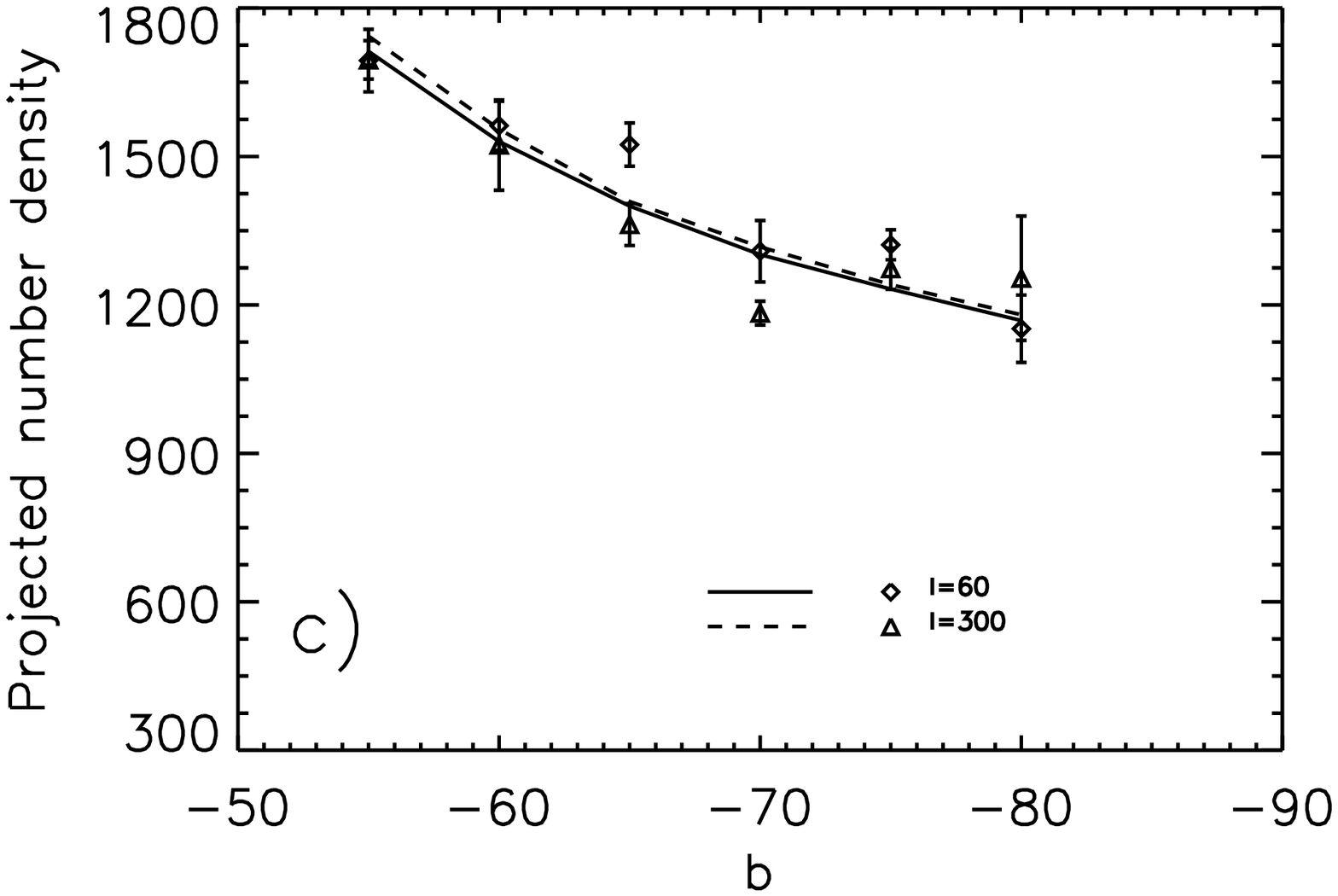}
\hfill
\includegraphics[scale=0.4]{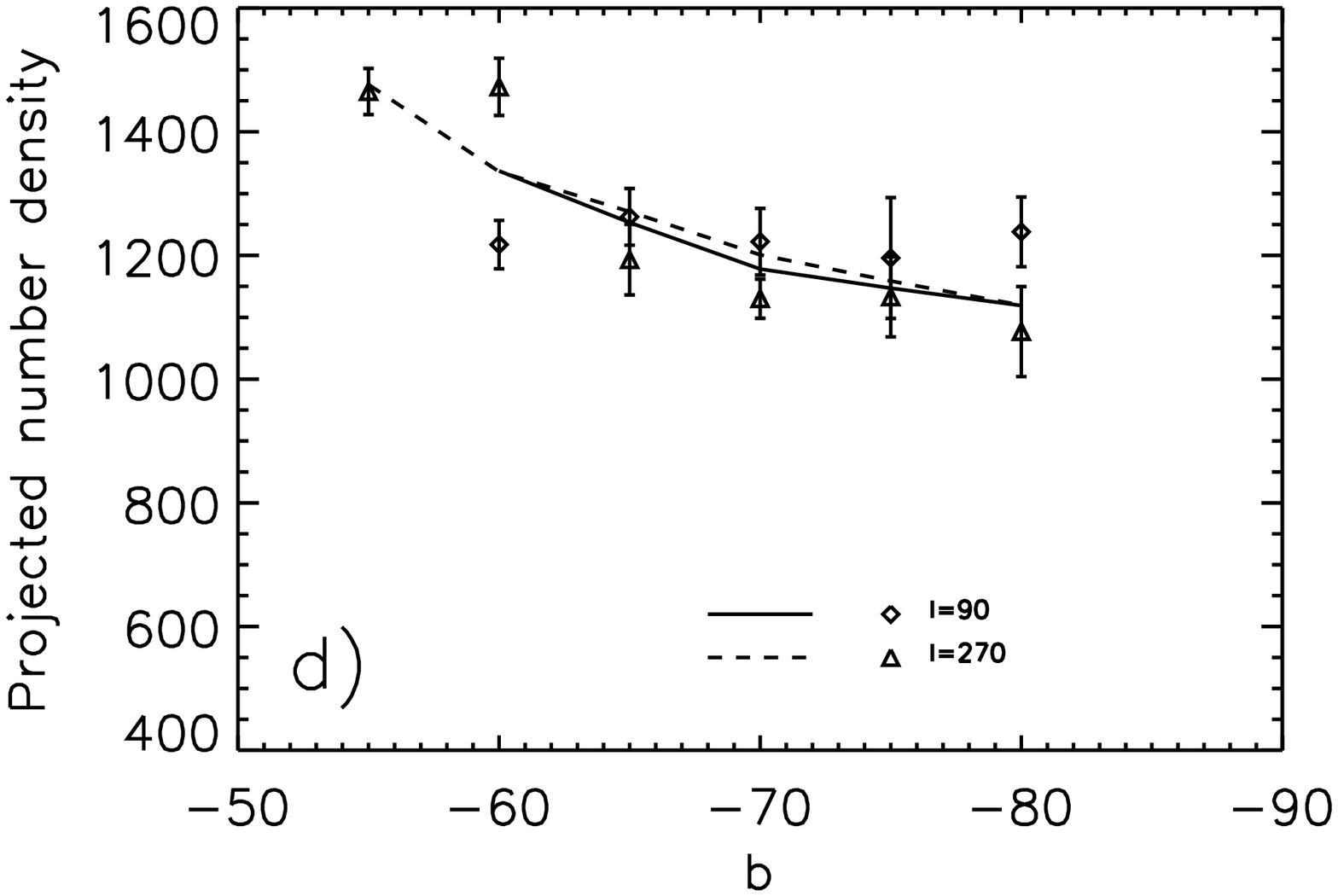}
\vfill
\includegraphics[scale=0.4]{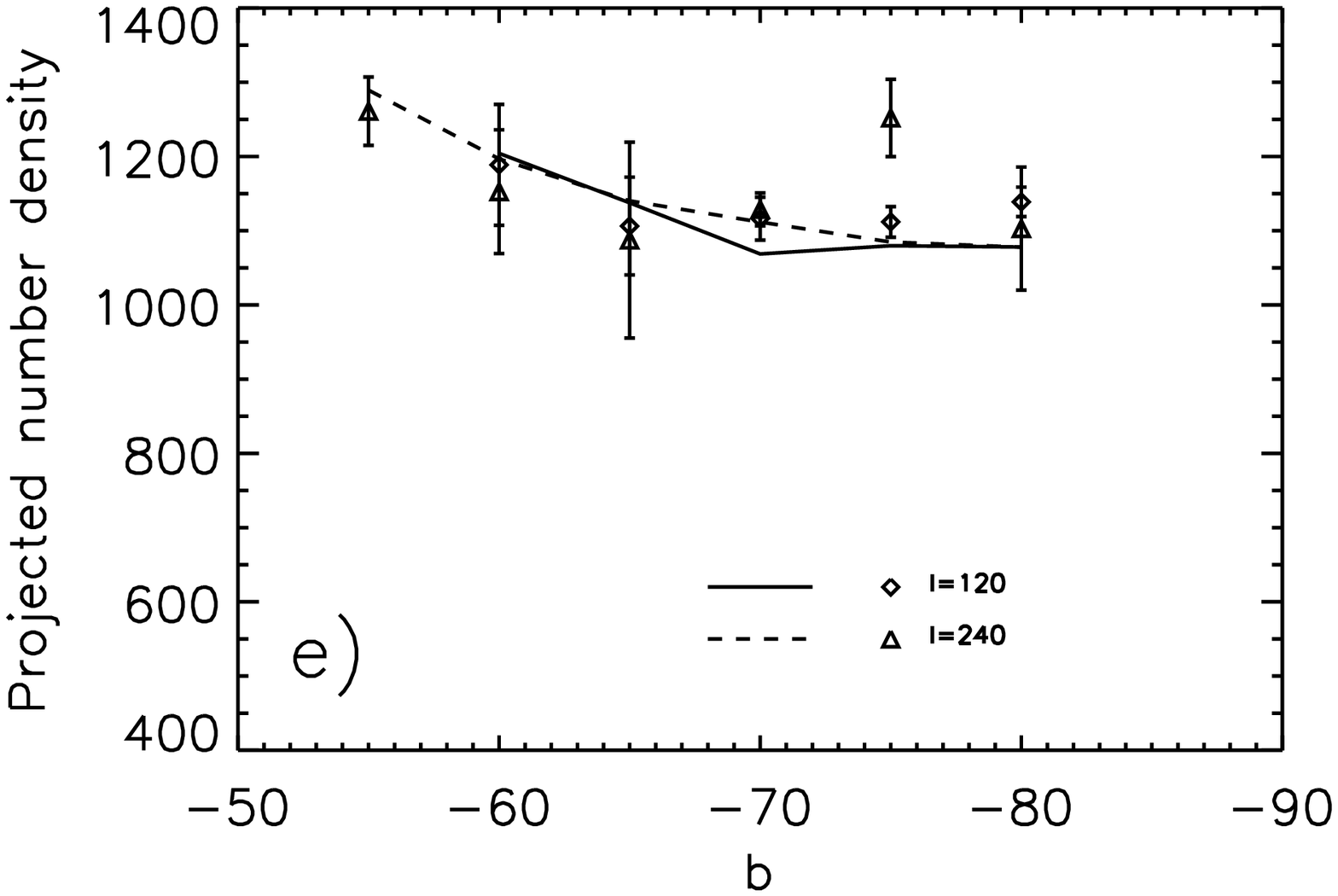}
\hfill
\includegraphics[scale=0.4]{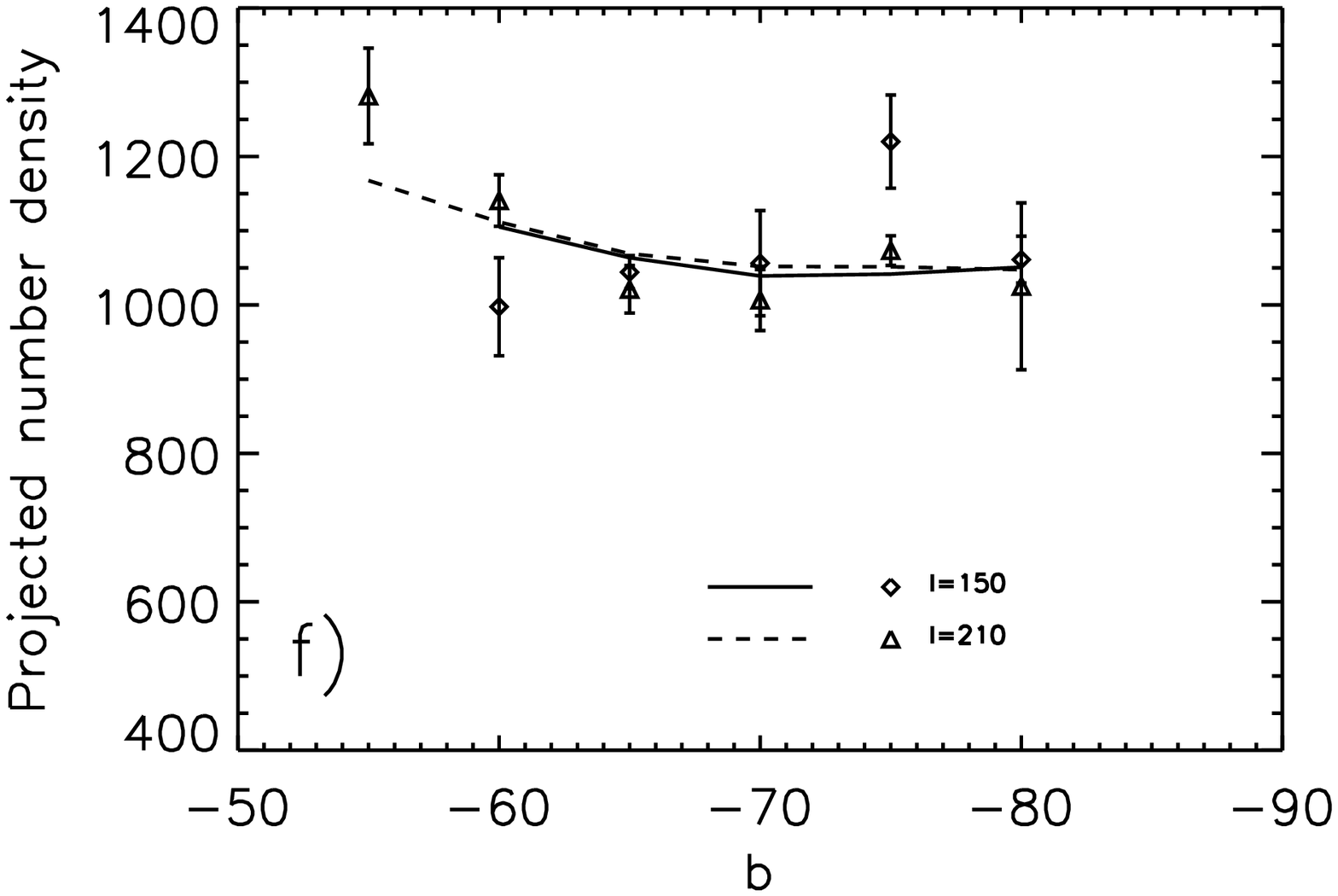}
\vspace{20pt} \caption{The same as figure~\ref{fig6}, but for $R_F$
band results.}\label{fig7}\end{figure*}

\begin{figure*}
\includegraphics[scale=0.45]{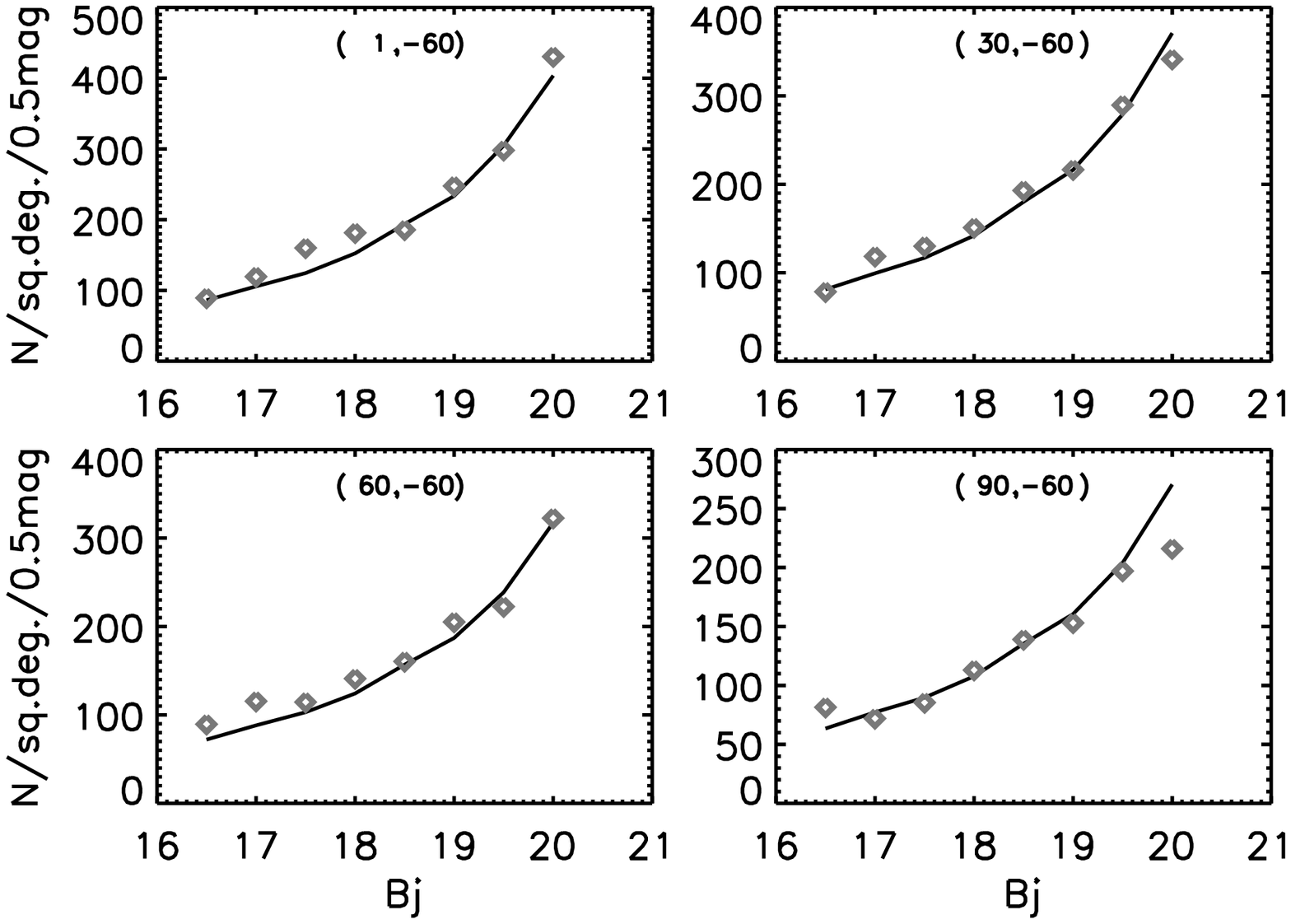}
\hfill
\includegraphics[scale=0.45]{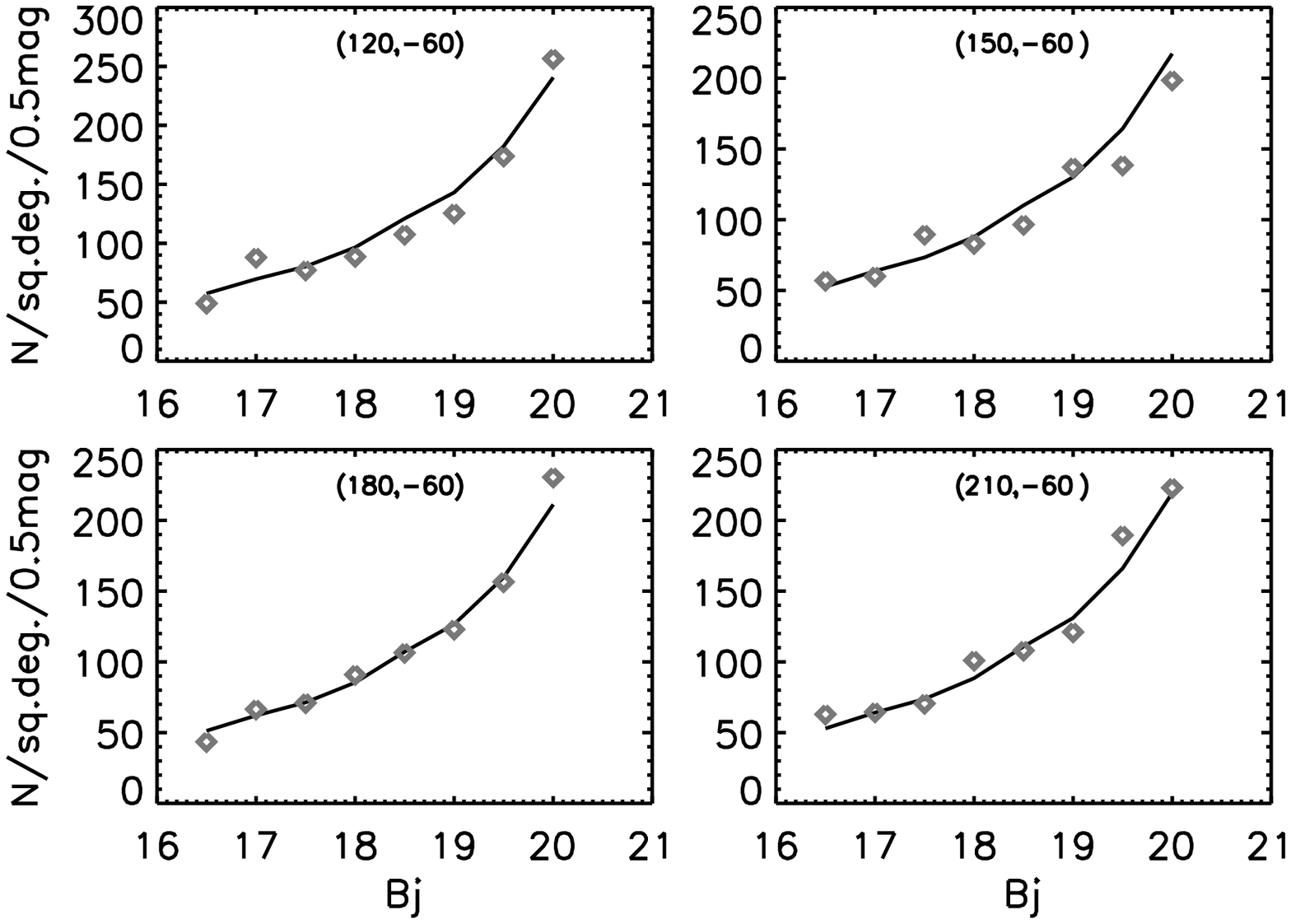}
\vfill
\includegraphics[scale=0.45]{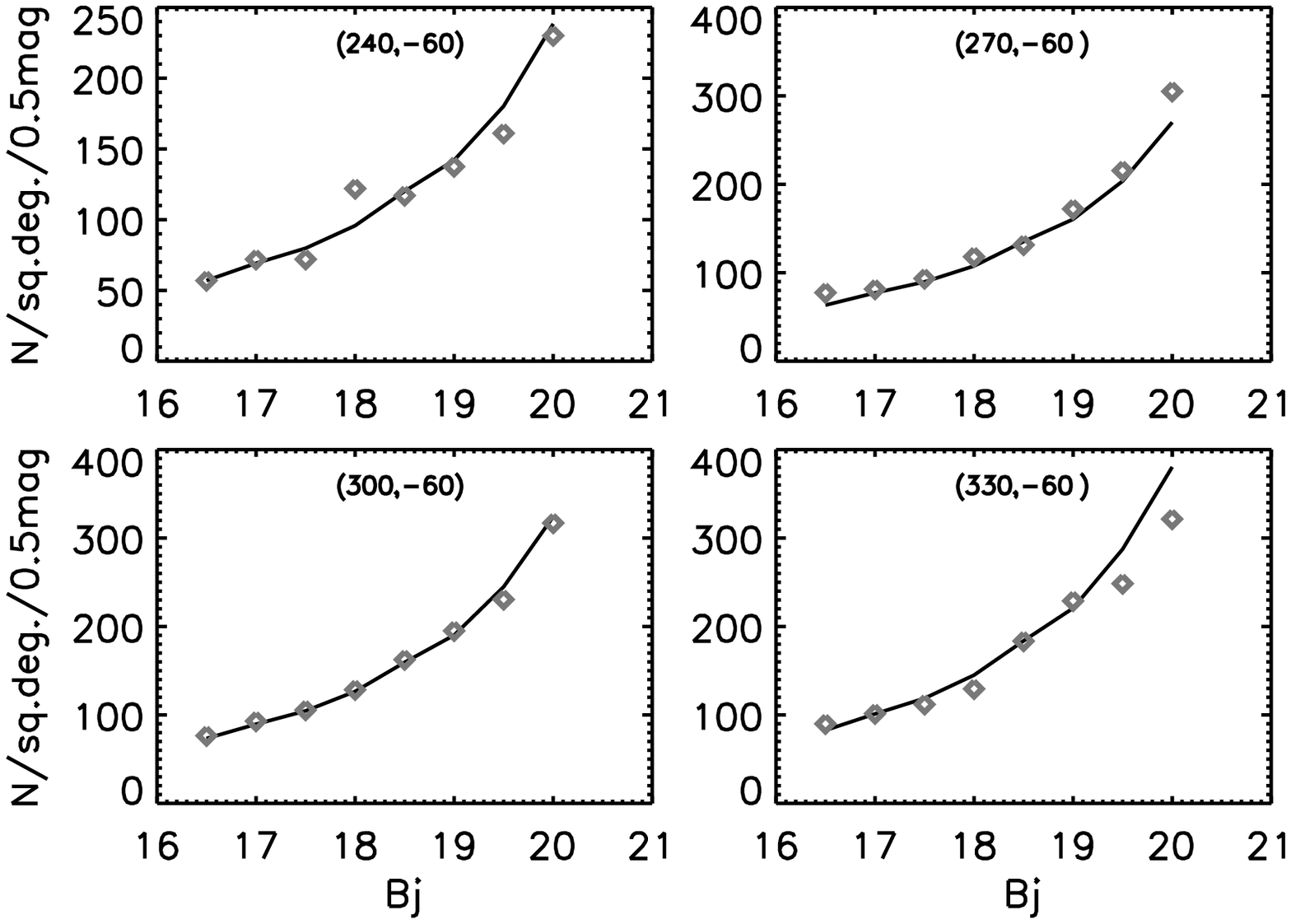}
\vfill \caption{Model fitting to $B_J$ star counts in a selected sky
areas whose Galactic coordinates are indicated in each panel. The
grey dots are observational star counts, and the solid lines are the
theoretical predictions.}\label{fig8}
\end{figure*}

\begin{figure*}
\includegraphics[scale=0.8]{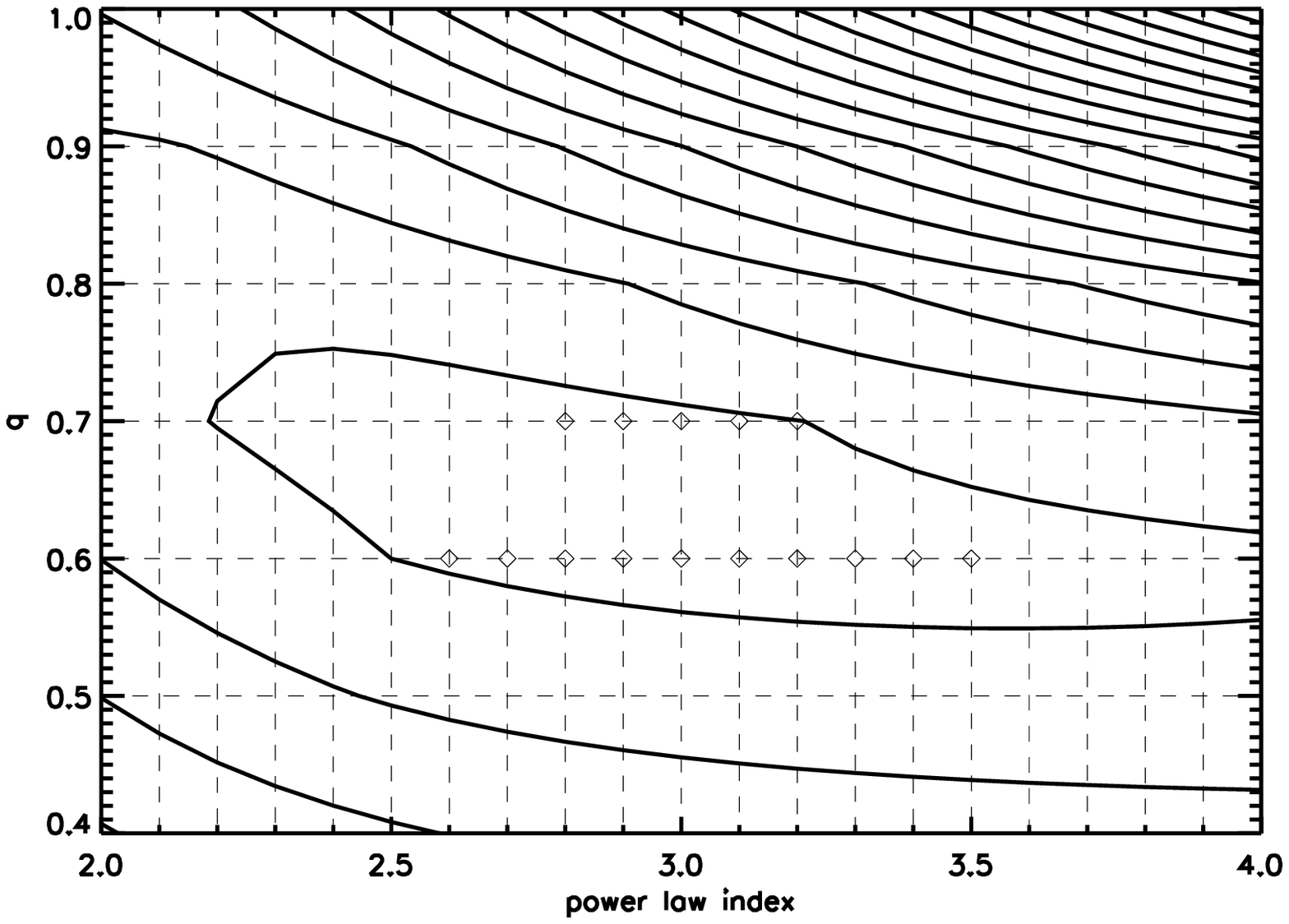}
\hfill
\includegraphics[scale=0.8]{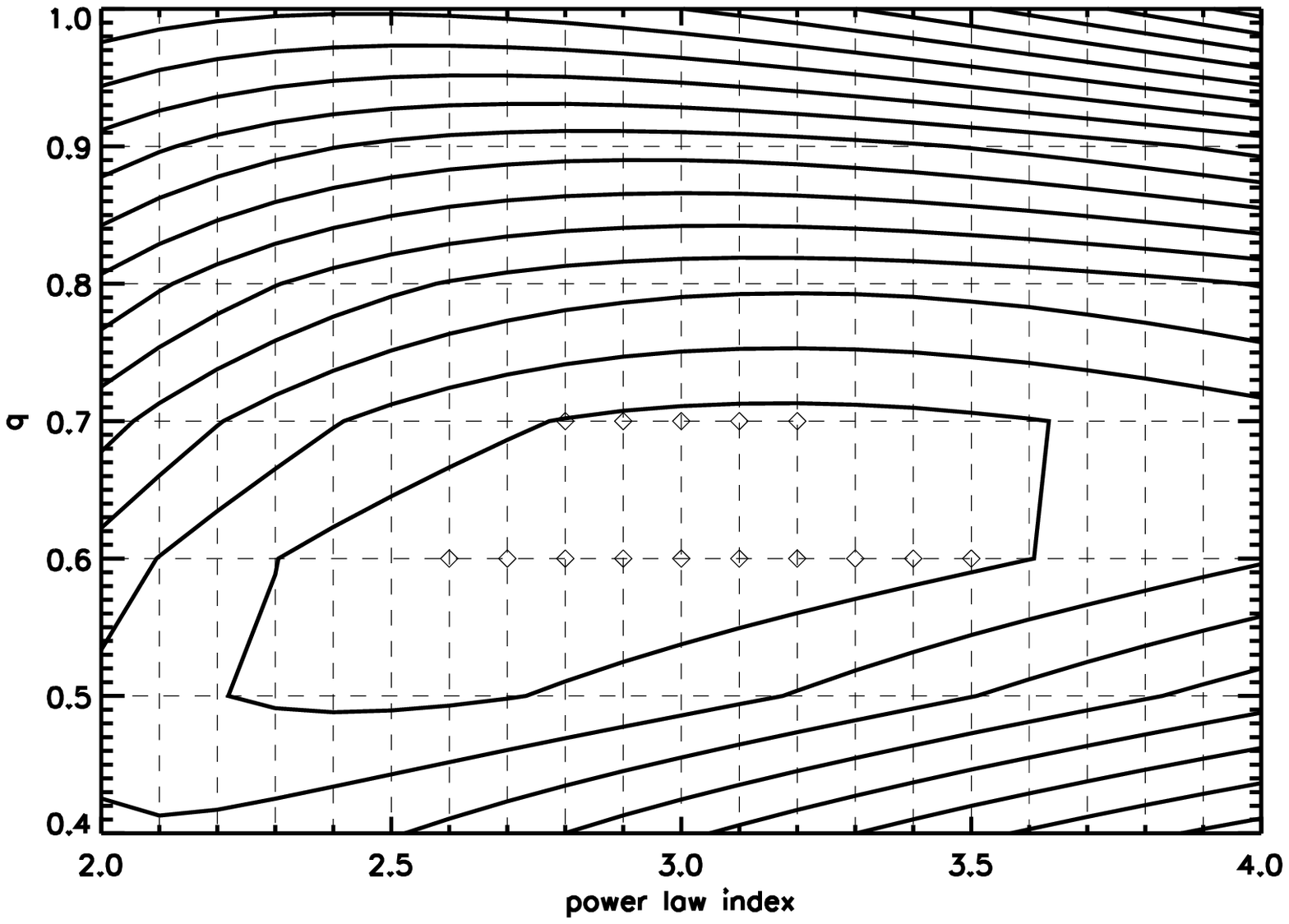}
\vspace{20pt} \caption{The upper panel: Contours of $\chi^2$ of the
theoretical models in power law index (horizontal) and axial ratio
(vertical) plane. The lower panel: Contours of
$\overline{\chi_{bin}^2}$ of the models in the same plane. The
overlapped grids of the minimum contour level are labeled by
diamonds, which define the best fitting model
parameters.}\label{fig9}
\end{figure*}

\begin{figure*}
\includegraphics[scale=0.8]{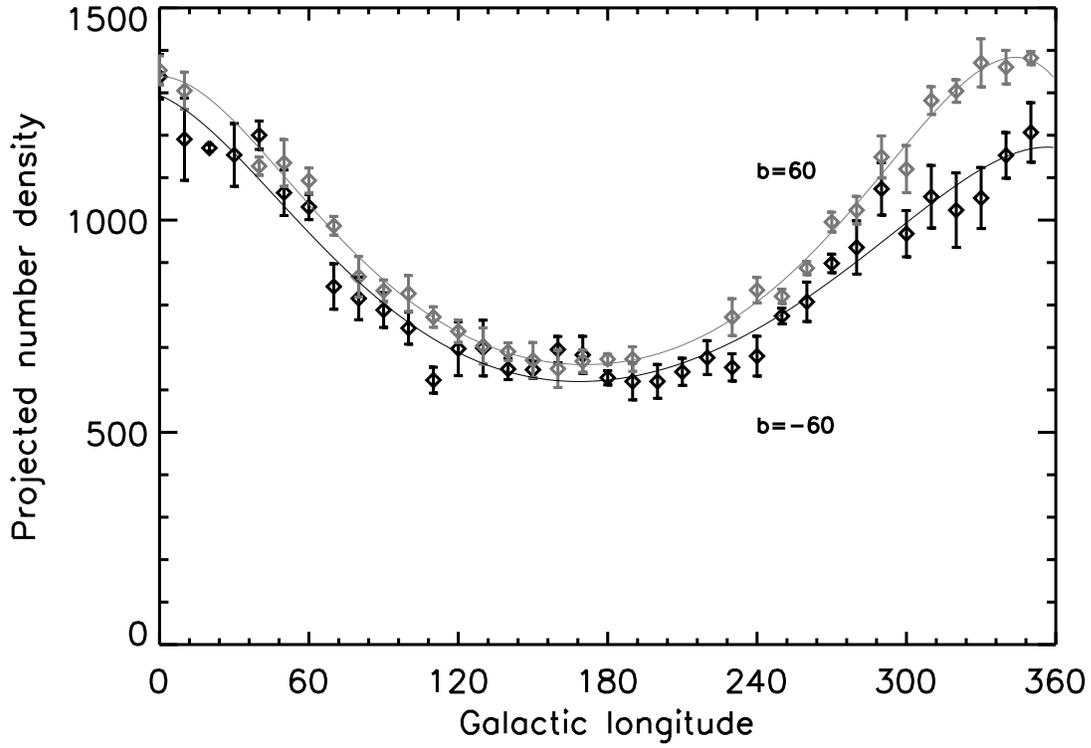}
\caption{A direct comparison between the distributions of the
surface number densities of $b=60^\circ$ sky areas and the
$b=-60^\circ$ ones, with $16^m.5<B_{JSDSS}, B_J<20^m.5$,
$16^m.5<R_{FSDSS},R_F<19^m.5$. The black points and line represent
SuperCOSMOS data of $b=-60^\circ$ and a polynomial fitting curve.
Gray points and line are the corresponding SDSS ones.}\label{fig10}
\end{figure*}

\begin{figure*}
\includegraphics[scale=0.8]{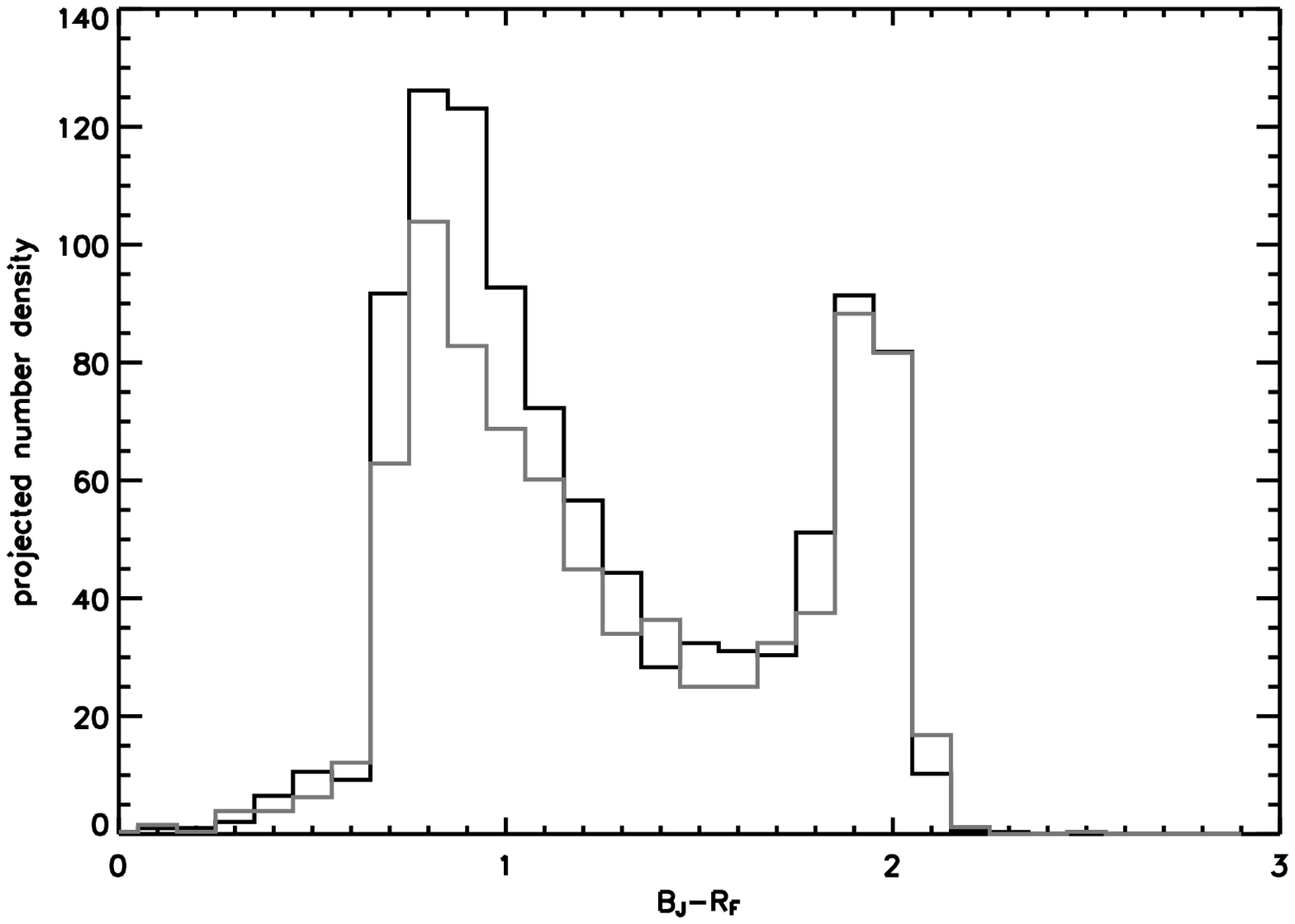}
\hfill
\includegraphics[scale=0.8]{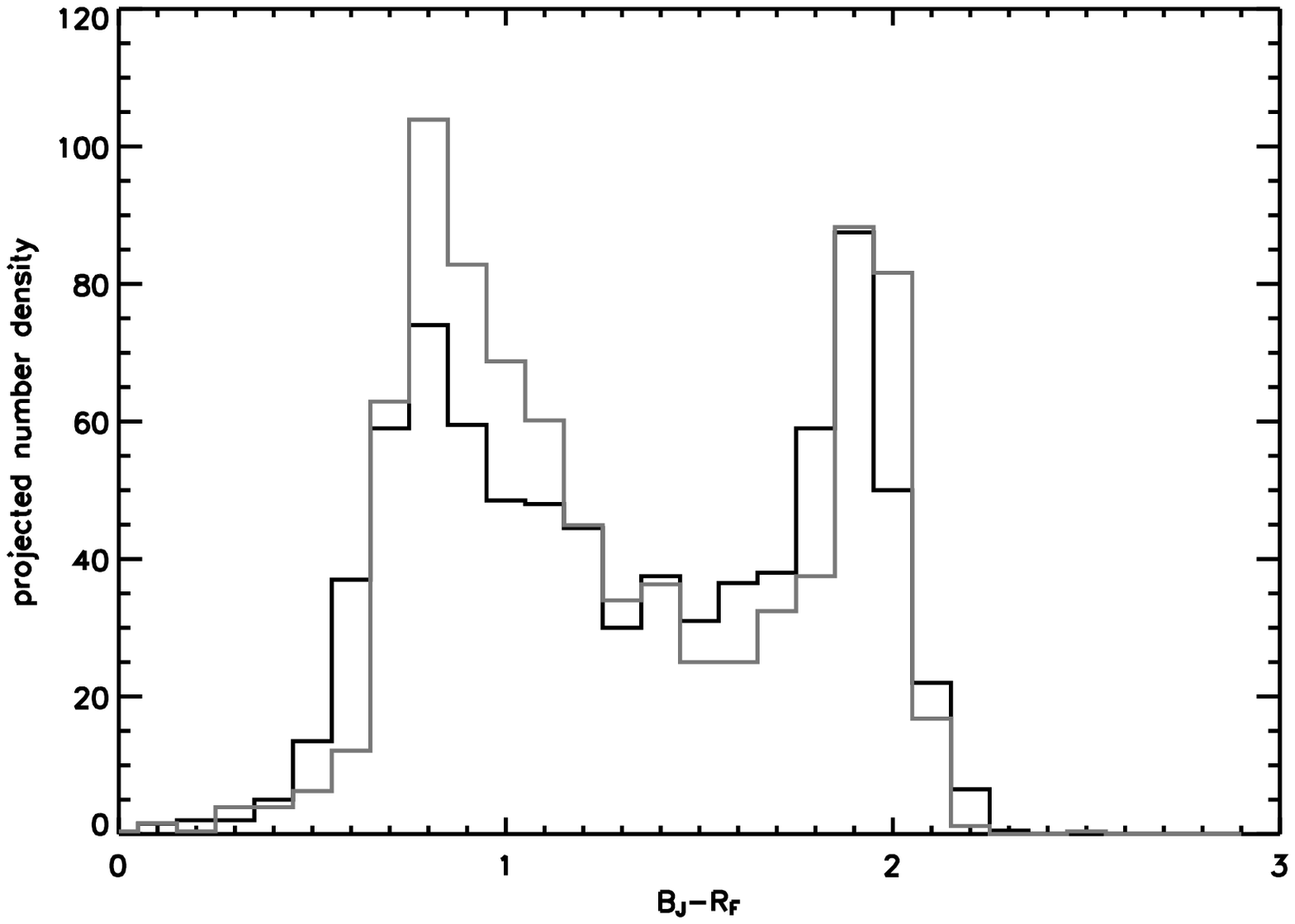}
\vspace{20pt} \caption{The upper panel: The color distribution of
stars in $(l,b)=(90^\circ,60^\circ)$ (the gray histogram) and that
in $(l,b)=(270^\circ,60^\circ)$ (the black histogram) for SDSS
downgraded data with $16^m.5<B_{JSDSS}<20^m.5$,
$16^m.5<R_{FSDSS}<19^m.5$. An overdensity due to halo stars at
$(l,b)=(270^\circ,60^\circ)$ is shown. The lower panel: The same as
the upper panel, but for $(l,b)=(90^\circ,60^\circ)$ (the gray
histogram) of the downgraded SDSS data and
$(l,b)=(90^\circ,-60^\circ)$ of SuperCOSMOS (the black histogram).
An overdensity also due to halo stars in the north
($(l,b)=(90^\circ,60^\circ)$) compared to its symmetric field in the
south ($(l,b)=(90^\circ,-60^\circ)$) is clear visible.}\label{fig11}
\end{figure*}

\begin{figure*}
\includegraphics[scale=0.8]{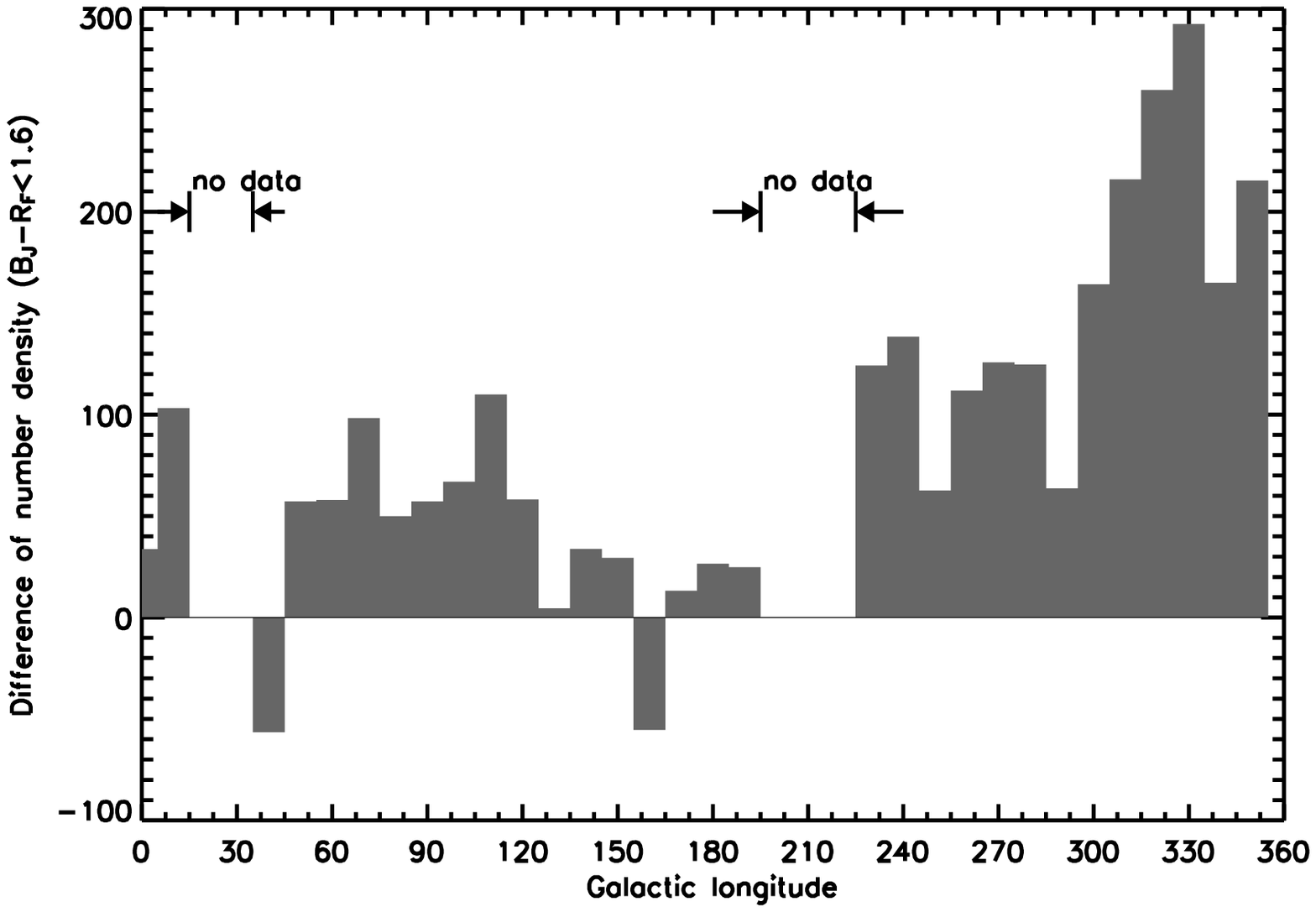}
\includegraphics[scale=0.8]{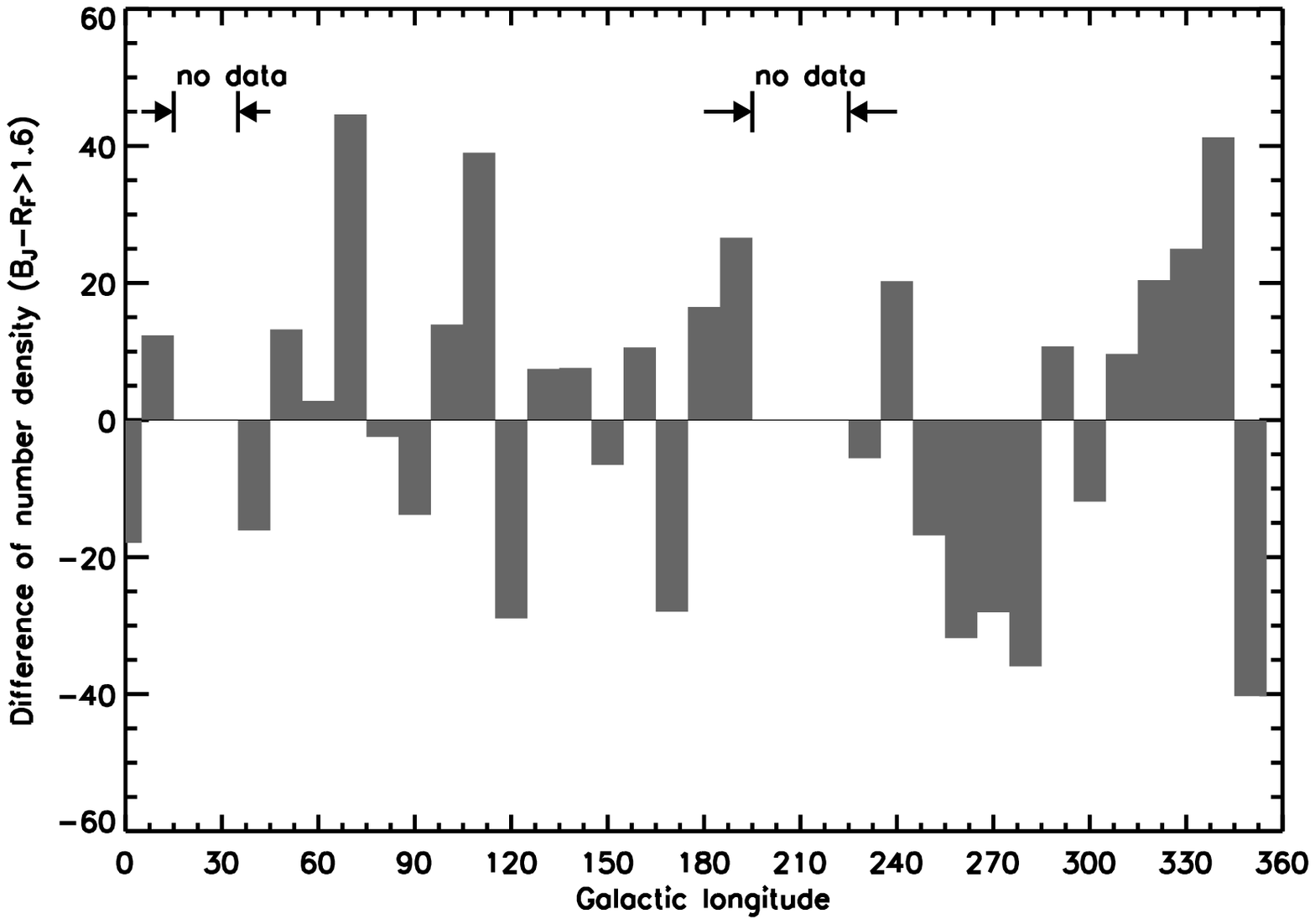}
\vspace{20pt} \caption{The difference of projected surface number
densities between the sky areas along the circles in the north
($b=60^\circ$) and south ($b=-60^\circ$). Stars are selected with
$16^m.5<B_{JSDSS},B_J<20^m.5$, $16^m.5<R_{FSDSS},R_F<19^m.5$. The
upper panel: Data points are constrained by
$0<B_{JSDSS}-R_{FSDSS},B_J-R_F<1.6$ which roughly represents the
halo population. The lower panel: Data points are constrained with
$1.6<B_{JSDSS}-R_{FSDSS},B_J-R_F<3.0$ which roughly represents the
disk population.}\label{fig12}
\end{figure*}

\end{document}